\documentclass{iopart}
\usepackage{graphicx} % Required for inserting images
\usepackage{color}
\usepackage{hyperref}

\usepackage[utf8]{inputenc}
\usepackage[T1]{fontenc}
\usepackage{indentfirst}
\usepackage{physics}
\usepackage{csquotes}

\usepackage{bm} % MB: added for some bold math symbols in odd-odd.tex

\usepackage{ulem} % BB: added for \sout, can be removed in case of problems.

\makeatletter
\long\def\@makefntext#1{\parindent 1em\noindent
 \makebox[1em][l]{\footnotesize\rm$\m@th{{\arabic{footnote}}}$}%
 \footnotesize\rm #1}
\def\@makefnmark{\hbox{${^\arabic{footnote}}\m@th$}}
\def\@thefnmark{\arabic{footnote}}
\makeatother

\usepackage{har2nat}
\hypersetup{breaklinks=true,colorlinks=true,linkcolor=blue,citecolor=blue,filecolor=magenta,urlcolor=cyan}
\newcommand{\newblock}{}
\renewcommand{\harvardurl}[1]{\url{#1}}

\newcommand{\hide}[1]{}

\newcommand{\caH}{\mathcal{H}}

\usepackage{orcidlink}
\usepackage{textcomp}

\begin{document}

\title[Odd and odd-odd nuclei by energy density functionals]{Prospects for modelling odd and odd-odd nuclei using energy density functionals}

\author{A. V. Afanasjev\orcidlink{0000-0002-4220-2386}}
\address{Department of Physics and Astronomy, Mississippi State University, MS 39762, USA}
\author{B. Bally\orcidlink{0000-0002-1802-6802}}
\address{Technische Universit\"at Darmstadt, Department of Physics, 64289 Darmstadt, Germany}
\author{M. Bender\orcidlink{0000-0001-8707-3410}}
\address{Universit{\'e} Lyon 1,
         CNRS,
         IP2I Lyon,
         UMR 5822, 69100 Villeurbanne, France}
\author{J. Bonnard\orcidlink{0000-0002-5164-3343}}
\address{School of  Physics, Engineering and Technology, University of York,  Heslington, York YO10 5DD, United Kingdom}
\author{J. Dobaczewski\orcidlink{0000-0002-4158-3770}}
 \address{School of  Physics, Engineering and Technology, University of York,  Heslington, York YO10 5DD, United Kingdom}
 \address{Institute of Theoretical Physics, Faculty of Physics, University of Warsaw, ul. Pasteura 5, PL-02-093 Warsaw, Poland}
 \author{M. Kortelainen\orcidlink{0000-0001-6244-764X}}
 \address{Department of Physics, University of Jyv\"askyl\"a, P.O. Box 35, FI-40014 Jyv\"askyl\"a, Finland}
 \author{K. Nomura\orcidlink{0000-0001-7663-6775}}
 \address{Department of Physics, Hokkaido University, Sapporo 060-0810, Japan}
 \address{Nuclear Reaction Data Center, Hokkaido University, Sapporo 060-0810, Japan}
%% %
 \author{S. Péru\orcidlink{0000-0002-5858-0314}}
 \address{CEA, DAM, DIF, 91297 Arpajon, France}
 \address{Université Paris-Saclay, CEA, Laboratoire Matière en Conditions Extrêmes, 91680 Bruyères-le-Châtel, France}
 \author{N. Pillet\orcidlink{0000-0002-4465-5549}}
 \address{CEA, DAM, DIF, 91297 Arpajon, France}
 \address{Université Paris-Saclay, CEA, Laboratoire Matière en Conditions Extrêmes, 91680 Bruyères-le-Châtel, France}
 \author{L. M. Robledo\orcidlink{0000-0002-6061-1319}}
 %\email{luis.robledo@uam.es}
 \address{%
 Center for Computational Simulation, Universidad Polit\'ecnica de
 Madrid, Campus Montegancedo, 28660 Boadilla del Monte, Madrid, Spain}%
 \address{Departamento  de F\'{\i}sica Te\'orica and CIAFF,
 Universidad Aut\'onoma de Madrid, 28049-Madrid, Spain}
%% %
 \author{R. Rodr\'{\i}guez-Guzm\'an\orcidlink{0000-0002-4676-5688}}
 %\email{guzman.rodriguez@nu.edu.kz}
 \address{Department of Physics, Nazarbayev
University, 53 Kabanbay Batyr Ave., Astana 010000, Kazakhstan}
 \author{W. Ryssens\orcidlink{0000-0002-4775-4403}}
 \address{Institut d’Astronomie et d’Astrophysique, Université Libre de Bruxelles, Brussels, Belgium}
 \address{Brussels Laboratory of the Universe - BLU-ULB}
 \ead{wouter.ryssens@ulb.be}

 \author{W. Younes\orcidlink{0009-0003-8201-8975}}
 \address{Lawrence Livermore National Laboratory}
 \address{Livermore, CA 94551, USA}

\vspace{10pt}

\begin{abstract}
About half of particle-bound nuclei have an odd number of nucleons. About one-quarter more are odd-odd, meaning they have an odd number of neutrons and an odd number of protons. In this context, it is about time to shift the 70-odd-year focus of theories using nuclear energy density functionals (EDFs) away from the remaining one-quarter of even-even nuclei. This White Paper aims to present open, unresolved problems and promising research directions in nuclear EDFs that were discussed during the 2024 ESNT workshop in Saclay, France, on ``The microscopic modelling of odd-mass and odd-odd nuclei''.

\end{abstract}

\submitto{\JPG}

\maketitle

\tableofcontents{}

\newpage

\section{Introduction}

Odd nuclei, that is, those containing an odd number of nucleons, have long posed a challenge for a microscopic description of their dynamical properties. They lack the energy gap in the spectrum that pairing correlations produce in even-even systems. As a consequence, single-particle excitations intermix with collective ones (rotational and vibrational), complicating the interpretation of the low-energy nuclear spectra. In addition, time-reversal symmetry is not conserved in their wave functions, and the traditional picture of nuclei as composed solely of Cooper pairs also does not hold. These peculiarities explain the lag in modelling the dynamics of odd systems compared with neighbouring even-even ones. Regarding the nuclear force, there is no fundamental difference between a nucleus with an odd number of protons and one with an odd number of neutrons. However, the situation where both proton and neutron numbers are odd (odd-odd nuclei) is far more complicated, as it is shown in the symbolic equation below, which represents the feelings of nuclear scientists when confronted with odd-odd nuclei
\begin{equation}
    \mbox{challenges in odd-odd nuclei} = \left(\mbox{challenges in odd nuclei}\right)^2
\end{equation}
In addition, most of the difficulties discussed in this study also appear in a self-consistent treatment of multi-quasiparticle excitations.

The White Paper summarises the discussion sessions held during the ESNT workshop in Saclay in 2024. In those sessions, contributors presented their views on different aspects of odd nuclei, open problems, and the methods (and tricks) used to perform calculations. The paper is organised by the workshop topics, with their associated sections. Each section or subsection was drafted by one or several authors; therefore, the content and vocabulary may reflect their writing styles and preferences. A typical example of the diversity found throughout the text is the interchangeable use of the terms nuclear density functional theory (DFT) and energy density functional (EDF).
The authors did not attempt to unify the terminology; instead, the text highlights the richness of these differences.

The White Paper avoids extensive review information, which can be found elsewhere, and focuses on prospects for future research. This reflects the contributors' personal perspectives on how to overcome the difficulties encountered and achieve progress. Readers should take these more as a wish list than a personal to-do list. 

\section{Constructing mean-field reference states\label{sec:mfconstruction}}

Within an EDF-based approach, accurate and fully consistent modelling of odd-even and odd-odd nuclei requires constructing one or more mean-field reference states tailored to these systems. A perturbative analysis based on calculations of even-even systems, proposed in earlier works~\cite{(Dug01a),(Dug01b)}, can serve as guidance. The precise number and physical content of such reference states depend on the complexity of the many-body framework employed. Regardless of the method, one must consider Bogoliubov vacua with the correct structure, that is, the appropriate number parity. We briefly recall that a fermionic system with an even (odd) number of particles is described by a Bogoliubov transformation with even (odd) number parity. Depending on the symmetries imposed on the reference states, the number parity is conserved either only for the full wave function of all nucleons or separately for its proton and neutron components. Traditionally, a reference state for an odd-even or odd-odd system is pictured as having one or more quasiparticle excitations, self-consistently blocked, on top of an intermediate, easy-to-handle, fully paired vacuum with even parity~\cite{(Rin80b),(Sch19b)}.

Despite several decades of mean-field calculations based on EDFs, constructing such states remains nonroutine: blocked mean-field calculations can be notoriously difficult to converge. The main issue arises from the blocking procedure itself: the need to select, at appropriate moments in the self-consistency cycle, the quasiparticle content of the Bogoliubov state. This selection is not straightforward for odd and odd-odd nuclei because their level density at equal excitation energy is much higher than that of even-even systems.
This difficulty is compounded by the non-linearity of the mean-field equations: quasiparticle spectra can shift in nontrivial ways from one iteration to the next, especially when accounting for the polarising effects of time-odd mean fields; see section~\ref{sec:eff_int_time_odd}. Neglecting the time-odd mean fields using the equal-filling approximation~\cite{perez-martin2008,(Sch10b)} can increase stability, but it runs counter to the purpose of studying their effects.

In section~\ref{sec:blocking}, we provide a brief overview of the challenges in constructing reference states for odd and odd-odd nuclei. Sections~\ref{sec:directdiag} and~\ref{sec:gradientmethod} introduce the two currently available iterative strategies for blocked mean-field calculations: direct diagonalisation and the gradient method, respectively. Finally, in section~\ref{sec:outlook}, we discuss why neither current implementation of these options is fully satisfactory and what alternatives could be developed.

\subsection{Self-consistency and blocking}\label{sec:blocking}

The generation of the reference states of interest is typically achieved through self-consistent mean-field calculations based on the HFB formalism~\cite{(Rin80b), Bender2003a}. A complete and successful HFB calculation requires three standard interlocking steps~\cite{(Rys19b)}:

\begin{enumerate}

\item diagonalize a given (fixed) HFB Hamiltonian $\mathcal{H}$,
\item construct a Bogoliubov vacuum based on the eigenvectors of $\mathcal{H}$,
\item update the mean-field potentials and construct a new $\mathcal{H}$.
\end{enumerate}

Provided that appropriate numerical strategies are adopted in each case, repeating these steps will lead to a self-consistent solution. Understanding the (non-)convergence of such an iterative process for the nonlinear HFB equations is not easy and generally requires numerical experimentation. Only under the assumption that one is reasonably `close' to the solution can linear-response arguments be used and shed light on the problem~\cite{payne1992}. Nevertheless, several public codes are currently available~\cite{(Sto05),pei2014,(Mar14),DIRHB-code.14,(Rys15),(Dob21f),(Mar22a),chen2022} that can relatively robustly, that is, without much human tweaking of numerical parameters, converge time-reversal-conserving configurations for even-even nuclei.

Similar robustness has so far not been achieved for odd and odd-odd nuclei~\cite{AO.13,(Sch10b)}, mostly because executing step (ii) for a fixed $\mathcal{H}$ is much more difficult than it appears~\cite{Bertsch2009}. In fact, most of the existing literature focuses on the (possibly numerically demanding) step (i) but omits discussion of step (ii). More precisely, the eigenvectors of $\mathcal{H}$ come in pairs with opposite eigenvalues/quasiparticle energies ($E^{\rm qp}_i, -E^{\rm qp}_i $); constructing a Bogoliubov transformation or vacuum requires choosing one as a quasiparticle excitation. Because this choice must be made for each pair of eigenvectors, many possible Bogoliubov vacua can be constructed. Selecting the positive ones, $E^{\rm qp}_i>0$, is a natural choice and usually yields the lowest total energy, but this is neither guaranteed nor does it always lead to a self-consistent solution. For even-even nuclei, nuclear pairing ensures that the quasiparticle energies have a minimal size -  typically one has $\left|E^{\rm qp}_i\right|$ greater than about 1\,MeV -- such that the unique energy-optimal Bogoliubov state is separated from all other possibilities by a sizable energy gap. Even if this holds for most applications to even-even nuclei, it can break down when time-reversal is broken~\cite{goodman1980a,tanabe1982, Bertsch2009,allard2024a}.

The situation is fundamentally different for odd and odd-odd nuclei: there is no lower bound on the absolute values of quasiparticle energies, and thus one must choose a Bogoliubov vacuum from among a larger set of nearly degenerate candidates. A suboptimal selection nearly always causes the iterative process to fail to converge; the most striking example is flip-flopping iterations, in which the computer program alternates between two nearly degenerate blocked configurations, even though the nature of the blocked quasiparticle differs in each. These convergence difficulties are entirely analogous to those encountered in the density functional theory for molecules and condensed matter with (some degree of) spin polarisation~\cite{woods2019}.

Two main strategies exist for solving the HFB equation: one is based on the ``fixed point theorem'' of nonlinear mathematics, which supports an iterative procedure in which the solution from the previous iteration is used to construct the relevant elements of the current iteration. We will call this the direct diagonalisation procedure; it has been widely used within fixed-point iterative algorithms. The other strategy is to use the variational principle to find solutions to the HFB equations by minimising the HFB energy in the multidimensional variational space spanned by all possible HFB states; the simplest approach is the gradient method. In practice, both approaches work reasonably well for time-reversal-conserving configurations typical of even-even nuclei, provided the method parameters are chosen wisely. In favour of the gradient method, one can cite its ability to handle multiple constraints. In what follows, we discuss the (dis)advantages of both methods for odd and odd-odd systems. 

\subsection{Direct diagonalisation}\label{sec:directdiag}

Of the two methods, direct diagonalisation is the most widely used~\cite{(Sto05),A250, AO.13,(Dob21f),(Mar22a),ryssens2016,dubray2025,(Kor26a)}: at
each iteration, one (i) performs a complete diagonalisation of the HFB Hamiltonian $\mathcal{H}$ and (ii) explicitly constructs an appropriate Bogoliubov vacuum. Assuming computational resources suffice, the diagonalisation is straightforward. As discussed above, constructing the Bogoliubov transformation is not straightforward, and various codes use several strategies for selecting quasiparticles. Nearly all of these rely on calculating overlaps between a reference state and the quasiparticle wavefunctions; by blocking the quasiparticle state(s) with maximal overlap, one hopes to obtain a stable iterative scheme. Several choices of reference states have been documented, examples include (i) Nilsson states with associated quantum numbers~\cite{A250,AO.13}, (ii) the quasiparticle wavefunctions from the previous iterations~\cite{A250,AO.13}, (iii) the states that diagonalise the single-particle Hamiltonian~\cite{(Dob09g),ryssens2016} and (iv) single-particle states obtained from a (fully converged) calculation of a polarised semi-magic even-even nucleus in the same region of the nuclear chart referred to as `tagging'~\cite{(Bon23c),(Dob26g)}.

The possibility of selecting the desired quasiparticle(s) is the main advantage of direct diagonalisation: it allows the practitioner to construct a variety of mean-field configurations for the same odd or odd-odd nuclei, which enables, e.g., a detailed investigation of low-lying spectra~\cite{(Sch10b),(Dob26g)} or the deployment of the projected generator coordinate method (PGCM)~\cite{(Bal14d),bally2021,(Bal23)}.

Unfortunately, the quasiparticle selection performed at each iteration is also a significant source of iterative instability. None of the selection strategies known to us has proven to be truly robust. Often, their application is only possible at the cost of significant human fine-tuning of numerical parameters on a case-by-case basis. Very little is known about their applicability to multi-quasiparticle excitations, e.g., to 3qp or 2qp excitations in odd or odd-odd nuclei, respectively. Nevertheless, the direct diagonalisation method can still be deployed in systematic studies: careful simulation setup can yield large numbers of simulations~\cite{(Sch10b),AS.11,DABRS.15,A250} with success rates of up to 96\%~\cite{(Dob26g)}. As far as stabilising measures go, we mention in particular (i) imposing self-consistent symmetries -- axial symmetry in particular -- such that quasiparticles can be labelled with additional quantum numbers, (ii) including a tiny cranking frequency to lift the degeneracy between time-reversal partners~\cite{(Sch10b)} and (iii) using the tagging technique~\cite{(Dob26g)}.

\subsection{The gradient method}
\label{sec:gradientmethod}

Instead of looking for fixed points of the HFB equations, for odd systems (and multi-quasiparticle configurations of even ones), one can find the minima of the HFB energy of the target system~\cite{Robledo2013,Robledo2024}). Finding minima of multidimensional functionals is a well-known procedure in numerical analysis (gradient descent in mathematical optimisation) that requires computing the energy gradient at successive points in the multidimensional space. Given the gradient and the initial point, the next point on the path towards the minimum is obtained using the Thouless theorem after choosing a convenient, usually small, step size for the gradient. The gradient step is fully determined if the second derivatives of the functional are available (Newton method). However, the matrix of second derivatives (the Hessian) is the same as the matrix that enters the linear response and is therefore costly to compute. Instead, one may approximate the Hessian evaluation, such as with the conjugate gradient method~\cite{(Egi95)}, or use only the diagonal part of the matrix~\cite{robledo2011}. A much cheaper alternative is to use a small variable step size and adjust it during the iterative process, depending on the evolution of the energy. In addition, one can supplement the first-order gradient method with a momentum term to accelerate convergence~\cite{ryssens2019}.

Compared to direct diagonalisation, the gradient method is less widely used. To the best of our knowledge, only two regularly used computer codes employ itfor blocking calculations: the Gogny solver ATB~\cite{Robledo2013} and the Skyrme solver MOCCa~\cite{ryssens2016}. Although ATB can be obtained via a reasonable request to its author, only MOCCa is publicly available at the moment~\cite{Ryssens2026}. We also note that (i) a first calculation using a Gogny EDF with the open-source code TAURUS was recently reported by~\citeasnoun{(Fue26)}, although this version is not yet publicly available and (ii) HFB3 can use the gradient descent method, but is restricted to the equal filling approximation~\cite{dubray2025}.
 
The strength of the gradient method is that the gradient update relates to the Thouless theorem, which guarantees that the transformed and original vacua are continuously connected and -- provided the step size is not too large -- cannot differ dramatically.
One consequence is that the gradient method self-consistently preserves the number parity of the Bogoliubov transformations. This means that throughout the iterative procedure, all vacua constructed have the same number parity as the original vacuum used to initialise the calculation. Therefore, the blocking procedure has to be performed only once at the start of the iterative minimisation. This is in contrast to the diagonalisation method in which such blocking has to be performed explicitly at each iteration.

This robustness means that blocked calculations for many nuclei can be fully automated, to the point where fitting a Skyrme parametrisation to the masses of thousands of nuclei, including time-reversal breaking, is feasible~\cite{(Rys22b)}. The gradient method requires more iterations for odd nuclei than for even-even systems, but future improvements are likely and much needed.

The downside of the gradient method is its variational nature: the resulting configuration is the lowest-energy one that is not orthogonal to the initial Bogoliubov vacuum. In other words, many different initial choices of the Bogoliubov vacuum will lead to an identical final state, so the diversity of reference states that can be constructed is limited. A convenient partial remedy to this issue is to impose self-consistent symmetries: the gradient method can yield one blocked, energy-optimal mean-field state for each combination of quantum numbers.

If this is not enough, one can explicitly impose orthogonality between the iterated Bogoliubov vacuum and one or more solutions obtained in earlier runs. We remark that the orthogonality condition can only be enforced approximately, up to a given numerical accuracy. Therefore, we must maintain the constraint throughout the entire minimisation procedure, not only apply it once when setting up the calculation. Although such a procedure does not present a formal problem, it does become unwieldy in large single-particle bases because (i) a tiny overlap between Bogoliubov vacua can be made out of many small overlaps between single-particle states and (ii) the memory cost of storing multiple solutions can become prohibitive.

An alternative approach could be to use the gradient method solely to generate the self-consistent minimum at each point of the (collective) multidimensional energy surface and then construct other quasiparticle configurations through perturbative blocking on top of these minima. This approach might be advantageous in multi-reference frameworks such as the PGCM, where the configuration mixing of symmetry-projected states does not necessarily favour states with lower HFB energy. In fact, this method has already been applied to even-even nuclei and two-quasi-particle excitations~\cite{Chen2017a}. A similar situation, using schematic interactions, appears in the projected shell model (PSM) approach \cite{Sun_2016}.

One appealing characteristic of the gradient method is its ability to handle constraints, including those arising from overlaps. This characteristic opens the possibility of increasing the number of blocking configurations generated by imposing large overlaps with target configurations \citeasnoun{carpentier2024,trilogy1,trilogy2,trilogy3,TPaul}.

\subsection{Outlook}
\label{sec:outlook}

The characteristics of direct diagonalisation and the gradient method have important consequences for practical applications. The gradient method provides a smooth iterative evolution and can be made extremely robust, but it is limited in its versatility. Direct diagonalisation can theoretically construct a far richer set of reference states for odd and odd-odd nuclei, but in practice it is hindered by convergence problems that can be alleviated, though not eliminated.

This means neither strategy can meet the community's need to take the treatment of odd and odd-odd systems to the next level. An iterative scheme that allows for obtaining a rich diversity of mean-field states in a way that is sufficiently robust for an automated deployment would benefit essentially all research lines proposed in this White Paper, from deploying advanced many-body techniques (section~\ref{sec:configurationmixing}), to fitting parameterisations sensitive to time-odd observables (section~\ref{sec:eff_int_time_odd}), building reduced spaces for collective modelling (section~\ref{sec:collective}) and global comparisons of results with experimental data (sections~\ref{sec:observables} and~\ref{sec:oddodd}).

We thus call for the development of new algorithms that (i) rival the robustness of the gradient method yet (ii) can generate a diverse set of states comparable to that of direct diagonalisation. Ideally, these techniques would further accelerate the iterative procedure and alleviate the numerical burden of blocked HFB calculations. The community might look to the quantum chemistry literature, where innovative solutions to similar problems have been proposed~\cite{hait2020,rossi2026a}.

Finally, we note that only four publicly available codes -- {\sc hfodd}, {\sc hfbtho}, {\sc HFB3} and {\sc MOCCa} -- offer blocking capabilities~\cite{(Dob21f),(Mar22a),Ryssens2026,dubray2025}. These codes represent only a fraction of the blocking techniques in regular use; for the community to advance, both current and future implementations must be publicly accessible.

\newpage
\section{The role of symmetries}

\subsection{Impact of the orientation of the angular momentum of blocked nucleons relative to the nuclear shape\label{alispin}}

In an HFB solver that imposes axial symmetry around the $z$ axis, the wave 
functions of all single-particle and quasiparticle states are by 
construction eigenstates of $\hat{\jmath}_z$ and therefore aligned 
with the symmetry axis. This leaves no possibility for the orientation of the observable many-body angular momentum of the blocked quasiparticle vacuum other than pointing along the symmetry axis as well. 

The situation is much more complicated in the full Cartesian geometry 
of three-dimensional (3D) codes. In a completely symmetry-unrestricted calculation, the many-body 
angular momentum vector $\bm{J}$ of a blocked and possibly also 
cranked quasiparticle state can point into any direction relative to a
deformed nuclear shape. The minimisation of the total energy then favours 
one specific orientation of $\bm{J}$ that might not be aligned with 
any of the major axes of the density distribution.

\begin{figure}[t!]
\centerline{\includegraphics[width=0.5\textwidth]{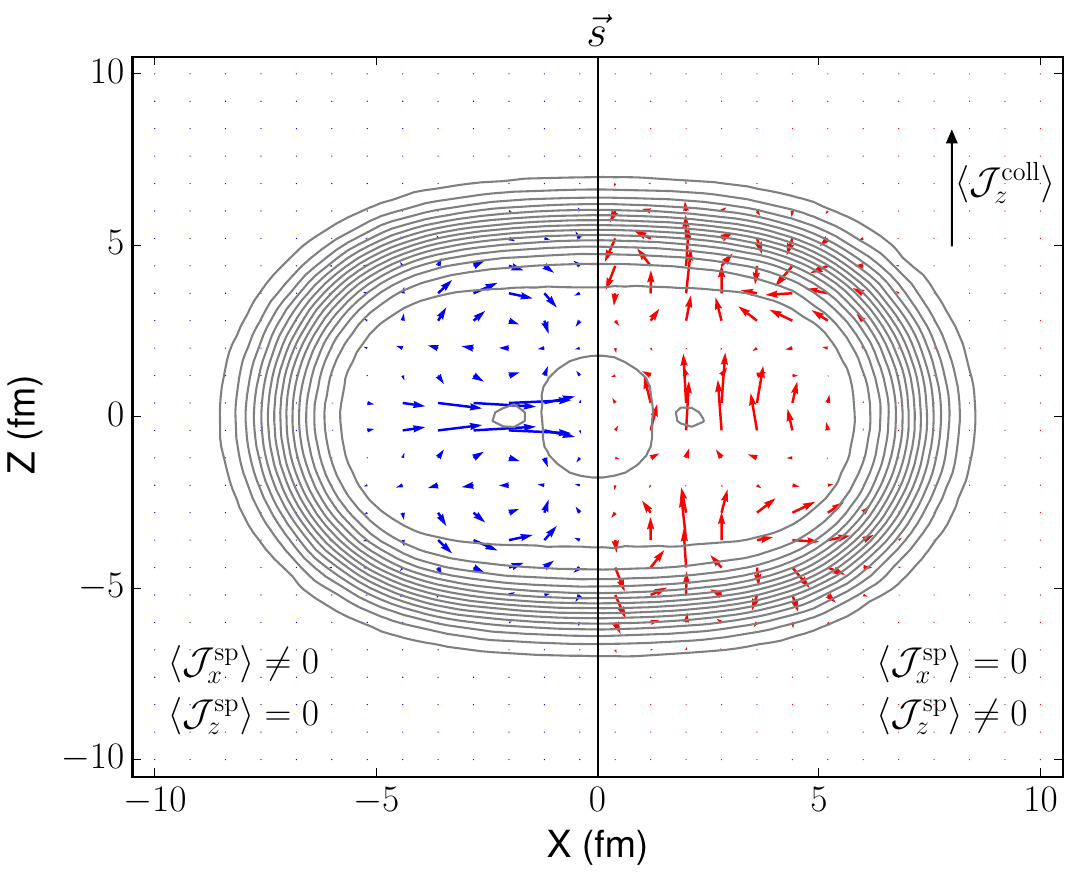}}
\caption{\label{fig:alispin1}
Caption: Cross-section through the normal density (indicated by isolines) 
and spin density (indicated by the arrows) of two signature-conserving 
self-consistent solutions for the lowest blocked one-quasiparticle 
configuration of an odd nucleus with different orientation of angular
momentum relative to the long axis of the prolate nuclear shape that 
is arbitrarily set to be the $x$ axis. The left and right panels show densities for signature symmetry conserved with respect to the $x$ and $z$ axes, respectively. The spin distribution exhibits quite different patterns, whereas the normal densities remain nearly identical.
}      
\end{figure}

The 3D solvers, however, in general impose some point-group symmetry on the 
single-particle states~\cite{(Dob00a),(Dob00b)}, which again limits the 
possible orientations of $\bm{J}$ relative to the shape. Unlike the case of axial symmetry, a 3D code can still find several different solutions
that are not equivalent, and that do not necessarily correspond to true
minima that would be found in a symmetry-unrestricted calculation. The issue is most straightforwardly illustrated for a combination of point-group 
symmetries that impose three plane-reflection symmetries on the one-body 
densities and currents as often done. Being a pseudovector density, some 
of the components of the one-body spin density $\bm{s}(\bm{r})$ 
change sign under some, but not all, plane reflections, which for the specific choice of symmetry operators used in~\citeasnoun{RyssensN2LO} leads 
to the relations
\begin{eqnarray}
s_x(x,y,z) & = & - s_x(-x,y,z) = + s_x(x,-y,z) = - s_x(x,y,-z) \, ,  \\
s_y(x,y,z) & = & + s_y(-x,y,z) = - s_y(x,-y,z) = - s_y(x,y,-z) \, ,  \\
s_z(x,y,z) & = & + s_z(-x,y,z) = + s_z(x,-y,z) = + s_z(x,y,-z) \, . 
\end{eqnarray}
Analogous relations, but with opposite sign, apply to the components of the 
vector current density $\bm{j}(\bm{r})$. This has to be contrasted
with the normal density that is reflection-symmetric in all three planes. 
Together, these relations imply that, within such solvers, a 
time-reversal-breaking quasiparticle state obtained for one specific 
orientation of the nucleus cannot be represented in a different 
orientation, because the pattern of the time-odd densities is not invariant 
under axis exchange. An illustration is given in figure~\ref{fig:alispin1} 
for a prolate axial shape and a blocked one-quasiparticle state obtained for 
signature symmetry\footnote{The signature operator is defined as rotation by $\pi$ about a given axis.} conserved with respect to the prolate axis. In this configuration, the angular momentum vector is aligned with the prolate axis, as in the cut used to prepare the normal and spin densities shown in the left panel. 
The right panel illustrates the solution obtained when the signature symmetry is conserved with respect to the axis
perpendicular to the long axis. In this case, the angular momentum vector aligns with that perpendicular axis, resulting in a very different spin-density pattern.

As pointed out by~\citeasnoun{(Sch10b)}, this issue can be associated with the concept
of \textit{alispin} that establishes a unitary transformation that can be used to characterise
how the observables of a nuclear state are modified when changing the direction in which the angular momentum vector is aligned. 

That the orientation of angular momenta relative to the nuclear shape is fixed by the setup of numerical calculations has several practical consequences that are specific 
to time-reversal-breaking calculations.
Most importantly, in a 3D code that imposes point-group symmetries, orienting the shape differently within the numerical box may, at convergence, lead to different angular momentum vs.\ shape orientations.
Then, all observables calculated from time-odd densities take 
different values. For reasons of polarisation effects induced by the time-odd mean fields, 
all other observables are slightly affected too. For blocked one-quasiparticle states, depending on the size and sign of time-odd coupling constants and the 
properties of the blocked quasiparticles, total binding energies can change on the 
order of a few tens to even a few 100\,keV, see examples in~\citeasnoun{(Sat12a)}.

There seems to be no general rule which relative orientation yields the largest binding energy. For 
multi-quasiparticle states or cranked blocked configurations, the lowest configuration 
might not even be representable within the symmetry choices offered by a given solver. 
In particular, a conserved signature, 
$\hat{R}_\mu \equiv e^{\mathrm{i} \pi \hat{J}_{\mu}}$,
$\mu = x$, $y$, or $z$, has the practical consequence that angular momentum points 
into the $\mu$ direction, which coincides with a major axis of the 
nuclear shape. Even when looking in a signature-conserving code for the minimum 
of triaxial configurations within a range of 180$^\circ$ in Bohr $\gamma$ deformation instead of 
the 60$^\circ$ that are sufficient for time-reversal-conserving states of 
even-even nuclei, one only covers the three distinct cases of aligning $\bm{J}$
with the major axes of a triaxial shape, but still excludes those configurations 
for which $\bm{J}$ is not aligned with a major axis of the shape~\cite{(Shi13)}.

In practice, however, most calculations are usually limited to a single choice 
for the relative orientation, which often also depends on the physics case. For 
the calculation of the ground state of low-lying non-collective quasiparticle 
excitations of axial or near-axial nuclei in a 3D solver that offers 
principal-axis cranking, it is often advantageous to align the symmetry axis with 
the direction of the cranking constraint, which can then be used to suppress the 
possible mixing of quasiparticles with the same conserved quantum numbers of 
the 3D code, but a different projection of angular momentum on that axis. 
By contrast, for principal-axis cranking calculations of collective rotational bands of 
odd nuclei, it is mandatory to align one of the other axes  with the axis of
collective rotation that, at low total angular momentum, is often perpendicular to the angular momentum of the blocked quasiparticle.

Using symmetry-unrestricted solvers that offer the possibility to orient $\bm{J}$ into any direction and letting the variational principle decide instead 
generates several new practical problems, the most obvious of which is that such solvers are more laborious to set up, more difficult to converge, and more 
time-consuming to run.

For fixed phenomenological mean-field potentials such as Nilsson or Woods-Saxon~\cite{(Rin80b)}, researchers can readily implement the relative orientation of the cranking axis and the nuclear shape, as previously studied~\cite{(Fri87a),(Sat05a)}. As discussed above, to explore all possible principal-axis orientations, the cranking axis is usually fixed in space along one of the Cartesian axes, and the nuclear shapes are varied by exploring a range of Bohr $\gamma$ deformations spanning 180$^\circ$, for instance, $-120^\circ\leq\gamma\leq60^\circ$. For odd and odd-odd nuclei, the orientation of the angular momentum is then fixed by the blocked quasiparticle(s). A conserved signature symmetry along the cranking axis enforces parallel cranking and angular-momentum directions, with the cranking axis aligned with one of the shape's principal axes. For a broken signature, one can then consider the so-called tilted-axis cranking~\cite{(Fra93b)}, in which the cranking axis departs from one of the shape's principal axes. Technically, in either case, obtaining the relevant solutions is relatively straightforward, since the shape's orientation and the cranking axis's orientation can be controlled independently.

Within self-consistent approaches discussed above, the problem becomes more complicated because (i) fixing the angular momentum orientation with respect to the shape can only be controlled by constraints, and (ii) for broken symmetries, the entire nucleus can be differently oriented in space with no restoring force. As shown by~\citeasnoun{(Ker81b)}, stationary cranking rotation occurs only when the angular momentum aligns with the cranking axis, which can be tilted relative to the shape. Numerical solutions of that type have been obtained by~\citeasnoun{(Shi13)}. However, it is still unclear how this theorem extends to odd and odd-odd systems in which non-zero angular momentum is present at zero cranking frequency. In addition, similarly to phenomenological models \cite{PhysRep-SBT,F-rev.01},
a conserved signature symmetry requires exploring the full 180$^\circ$ range of $\gamma$ deformations, which was not performed in the triaxial applications to date~\cite{(Bal14d),(Bor16),(Bor17),(Bal22a),(Bal23),(Rys22b)}. Extension of triaxial calculations in this direction is warranted.

\subsection{Impact of symmetry restoration in a variation after projection framework}
\label{sec:sym:vap}

The low-energy spectrum of odd nuclei (the same applies to odd-odd nuclei and quasiparticle excitations) often contains many non-collective states within a narrow energy window of a few hundred keV. Because their excitation energies are comparable, the ordering of the spectrum is very sensitive to fine details in both the interaction and the many-body method used for the calculations. One aspect that has received little attention in the literature is the impact on the spectrum of considering correlation energies associated with symmetry restoration. To understand the situation, one can use the rotational energy correction in the strong deformation limit $E_\textrm{Rot}=\langle \Delta \vec{J}^2 \rangle / ( 2 \mathcal{J}_\textrm{Yocc} )$, a quantity that can be as large as several MeV and that strongly depends upon the Yoccoz moment of inertia. Different blocked configurations will have different moments of inertia (the magnitude of the moment of inertia is strongly correlated with the amount of pairing correlations) and therefore different values of $E_\textrm{Rot}$ that, when considered, will lead to a (often substantial) reordering of the spectrum.  

\begin{figure}[t]
\centerline{\includegraphics[width=0.78\textwidth]{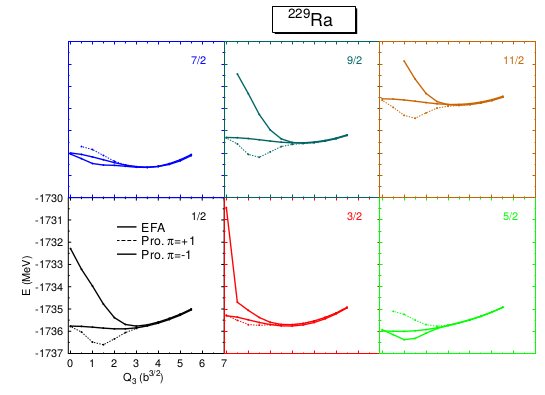}}
\caption{HFB energy of blocked configurations (at the EFA level) as a function of octupole moment. Axial symmetry is preserved; therefore, each blocked configuration can be labelled by the $K$ quantum number. Parity-projected energies (both positive and negative parity) are also depicted. The results shown correspond to a calculation with Gogny D1S in $^{229}$Ra.\label{229RaPESProj}}
\end{figure}

Projection can be carried out in the variation-after-projection (VAP) or projection-after-variation (PAV) setup. In the former, the mean-field intrinsic state is determined by minimising the projected energy, whereas in the latter, it is determined by minimising the energy of the unprojected mean-field intrinsic state and then performing the projection. The VAP is far more advanced than the PAV in capturing correlations beyond the mean field, but is much more difficult to implement. To better understand the differences between the two approaches, one can consider parity projection in heavy systems~\cite{(Rob26)}. Here, instead of the full VAP, a simpler-to-implement restricted VAP (rVAP) was used. 

In the rVAP, we obtained the lowest quasiparticle configurations as a function of the relevant driving coordinate (in this case, the octupole moment), as well as the parity-projected energies. One obtains the rVAP solution by looking at the minima of the different curves. In the past, an analogous approach was used by~\citeasnoun{(Zdu07c),(Sab07),(Rob11a),(Bal14d),(Bal23),(Bor16)}, see also section~\ref{sec:MAP}. We also note that, for parity-symmetry restoration, the exact VAP method is relatively easy to implement~\cite{(Oht04)}, making it an interesting area for future research.

\begin{figure}[t]
\includegraphics[width=0.48\textwidth]{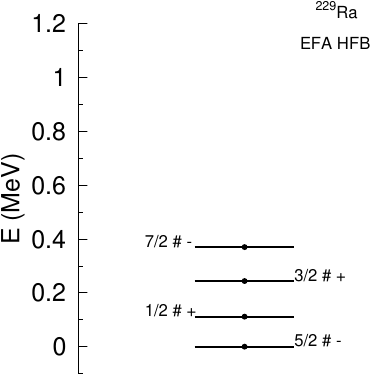}%
\includegraphics[width=0.48\textwidth]{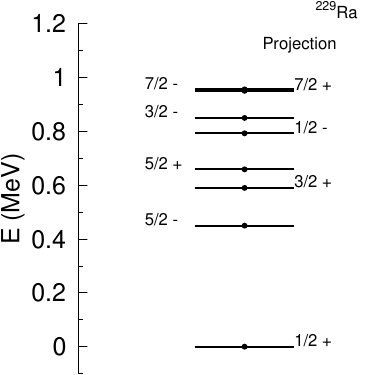}%
\caption{Low-energy spectrum of $^{229}$Ra computed within the EFA HFB mean-field framework (left) and including parity projection (right). In the left EFA HFB panel, states with parities preceded by a hash symbol are not reflection-symmetric, and parity is not a good quantum number. Only its mean value is defined.\label{229RaSPEC} }
\end{figure}

An example is shown in Fig \ref{229RaPESProj} where the blocked EFA HFB energies for different values of the $K$ quantum number in an axially symmetric calculation are plotted as a function of the octupole moment. In the same plot, the projected energies to positive and negative parities are shown. In the plot, the energy gain from parity projection depends on the blocked configuration and the parity onto which the state is projected. As a consequence, the projected energy curves shift relative to the intrinsic energies, altering the nucleus's spectrum, as shown in figure~\ref{229RaSPEC}. In the blocked HFB calculation, the ground state has spin and parity $J=5/2$ and a marked negative mean value of the parity operator. The first excited state is a $J=1/2$ state with a positive mean value of the parity operator. When rVAP parity projection is included, the spectrum changes substantially, with $J=1/2^+$ predicted to be the ground state and $J=5/2^-$ the first excited state. Figure~\ref {229RaSPEC} shows the spectrum obtained in this case.

\begin{figure}[t]
\includegraphics[width=0.99\textwidth]{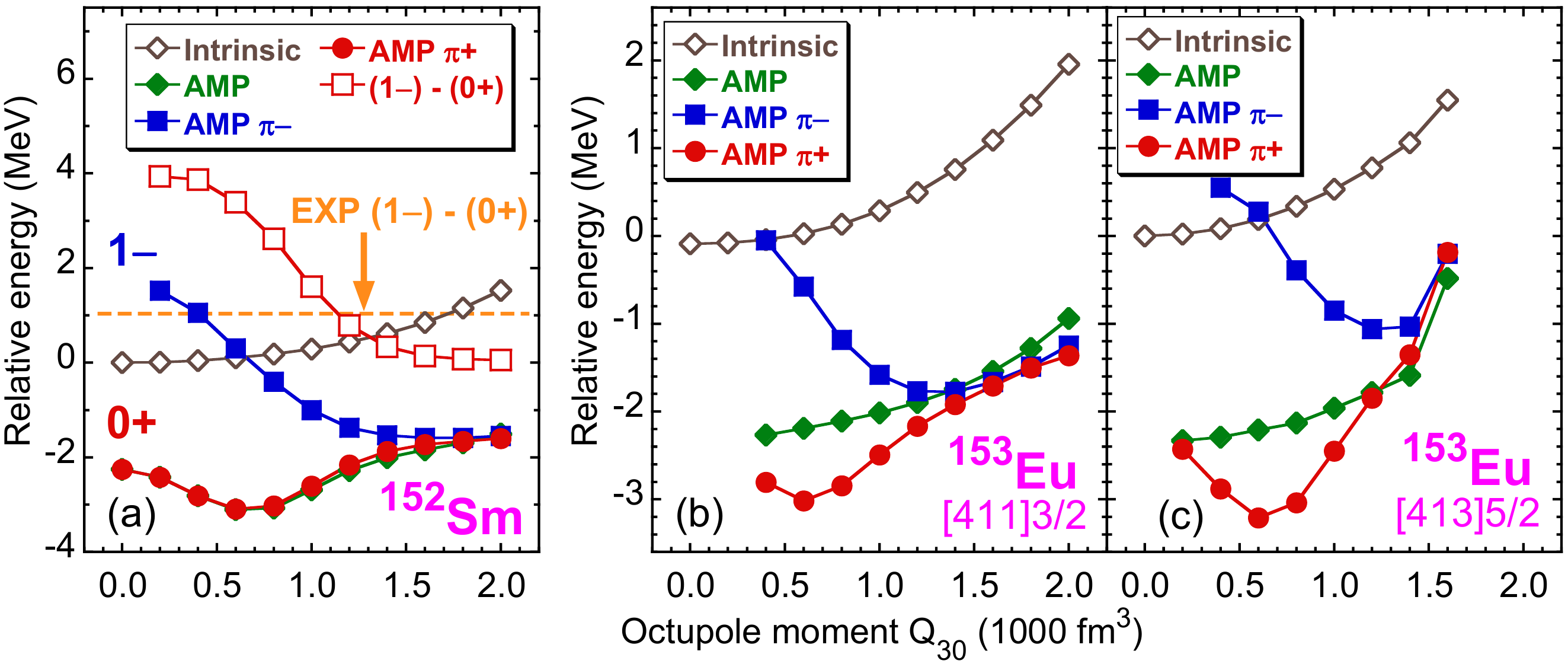}%
\caption{Intrinsic, AMP, AMP\,$\pi+$, and AMP\,$\pi-$ energies calculated for $^{152}$Eu (a), and for the [411]3/2 (b) and [413]5/2 (c) configurations of $^{153}$Eu. Calculations were performed for the UNEDF1 Skyrme functional~\cite{(Kor10c)}.\label{eu153}}		
\end{figure}

In fact, the considerations presented in this section are not limited to octupole-deformed nuclei such as $^{229}$Ra discussed above; they should be addressed across the entire nuclear chart. In figure~\ref{eu153}, we show an example of PAV calculations performed in $^{152}$Eu and $^{153}$Eu using the code {\sc hfodd}~\cite{(Dob21f)} with time-reversal, parity, signature, and rotational symmetry broken, with axial symmetry conserved and s.p.\ angular momenta aligned along the axial-symmetry axis.

In function of the total intrinsic octupole moment $Q_{30}$, intrinsic energies of $^{152}$Eu and of the two lowest proton configurations, [411]3/2 and [413]5/2, in $^{153}$Eu have minima at $Q_{30}=0$, figure~\ref{eu153}, so these nuclei are classified as not octupole deformed. However, their angular-momentum- and parity-projected energies, AMP\,$\pi+$ and AMP\,$\pi-$, exhibit minima at non-zero $Q_{30}$ deformations.

In $^{152}$Eu, the energy difference between states $1-$ and $0+$ decreases from about 4\,MeV to 0, and reaches the experimental value of 0.963\,MeV~\cite{(Gar05a)} near $Q_{30}=1.2$\,$b^{3/2}$, which is between the minimum values of the $1-$ and $0+$ energies. It appears that future octupole-collective mixing of the $1-$ states, and of the $0+$ states, would be warranted here to be benchmarked against the experimental value of the B(E1) transition rate of 4.2\,$10^{-3}$\,Wu.

In $^{153}$Eu, the ordering of intrinsic energies at $Q_{30}=0$ agrees with the order of quasiparticle energies. The intrinsic [411]3/2 configuration is the ground state, and the [413]5/2 configuration lies only 88\,keV above. The angular-momentum projection (AMP) shifts both states, $I=5/2$ and $I=3/2$, uniformly downward by about 2.2\,MeV, but only the parity projection AMP\,$\pi+$ brings the 5/2+ state below 3/2+, in agreement with experiment~\cite{(Pea94)}. Considering the large overall uncertainties in DFT energies, the calculated 3/2+ excitation energy of 193\,keV matches the measured value of 103\,keV perfectly. The example of $^{153}$Eu is likely to represent a fortuitously favourable situation and illustrates the degree of fine-tuning that symmetry restoration can bring to the description of spectra. 

In summary, correlation energies arising from symmetry restoration affect the low-energy spectra and must be considered at the VAP, rVAP or MAP (section~\ref{sec:MAP}), or PAV level to better compare with experimental data. This also applies to determining the ground-state spins and parities of nuclei across the nuclear chart. Achieving this goal requires continued effort to develop computer codes that restore symmetries in time-breaking frameworks, such as in~\citeasnoun{(Dob21f),bally2021,(Mar22a),(Bor15)}.

\clearpage
\section{Configuration mixing\label{sec:configurationmixing}}

\subsection{Configuration mixing of symmetry-projected reference states\label{sec:mix:pgcm}}

Compared with projected generator coordinate method (PGCM) calculations for the collective states of even-even nuclei, odd nuclei present several conceptual and practical complications. First and foremost, this requires considering an additional discrete coordinate associated with the quasiparticle content of the reference states\footnote{We remark in passing that the possibility to take into account non-collective modes was already envisaged in the seminal paper by Griffin~\cite{(Gri57)}.} in parallel with the traditional exploration of continuous collective degrees of freedom such as the multipole deformations or pairing. Given that in practice the collective coordinates are generally discretised, this new coordinate does not introduce any conceptual novelty and 
the PGCM ansatz takes the usual form of a finite sum of weighted symmetry-projected reference states.

In principle, (multi-)quasiparticle excitations can also be included in the calculation of even-even systems to improve the description of certain excited states ~\cite{Chen2017a}. The key difference is that modelling odd nuclei requires exploring different quasiparticle configurations, as they exhibit a much denser and richer low-energy spectrum. In particular, the presence of several states with the same quantum numbers at similar energies in the symmetry-projected spectrum of the various reference states favours a non-trivial configuration mixing. For example, one can imagine that the $9/2^+$ state that represents the lowest projected state in the decomposition of a given one-quasiparticle excitation can mix with another $9/2^+$ state originating from the rotational band of a lower-lying positive-parity bandhead with smaller angular momentum and appearing in the decomposition of a different one-quasiparticle excitation.

The calculations in~\citeasnoun {(Bal23)} for the heavy nucleus $^{197}$Au also provide an excellent example: the energy gap between states with different spin and/or parity changes by several hundred keV as one moves from the pure mean-field description to symmetry-projected calculations in the full PGCM. These changes are of the same order of magnitude as the typical excitation energies in this nucleus. In the end, we can conclude that both the symmetry projection and configuration mixing have a large impact on the properties and energy ordering of many-body states such that the ground state, excitation energies as well as other observable properties of the final correlated states often cannot be guessed only by looking at the properties of the initial mean-field reference states (see also the discussion in section~\ref{sec:sym:vap}).

The necessity to explore a new degree of freedom associated with the quasiparticle content of the reference state also makes PGCM calculations for odd nuclei intrinsically more cumbersome and computationally intensive than for their even-even counterparts. First, at the mean-field level, this implies generating more Bogoliubov vacua, which are also more difficult to converge, as explained in section~\ref{sec:mfconstruction}. This problem is further reinforced by the fact that it is not always easy to guess the importance of any given reference state in the subsequent PGCM calculation. Second, at the beyond-mean-field level, the PGCM scales quadratically with the number of reference states included in the configuration mixing, so adding a new coordinate is always numerically very costly (assuming it is systematically explored). To address these problems, we need more robust HFB solvers and efficient, automated strategies for selecting the reference states that enter PGCM calculations.

Another difficulty is the frequent collapse of pairing correlations in blocked mean-field calculations. First, this implies that beyond-mean-field calculations do not fully account for pairing correlations. Second, this can lead to pairs of orthogonal Bogoliubov vacua, for which the Generalised Wick Theorem~\cite{(Bal69)} does not apply, within the selected set of reference states. A practical, but not fully satisfactory, solution to the latter problem was employed, for example, in~\citeasnoun{(Bal22a),(Bal23)} where the average particle numbers of the reference states were constrained to non-integer values to enforce a minimal presence of pairing. Reference states optimised within a scheme that includes the VAP on the particle numbers would tackle both problems simultaneously. We note that such an approach has already been employed in~\citeasnoun{Borrajo2018a} and in calculations based on \textit{ab initio} chiral interactions~\cite{Zhou2025a}.

In fine, the quality of the results obtained from a full PGCM calculation inherently depends on the quality of the EDF employed. For example, for the light nucleus $^{25}$Mg, calculations using the well-known Gogny D1S functional~\cite{Borrajo2018a} better reproduce the level ordering of the rotational band based on the $1/2^+$ excited state than calculations based on the more schematic pseudopotential SLyMR0~\cite{(Bal14d)}. Therefore, we must use functionals that can accurately describe the complex energy spectra of odd nuclei (see also section~\ref{sec:eff_int_time_odd}). The difficulty, however, is that PGCM calculations require functionals derived from the expectation value of a pseudopotential operator. Otherwise, this leads to ill-defined and potentially unreliable calculations~\cite{(Lac09),(Ben09),(Dug09)} (see also section~\ref{singularities}). Unfortunately, as of today, very few such functionals exist~\cite{(Bei75b),(Sad13),(Sat14g),(Ben14b)} and they tend to be less robust than the traditional (pure) functionals. The development of pseudopotentials of good spectroscopic quality is therefore a desired endeavour. 

Despite these limitations, the few available fully fledged PGCM calculations of odd nuclei based on triaxial states~\cite{(Bal14),Borrajo2018a,(Bal22a),(Bal23)} have demonstrated their relevance for describing the spectroscopy of deformed open-shell nuclei, which often have numerous low-lying, low-spin bandheads. Recently, similar calculations were even performed to tackle the structure of very heavy nuclei in the region of fermium isotopes~\cite{Warbinek2024a,Urquiza2026a}. As mentioned in section~\ref{alispin}, however, these calculations explored only one sextant of the $\beta$-$\gamma$ plane. For triaxial signature-conserving configurations, one should at least construct quasiparticle vacua that cover a 180-degree range in $\gamma$. This would still exclude tilted-axis configurations, whose relevance to PGCM calculations that mix different shapes has not yet been explored.

As a final remark, we note that past studies have been carried out only on one or very few individual nuclei at a time. Moving to large-scale triaxial calculations that cover a large area of the nuclear chart will be quite involved and probably not possible in the near future. By contrast, axial calculations are much less computationally intensive, as the angular-momentum projection is significantly simplified and the dimension of the configuration mixing is reduced, occurring only between reference states with the same value of the third component of the angular momentum, $K$. Even though such mixing must be performed for different values of $K$, the PGCM's quadratic scaling makes axial calculations substantially less computationally expensive. Therefore, we advocate for future large-scale axial PGCM calculations of odd nuclei. 

%================================================================================
%================================================================================
%================================================================================

\subsection{Configuration mixing of $K$ substates near sphericity\label{K-mixing-spherical}}

In general, quantum states of an odd nucleus can be built as products of antisymmetrised states of an even system and those of a single particle; that is, $\caH_{A+1}=\caH_{A}\otimes\caH_{\text{s.p.}}$. In the perturbative approach, we assume that such states are dominated by products of a few low-energy states of the even system and a few s.p.\ states near the Fermi energy of the even system. Thus, in the simplest case, the odd nucleus is described as a product of the $J=0$ wave function of the even core and the s.p.\ wave function carrying angular momentum and its projection $jm$, with the total angular momentum and its projection of the odd nucleus $JM=jm$.

In the nuclear-DFT description of nearly spherical odd nuclei, one must consider the occupation of individual magnetic substates of a spherical level $jm$ that, in the spherical even core, are degenerate. Since none of the $\ell>0$ substates has a vanishing quadrupole moment, the self-consistent polarisation of the spherical core generally induces symmetry breaking and a non-zero quadrupole moment of the odd nucleus. Traditionally, symmetry-broken projections of the angular momentum on the axis of axial symmetry are denoted by $\Omega$ and their moduli by $K$, that is, $K=|\Omega|=|m|$. Then, because of Kramers degeneracy, we need to consider only positive projections, $\Omega=K$. We conclude that, apart from occupied $\ell=0$ states, none of the self-consistent states of odd nuclei in nuclear DFT is spherical. However, the question that arises is: what are the consequences of occupying one of the magnetic substates $jm$?

\begin{figure}[ht]
\hfill{\includegraphics[width=0.98\textwidth]{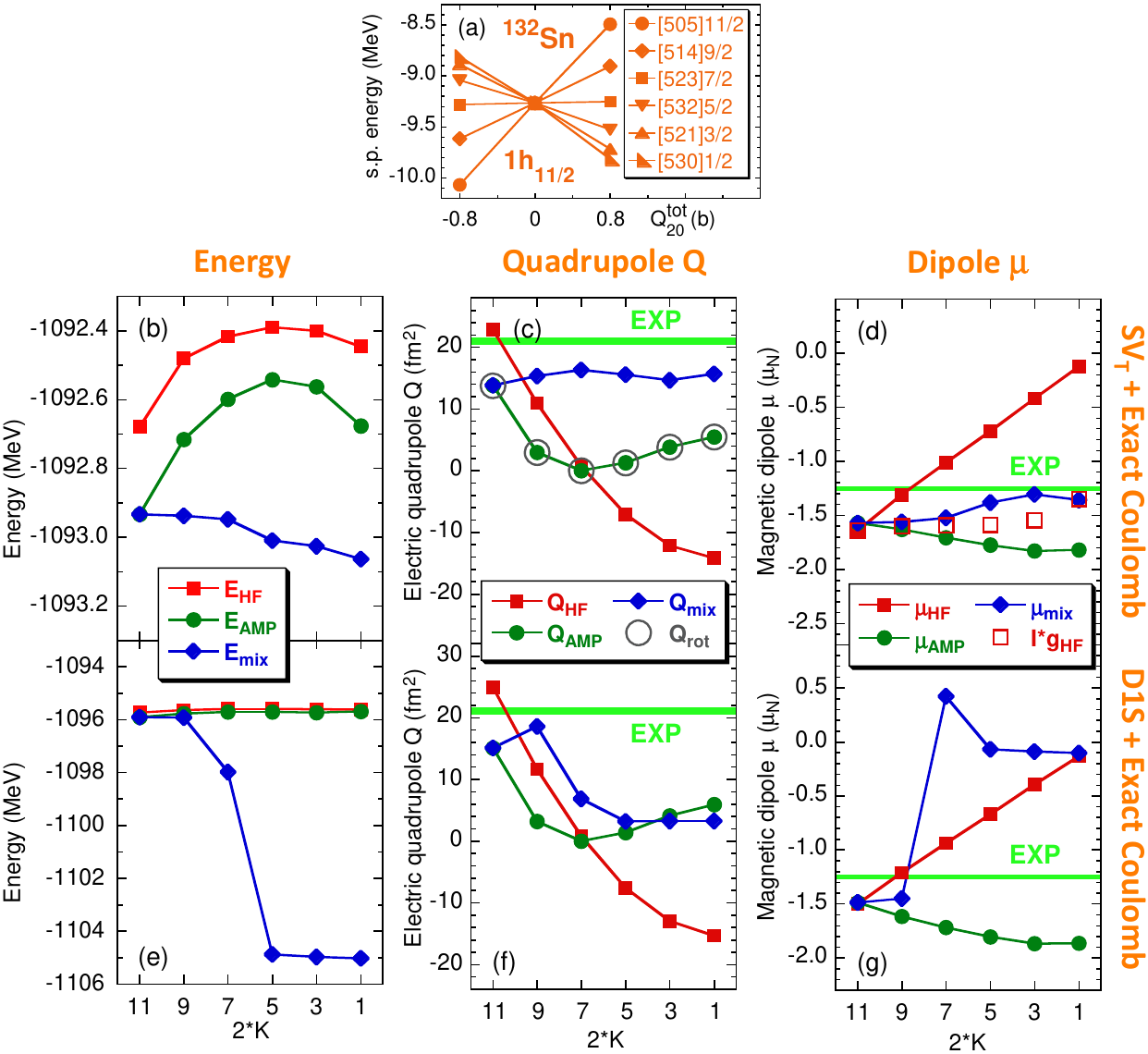}}
\caption{Neutron-deformed s.p.\ energies in unpaired $^{132}$Sn, panel (a), and results of calculations performed for the Skyrme pseudopotential SV$_{\text{T}}$~\protect\cite{(Bei75b),(Sat14g)}, panels (b)--(d), and for the Gogny functional D1S~\cite{(Ber91e)}, panels (e)--(g). The left, middle, and right panels show the calculated total energies (in MeV), electric quadrupole moments (in fm$^2$), and magnetic dipole moments (in $\mu_N$), respectively. Squares, circles, and diamonds show the HF, AMP, and $K$-mixing results, respectively; see text.\label{K-mixing}}
\end{figure}
In Figure~\ref{K-mixing}, we present preliminary results from~\citeasnoun {(Dob26c)}, who attempted to answer this question using the 1h$_{11/2}$ hole states in $^{131}$Sn as an example. Figure~\ref{K-mixing}(a) depicts the calculated neutron s.p.\ energies in unpaired $^{132}$Sn, obtained by constraining intrinsic quadrupole moments to $Q^{\text{tot}}_{20}=\pm0.8$\,b and quantising projections of the angular momentum along the $z$ axis with a small cranking frequency of $\hbar\omega_z=1$\,keV. Each line represents two projections with $K=|\Omega|=1/2,\ldots,11/2$. We then use the resulting prolate neutron s.p.\ wave functions to construct the corresponding self-consistent unpaired hole states in $^{131}$Sn via the tagging mechanism~\cite{(Wib25d)}.

Panels (b)--(d) show results for the Skyrme pseudopotential SV$_{\text{T}}$~\cite{(Sat14g)}, which is the SV density-independent Skyrme force from~\citeasnoun{(Bei75b)} with tensor terms included, and panels (e)--(g) display results for the density-dependent Gogny functional D1S~\cite{(Ber91e)}. In both cases, we calculated the direct and exchange terms of the Coulomb potential exactly. We compare these two versions because the former produces a functional that is an exact average of a two-body pseudopotential, avoiding the kernel singularities discussed by~\citeasnoun{(She21)} and in section~\ref{singularities}. In contrast, the latter is not of this type and suffers from singularities arising from density dependence and the resulting mismatch between time-even and time-odd mean fields.

As shown in figure~\ref{K-mixing}(b), the $^{131}$Sn HF energies\footnote{Note that the experimental binding energy of $^{131}$Sn is $E_{\text{exp}}=1095.489727$\,MeV} (squares) are weakly split, with the blocked $K=11/2$ substate having the lowest energy, consistent with the standard quadrupole splitting, panel (a), where the hole s.p.\ energy on the prolate side is the highest. Subsequently, we projected each of the $K=|{\Omega} |=1/2,\ldots,11/2$ blocked HF states (AMP) onto the total angular momentum $J=11/2$, resulting in an almost uniform downward energy shift (diamonds) of about 200\,keV. This shift reflects the small rotational-energy correction characteristic of these weakly deformed states. We observe that the energies of the $J=11/2$ states projected from different $K=|\Omega|=1/2,\ldots,11/2$ intrinsic states differ, indicating that the polarisation effects induced by occupying these states are different from one another.

Finally, the blocked AMP $J=11/2$ HF states were mixed\footnote{We note that the mixing discussed here corresponds to a genuine multi-reference mixing of Slater determinants, which differs from the mixing of $K$ components projected from a single Slater determinant, discussed by~\protect\citeasnoun{(Lu25)}.} in the reverse order, that is, in figure~\ref{K-mixing}, diamonds at $K=9/2$ denote states $K=11/2$ and 9/2 mixed, those at $K=7/2$ denote states $K=11/2,\ldots,7/2$ mixed, and those at $K=1/2$ denote all states $K=11/2,\ldots,1/2$ mixed. As shown, increasing the number of mixed $K$ components decreases the total energy by only another 100\,keV, indicating a high level of redundancy among the different $K$ states.

The electric quadrupole moments are shown in figure~\ref{K-mixing}(c). The intrinsic HF moments, $Q_{\text{HF}}$, follow the slopes of the s.p.\ energies shown in figure~\ref{K-mixing}(a), thereby reflecting the polarisation strengths of the different hole $K$ states. These are large and positive for $K=11/2$ and decrease uniformly towards smaller, negative values for $K=1/2$. The $J=11/2$ spectroscopic quadrupole moments, $Q_{\text{AMP}}$, are nearly equal to those predicted by the so-called large-axial-approximation, $Q_{\text{rot}}$, see~\protect\citeasnoun{(Rin80b)}, equation~(11.143), and~\cite{(Dob26g)}, that is,
\begin{equation}
\label{rot}\hspace*{-10mm}
Q_{\text{AMP}}(JK)\simeq{}Q_{\text{rot}}(JK)=C^{JJ}_{JJ,20}C^{JK}_{JK,20}Q_{\text{HF}}(K)=\frac{3K^2-J(J+1)}{(2J+3)(J+1)}Q_{\text{HF}}(K),
\end{equation}
which shows that this approximation is not limited to large deformations~\cite{(Dob25k)}. Most interestingly, the spectroscopic moments of the mixed states, $Q_{\text{mix}}$, differ little from that of the lowest $K=11/2$ state and vary by only about 2\,fm$^2$.

The intrinsic HF magnetic dipole moments in panel (d), $\mu_{\text{HF}}$, depend linearly on $K$ and are characterised by an approximately constant intrinsic $g$ factor $g_{\text{HF}}=\mu_z/K$ of about $-1.65\,\mu_N$. In contrast, the spectroscopic moments $\mu_{\text{AMP}}$ decrease from $K=11/2$ to $K=1/2$ by about $0.4\,\mu_N$, while those of the mixed states increase by roughly the same amount. This range of variation is consistent with typical differences between calculations and experimental data; see the review by~\citeasnoun{(Dob25k)}. An interesting subject for future investigation is whether the $K$ mixing discussed here could systematically improve agreement with measurements.

However, such a study must be postponed until better functionals, density-independent and generated by pseudopotentials, become accessible. This is clear from the results for the D1S functional, panels (e)--(g), where singularities in the mixing matrix elements cause the mixed-state energies to collapse, resulting in erratic, uncontrollable behaviour of the electric quadrupole and magnetic dipole moments.

\newpage
\section{Effective interactions and time-odd mean fields\label{sec:eff_int_time_odd}}
%%%%%%%%%%%%%%%%%%%%%%%%%%%%%%%%%%%%%%%%%%%%%%%%%%%%%%%
\subsection{Single-particle properties of EDFs -- dependence on basic assumptions}
%%%%%%%%%%%%%%%%%%%%%%%%%%%%%%%%%%%%%%%%%%%%%%%%%%%%%%%

 There are several basic assumptions in building 
the nuclear DFTs. These are (i) the selection of underlying framework (relativistic built on the Dirac equation or non-relativistic built on the Schr{\"o}dinger equation), (ii) the selection 
of the range of interaction (zero range for point-coupling (PC) relativistic and non-relativistic Skyrme functionals or finite range for meson-exchange (ME) relativistic and non-relativistic  Gogny functionals) and (iii) the selection of the density dependence of the functional. Each of these assumptions 
affects the description of experimental 
single-particle energies. For example, spin degrees of 
freedom, and thus the spin-orbit splittings emerge naturally in a relativistic framework, but they require adjustable parameters in a non-relativistic one. A substantial degree of arbitrariness also exists in choosing the interaction's range and its density dependence. The issues emerging from such a selection are illustrated by the examples discussed below.

%%%%%%%%%%%%%%%%%%%%%%%%%%%%%%%%%%%%%%%%%%%%%%%%%%%%%%%%%%%%%%%%%%%%%%%%%%%%%%%%%%%%%%
\begin{figure}[htb!]
\centering
\includegraphics[width=1.0\textwidth, angle=0]{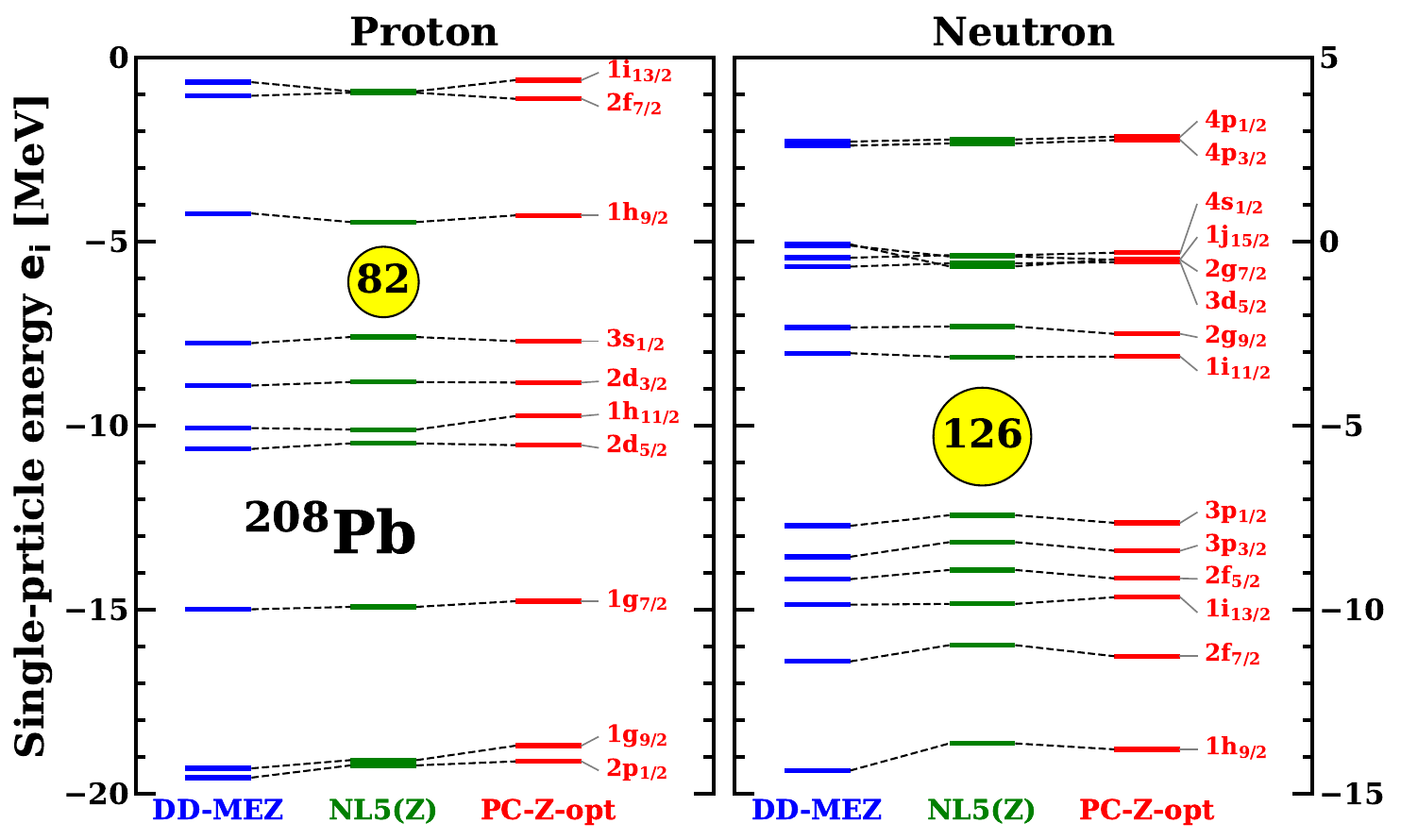}
\caption{Proton and neutron single-particle states of $^{208}$Pb obtained
with indicated covariant energy density functionals. 
}
\label{single-particle-Pb208}
\end{figure}
%%%%%%%%%%%%%%%%%%%%%%%%%%%%%%%%%%%%%%%%%%%%%%%%%%%%%%%%%%%%%%%%%%%%%%%%%%%%%%%%%%%%%%%%%%%%

 The energies of the single-particle states at spherical shape are 
frequently compared for different functionals in the publications. 
However, this kind of comparison is not fully informative since the
compared functionals have been generated using fitting protocols which 
frequently differ substantially. To reveal a true impact of the form of the 
functional on the single-particle properties and remove undefined 
relative biases
created by the choice of the fitting protocol, one should compare 
such properties for the functionals generated with the same fitting 
protocol. This is done in figure~\ref{single-particle-Pb208} for 
the density-dependent meson exchange DD-MEZ, non-linear meson 
exchange NL5(Z) and point-coupling PC-Z-opt covariant energy density
functionals (CEDFs). Note that these functionals were globally adjusted using the same
fitting protocol~\cite{NL5Z-DDMEZ-PCZ,ODA.26}. Although the general structure of 
the calculated single-particle structure is the same, there are substantial 
differences in the relative and absolute values of the energies of 
the single-particle states (frequently reaching 1.0 MeV) of the compared 
functionals, especially when ME functionals are confronted with the
PC ones. These  differences are the sources of the 
discrepancies between experimental data and theoretical results. 
They are also the sources of substantial systematic uncertainties in the 
predictions of the energies of the single-particle states. Note that 
statistical errors $\sigma(e_i)$ in the predictions of such energies are smaller 
($\sigma(e_i) \approx 0.1$ MeV for the states of interest) for 
covariant energy density functionals~\cite{AAT.19}.   

%%%%%%%%%%%%%%%%%%%%%%%%%%%%%%%%%%%%%%%%%%%%%%%%%%%%%%%%%%%%%%%%%%%%%%%%%%
\begin{figure}[htb!]
\centering
\includegraphics[width=01.0\textwidth, angle=0]{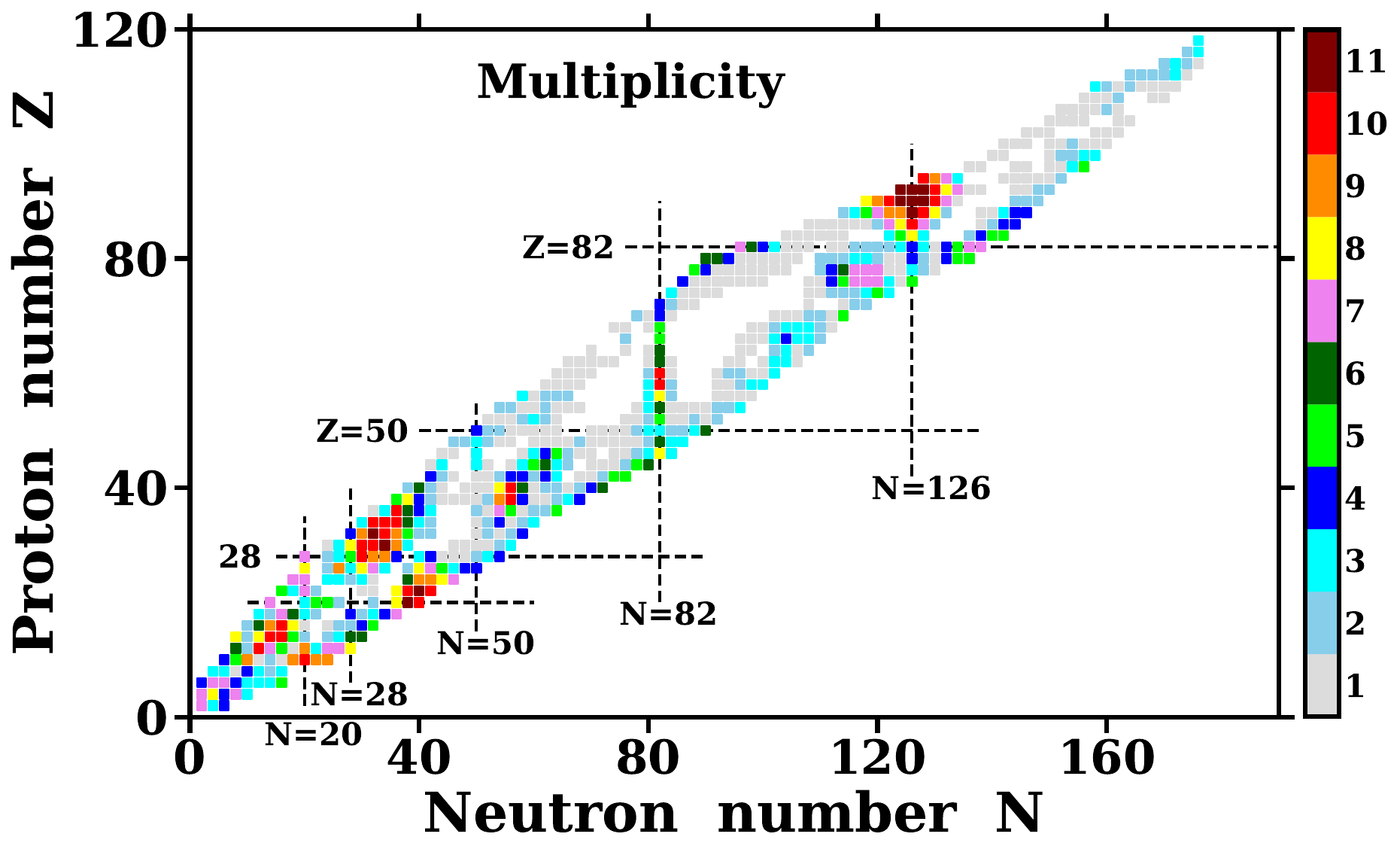}
\caption{The model multiplicity $m(Z,N)$ of the reproduction of binding 
energies in the set of 11 covariant energy density functionals. See
text for details.
}
\label{global-biases}
\end{figure}
%%%%%%%%%%%%%%%%%%%%%%%%%%%%%%%%%%%%%%%%%%%%%%%%%%%%%%%%%%%%%%%%%%%%%%%%%%

  It is also important to understand which kind of global biases are 
introduced in the description of physical observables of interest by 
selecting a specific type/form of the functional. Such biases are 
illustrated in figure~\ref{global-biases}, which is based on the global
calculations of binding energies with 11 CEDFs such as DD-ME2, DD-PC1, 
DD-ME$\delta$, NL3*, DD-MEY, NL5(Y), PC-Y, DD-MEZ, PC-Z and PC-PK1 
\cite{(Agb14a),TA.23,NL5Z-DDMEZ-PCZ,RHB-e-e-continuum.22}. With the 
exception of the PC-PK1 functional, all these calculations are carried 
out at the mean-field level, and the calculations with the $i$-th 
functional are characterized by global rms-error $\Delta B_i^{glob}$. 
Then, the number $k$ of the functionals for which the condition
$|B(Z,N)_i-B(Z,N)_{exp}| \geq 1.06 \cdot \Delta B_i^{glob}$ $(i=1,2, ..., 11)$
is satisfied for a given $(Z,N)$ nucleus is counted. It defines the model 
multiplicity $m(Z,N)=k$ which represents the number of the functionals 
which fail to describe the binding energy of the $(Z,N)$ nucleus with a 
required precision. Figure~\ref{global-biases} shows that there are
several regions (such as the one centred around $Z=92, N=126$) in which the 
multiplicity is consistently  close to $m(Z,N)\approx 10$ and the differences
between calculated and experimental energies can exceed 5.0\,MeV~\cite{(Agb14a),TA.23,NL5Z-DDMEZ-PCZ,RHB-e-e-continuum.22}.
This failure to reproduce binding energies accurately stems from deficiencies in the description of single-particle energies: for example, all covariant functionals defined at the Hartree level overestimate the $ Z=92$ shell gap~\cite {AOD.25}.

  The uncertainties in the description of the energies of the
single-particle states emerging from the basic assumptions of
nuclear DFT also exist in its non-relativistic versions. Both 
in relativistic and non-relativistic frameworks, they become especially pronounced on the extrapolation towards unknown regions such as superheavy nuclei~\cite{BRRMG.99,AANR.15} and the neutron drip line~\cite{Eet.12,AARR.15}, where they play a decisive role in the definition of the properties of the nuclei and the boundaries of regions. Global DFT investigations also reveal that even in the experimentally known part of the nuclear chart,
the structure of the ground state is properly described 
only in $\approx 40\%$ of odd nuclei~\cite{BQM.07,AS.11}.

  This discussion clearly reveals several open problems related to the definition of the functionals in the time-even channel which affect the description of the single-particle degrees of freedom. These are:

\begin{itemize}
    \item
     The DFT is an approximation to an exact solution of a many-body
     problem, and as a consequence there are intrinsic theoretical uncertainties in the description of the single-particle energies which are difficult to quantify. For example, 
     what are the consequences of the parametrisation of the
     spin-orbit interaction on the accuracy of the description
     of single-particle energies in non-relativistic DFT?
     Similarly, what are the consequences of the neglect of 
     exchange terms 
     in the 
     case of covariant DFT and whether their impact can be effectively and accurately taken into account at the Hartree level of the CDFT?

    \item 
      Presently used fitting protocols focused on the properties
  of even-even nuclei allow one to reliably define only a restricted 
  set of independent parameters (only 5 or 6 in the case of the CDFT
 ~\cite{TAAR.20}. However, one can construct EDFs with substantially
  larger physically motivated terms. Will the use of information
  on the structure and excitation energies of odd and odd-odd 
  nuclei in the fitting protocols  
  provide a better constraint on the parameters of the terms of the
  functionals which are otherwise poorly defined? Will such an approach allow us to reliably define missing terms (such as tensor interactions, $\delta$-mesons, etc.) in the functionals?
    
    \item
      How must a fitting protocol incorporating information on
    odd nuclei be designed? Should it include information 
    on the ground and excited states of odd nuclei or be 
    restricted only to the ground states? The use of beyond 
    mean-field approaches is required in the former case since
    excited states are strongly affected by particle-vibration
    coupling~\cite{GISF.73,ZBNLRS.22}. In the latter case it can 
    be restricted to the mean-field level since the ground states of odd nuclei are less affected by this type of coupling. 
    For example, very successful parametrisations 
    of the Woods-Saxon potential have been globally fitted 
    to only ground states of odd nuclei~\cite{DW.78,BDNO.89}.

\end{itemize}

%%%%%%%%%%%%%%%%%%%%%%%%%%%%%%%%%%%%%%%%%%%%%%%%%%%%%%%%%%%%%%%%%%%%%%%%%%%%%%%%%%%
\subsection{Fit of functionals to first excited states in odd nuclei}
%%%%%%%%%%%%%%%%%%%%%%%%%%%%%%%%%%%%%%%%%%%%%%%%%%%%%%%%%%%%%%%%%%%%%%%%%%%%%%%%%%%

Since the 1970s, nuclear energy density functionals have typically been adjusted to reproduce the properties of even-even nuclei, occasionally incorporating beyond-mean-field effects (see, for example, \cite{pilletreview} for the Gogny interaction). However, the properties of odd nuclei were far less frequently considered — whether these pertain to fundamental ground-state characteristics (such as mass, spin, and parity) or, even more rarely, to those of excited states. 
These last properties are crucial, however, as they enable the refinement of the functionals' properties. For instance, in odd systems, the first excited states often exhibit distinct characteristics and offer direct insight into the proton or neutron gaps — specifically, the energy difference between the orbital occupied by the odd particle and the next higher-energy orbital. The precise values of these gaps are particularly sensitive to the spin-orbit and tensor components of the functionals.
\begin{figure}[htb!]
\centering
\includegraphics[width=0.4\textwidth, angle=-90]{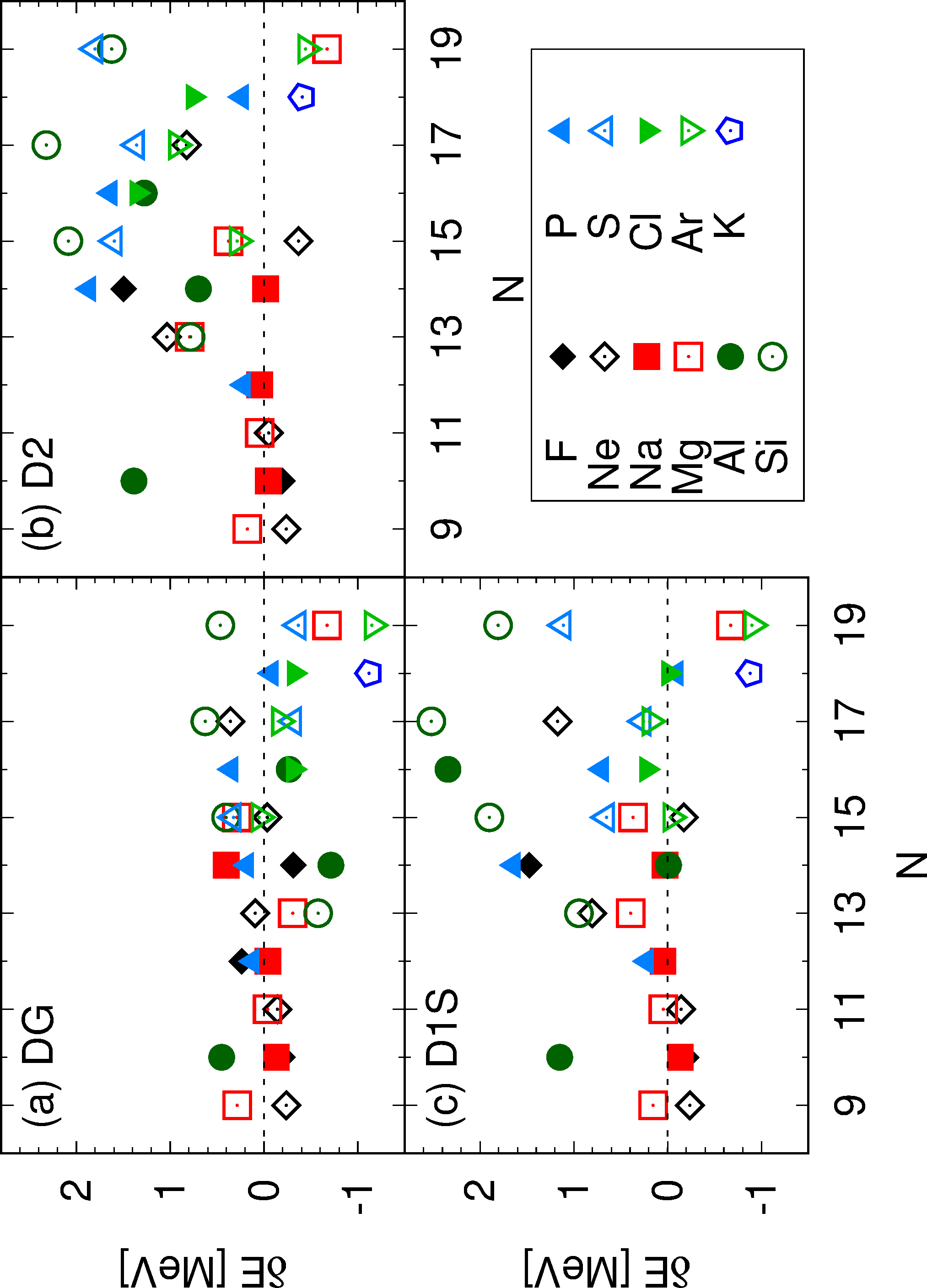}
\caption{Differences between the experimental and theoretical energies of the first excited states in odd $sd$-shell nuclei for the DG (panel (a)), D2 (panel (b)) and D1S (panel (c)) Gogny interactions, reprinted with permission from~\citeasnoun{DG}.}
\label{np1}
\end{figure}

For example, the recent fitting procedure of the DG Gogny interaction \cite{DG} has demonstrated the importance of incorporating constraints on odd nuclei. Specifically, the first experimental excited states of odd-$sd$-shell nuclei were used to guide the selection of parametrisation. Indeed, in these specific cases, they have more of a single-particle than collective nature. They are therefore good indicators of the proton and neutron gaps. In \cite{Robin2016,Robin2017}, low-lying spectra were calculated using the MCSCF approach, which accounts for beyond-mean-field correlations. This method determines gaps in a manner consistent with beyond-mean-field effects.
The global impact of this approach is illustrated in Fig.~\ref{np1}, which shows the difference between the energies of the first experimental excited states and their associated MCSCF counterparts for three Gogny parametrisations: DG (fully finite range with an added tensor term, panel (a)), D2 (finite-range density-dependent term, panel (b)) [\cite{D2}], and D1S (original analytical form, panel (c)). Notably, the DG interaction significantly reduces the bump observed near subshell closures, although there is still room for improvement. A global spin-orbit effect and a local tensor term drive the present result.

From a broader perspective, this raises at least two key challenges:
\begin{itemize}
\item Incorporating odd nuclei (ground and excited states)—alongside odd-odd nuclei—into the functional fitting procedure on an equal footing with even-even nuclei, which requires extending many-body approaches to these systems.
\item Revisiting and extending the analytical form of the functionals. This implies, in particular, increasing the number of parameters, which can leverage the vast amount of data available on odd- and odd-odd nuclei. Selecting observables sensitive to the various terms of the functional will be particularly challenging.
\end{itemize}

\subsection{Time-odd mean fields in Skyrme EDFs\label{timeoddskyrme}}

Separate terms of the Skyrme EDF generate the time-even and time-odd mean fields; therefore, they are often referred to as the ``time-even'' and ``time-odd'' terms, respectively, even though they are both parts of the time-reversal-invariant functional. This possibly confusing, but widely used terminology, does not apply to finite-range EDFs, in which each term generates both time-even and time-odd mean fields. The time-odd mean fields only contribute to the total energy when evaluating the energy for a time-reversal-breaking configuration, 
or when evaluating off-diagonal kernels that enter the restoration of angular momentum or 
shape mixing in GCM. In general, they model spin-spin, current-current, and spin-current 
interactions, which, in relativistic DFTs, for similarity with electromagnetic interactions, are sometimes jointly called 
``nuclear magnetism'', see section~\ref{sec:covariantDFT}.

When the Skyrme force is used as a functional generator, that is, when the functional is equal to the average of the Skyrme force in a Slater determinant, each term of the force generates both time-even and time-odd terms, both fixed by the corresponding Skyrme coupling constant; see, e.g., Table I in~\citeasnoun{(Per04c)}. However, the time-even and time-odd parts of the quasilocal EDFs, such as Skyrme, can also be separated because such EDFs are built 
out of local densities that have a definite symmetry under time-reversal. 
By contrast, non-local one-body densities have no definite symmetry under time reversal, 
which prohibits such separation for EDFs with finite-range exchange terms.
The features of time-odd terms are therefore most transparently illustrated for the 
quasilocal Skyrme EDF, which is also the one for which arguably the largest number of studies 
of time-odd mean fields have been carried out. For the most widely used standard form of the 
Skyrme EDF with up to two gradients, the time-odd terms read
\begin{eqnarray}
\label{eq:Skyrme_to}
\mathcal{E}_{t, \rm o}(\bm{r})
& = &
            C^{s s}_t \, \bm{s}_t (\bm{r}) \cdot \bm{s}_t (\bm{r})
             \nonumber \\
&   &              
            + C^{j j}_t \, \bm{j}_t (\bm{r}) \cdot \bm{j}_t (\bm{r}) 
            + C^{j \nabla s}_t \, \bm{j}_t (\bm{r}) \cdot \nabla \times \bm{s}_t (\bm{r}) 
             \nonumber \\
&   &
            + C^{s T}_t \, \bm{s}_t (\bm{r}) \cdot \bm{T}_t (\bm{r})
            + C^{s \Delta s}_t \, \bm{s}_t (\bm{r}) \cdot \Delta \bm{s}_t (\bm{r})
             \nonumber \\
&   &
            + C^{s F}_t \, \bm{s}_t (\bm{r}) \cdot \bm{F}_t (\bm{r})
           + C^{\nabla s \nabla s}_t \, \big[ \bm{\nabla} \cdot \bm{s}_t (\bm{r}) \big]
                                         \big[ \bm{\nabla} \cdot \bm{s}_t (\bm{r}) \big]
\, ,
\end{eqnarray}
which depends on the isoscalar ($t=0$) and isovector ($t=1$)
spin density $\bm{s}_t (\bm{r})$, 
kinetic spin densities $\bm{T}_t (\bm{r})$
and  $\bm{F}_t (\bm{r})$,
and the current density $\bm{j}_t (\bm{r})$. 
These expressions contain all possible linearly independent 
bilinear terms with up to two gradients, including the 
rarely used terms with coupling constants $C^{s F}_t$
and $C^{\nabla s \nabla s}_t$ obtained from genuine 
tensor forces~\cite{(Hel12)}, where each term has a specific
interpretation. This Skyrme EDF is useful for the 
evaluation of the physics content of other types of functionals
as it represents the zero-range limit of the EDF from 
finite-range interactions~\cite{(Car10c)} and the 
non-relativistic limit~\cite{Sulaksano07a,Sulaksano11a} 
of relativistic EDFs. Up to now, however, such analyses
mainly focused on the time-even mean fields, and future, more intensive studies of the time-odd ones are warranted.

All coupling constants in (\ref{eq:Skyrme_to}) might be density 
dependent, but for the traditional Skyrme parameter sets this is 
only used for $C^{s s}_t$. Further extensions to higher-order terms 
in gradients~\cite{(Car08b),RyssensN2LO}, or from three-body forces with 
gradients~\cite{(Sad13e)} have 
also been 
proposed in the literature. Because the contributions
of conjugate pairs in the canonical basis almost cancel out 
when summing up the time-odd densities, the contribution of 
time-odd terms to the total energy of finite nuclei 
is in general much smaller than that of the time-even terms. 
However, they become essential in odd and odd-odd systems characterised by the presence of unpaired nucleons.
In addition, for polarised infinite matter, 
the volume spin terms with coupling constants $C^{s s}_t$, 
and $C^{s T}_t$ provide a dominant contribution to the binding 
energy and also completely determine the Landau parameters 
$g_\ell$ and $g_\ell'$~\cite{(Idi17)}.

There are several practical issues with the time-odd terms, in 
particular for the standard Skyrme EDF of equation~(\ref{eq:Skyrme_to}).

One issue is that the time-odd terms of Skyrme EDFs are often inconsistently 
used across the literature, and this even for the same parameterisation. 
The principal reason is that many authors modify some of the coupling constants 
in both the time-even and the time-odd parts of the Skyrme EDF. While for the 
time-even part this is always detailed in the publications proposing new 
parameter sets, the situation is often left open for the time-odd part, in 
particular when no data that are affected by the time-odd terms are included 
in the fit protocol.

For reasons of Galilean invariance of the (non-relativistic) Skyrme EDF~\cite{(Dob95)}, 
the coupling constants $C^{j j}_t$, $C^{j \nabla s}_t$, $C^{s T}_t$, and $C^{s F}_t$ of the terms containing densities with gradients cannot be chosen independently of the kinetic, 
spin-orbit, and spin-current terms in the time-even part of the Skyrme EDF,\footnote{Using Galilean-invariance breaking functionals in dynamical 
calculations leads to spurious excitations of the nuclear system. An 
illustrative example for a TDHF calculation with omitted time-odd spin-orbit 
terms that produces spurious results is discussed by \citeasnoun{Maruhn06a}.} 
but that still leaves the  spin-spin terms with coupling constants $C^{s s}_t$ 
and $C^{s \Delta s}_t$ up to a choice.

There are in fact three broad strategies to fix the values of the coupling 
constants of the EDF. One strategy consists in strictly relating them to the 
parameters $t_i$, $x_i$, etc. of the underlying (density-dependent) Skyrme 
generator, which is however not very often done. Another strategy is to treat 
all coupling constants of a general Skyrme EDF not constrained by 
Galilean invariance as independent, examples being~\cite{(Ben02a),(Sas22c)},
where the coupling constants of the time-odd spin-spin terms are adjusted to 
suitably chosen data, see also section \ref{sec:adjust}.
A third strategy is a hybrid where most coupling constants of the EDF are 
related to the $t_i$, $x_i$, etc. of a Skyrme generator, but a few specific 
terms are manipulated for reasons of phenomenology or computational 
convenience.

This is in fact already a popular choice also for time-even 
terms, where the coupling constant $-C^{s T}_t$ of the term that is 
bilinear in the spin-current tensor density $J_{t,\mu\nu}(\bm{r})$ 
is set to zero for the vast majority of standard Skyrme parameter sets. 
In all cases, however, the Skyrme parameterisations are often published 
as a set of coupling constants $t_i$, $x_i$, etc of a Skyrme generator 
to which is added a list of modifications that are to be made to the 
resulting EDF. This ambivalent representation of general and hybrid EDFs 
is a source of confusion and frequently provokes the inconsistent use 
of parameter sets in RPA or calculations of nuclear matter properties 
such as Landau parameters for which the expressions obtained from a 
Skyrme generator are used instead, leading to published values that
are incorrect. The situation would be more transparent if authors 
that adjust the parameters of general or hybrid EDFs were publishing 
directly the coupling constants of the EDF that are fixed by the fit 
protocol instead.

Another issue with time-odd terms is that for any EDF that considers them, the 
spin-spin interactions can be at the origin of instabilities that remain completely 
hidden when adjusting the parameters of the EDF only to the usual properties of 
even-even nuclei and spin-saturated matter. Two kinds of instabilities can be characterised through the model system of infinite matter. On the one hand, 
there are possible spin instabilities related to the transition from homogeneous 
spin-saturated matter to homogeneous spin-polarised nuclear matter that can be 
easily identified via the Landau parameters in the spin channels (as long as 
they are consistently calculated, see above). On the other hand, there are 
finite-size instabilities that are related to a transition from homogeneous 
matter to an inhomogeneous phase. For the local Skyrme EDF,
such instabilities are driven by the gradient terms in the EDF (\ref{eq:Skyrme_to})
and can be identified in RPA calculations of infinite nuclear matter~\cite{(Pas15a)}.

As it turns out, calculating the coupling constants $C^{s \Delta s}_t$ from
the $t_i$ and $x_i$ of a Skyrme generator leads for the vast majority of 
Skyrme parameterisations to a finite-size spin instability in either the isoscalar or
isovector channel, one of the reasons why these coupling constants are often set to zero by hand~\cite{(Pot10),(Hel12),(Wib25d)}. Using a parameter set that exhibits a finite-size instability in a code whose numerical representation
does not fully resolve it, then often leads to convergence problems such as the
ones reported by \citeasnoun{(Sch10b)}, or might lead to unusual results with large density oscillations when the solver's convergence criteria can still be met somehow. 
When used in a code that fully resolves its instability, a parameter set is 
simply unusable~\cite{(Hel12)}. 
Working with spin-unstable parameter sets can also cause several problems 
in RPA calculations for finite nuclei that are not always entirely obvious 
\cite{(Pas15b)}.

Finite-size instabilities are not limited to local EDFs with gradient terms, but 
can also emerge for finite-range interactions~\cite{(Mar19),Davesne21a}. Following 
\citeasnoun{(Pas13b)}, some recent standard Skyrme parameter sets have been 
adjusted with a constraint on the linear response of infinite matter to avoid 
such finite-size instabilities, but such a strategy might become too costly for 
more general functionals. Finding strategies to identify and avoid this issue 
deserves more attention to avoid the future publication of code-dependent 
results obtained with in principle unstable parameter sets.

In principle, there also is a time-odd part of the electromagnetic interaction 
energy, mainly in the form of magnetic interactions. These are usually neglected,
such that their role for nuclear structure remains largely unexplored. Also, 
when evaluated for a time-reversal-breaking quasiparticle vacuum, there 
are also time-odd contributions to the pairing EDF. For the like-particle 
pairing interaction, these are often masked by a notation that works with, in general,
complex pair densities that do not have a definite symmetry under time reversal
\cite{RyssensN2LO}. Contrary to the particle-hole EDF, however, because of the
symmetry properties of pair densities, {time-odd terms in the} pairing functionals do not have coupling 
constants that could be chosen independently of those of the time-even part.

Another issue with the often made choices for the coupling constants is that using 
general or hybrid functionals, and for that matter also functionals derived 
from a density-dependent generator, unavoidably leads to spurious self-interactions 
that affect the energy differences between different blocked quasiparticle states, 
see section \ref{self-interaction:spe}. Even worse, using such a functional in projection 
and GCM leads to spurious finite steps or even singularities in the results 
\cite{(Dob07),(Lac09),(Ben09),(Dug09)}, see section \ref{self-interaction:rot}.
As of today, the only viable strategy to safely bypass all of these pathologies is to work 
with pseudopotential-based functionals (including the pairing part) that are constructed 
as the strict many-body matrix elements of an underlying \text{density-independent} 
generator that comprises two-body and (at least) three-body interactions. While 
first exploratory studies~\cite{(Bal14d),(Bal22a),(Bal23)} using such many-body 
Skyrme generators~\cite{(Sad13),(Sad13e)} are encouraging, this form of the EDF 
is insufficient and more flexible forms are needed, which might only be achievable 
using suitably constructed finite-range generators~\cite{(Rai14b),guillon26}.

\subsection{Time-odd mean fields in covariant density functional theory 
(CDFT)\label{sec:covariantDFT}}

  The origin of time-odd mean fields is different in the CDFT compared 
to non-relativistic DFTs. They are due to a magnetic potential $\bm{V} (\bm{r})$
\begin{eqnarray}
\bm{V}(\bm{r})=g_{\omega }{\bm{\omega}}(\bm{r})+ g_{\rho}\tau_3{\bm{\rho}}
(\bm{r})+e\frac{1-\tau _{3}}{2}\bm{A}(\bm{r}),
\label{magnetic}
\end{eqnarray}
which originates from the space-like components of the vector mesons. 
It enters into the Dirac equation,
\begin{eqnarray}
\hat{h}_{D}=\hat{\bm{\alpha}}\cdot\left(-i{\bm{\nabla}}-{\bm{V}}(\bm{r})\right)~+~V_{0}(\bm{r})~
+~\beta (m+S(\bm{r})),
\end{eqnarray}
in which it has the structure of a magnetic field. The presence of
this potential leads to currents
\begin{eqnarray}
{\bm{j}}^{n,p}(\bm{r}) &=& \sum_{i=1}^{N,Z} (\psi_i(\bm{r}))^{\dagger}
\hat{\bm{\alpha}} \psi_i(\bm{r}) \label{current}
\end{eqnarray}
which enter into the Klein-Gordon equation for vector $\omega$ mesons
\begin{eqnarray}
\left\{ -\Delta
+m_{\omega}^{2}\right\}
\omega_{0}(\bm{r}) & = &g_{\omega} [\rho_v^n(\bm{r}) + \rho_v^p(\bm{r})],  
 \label{KGomega0} \\
\left\{ -\Delta
+m_{\omega }^{2}\right\}
~{\bm{\omega}}(\bm{r}) & = &g_{\omega} [{\bm{j}}^n(\bm{r}) + {\bm{j}}^p(\bm{r})].
\label{KGomegav} 
\end{eqnarray}
 Note that in these equations, the four-vector components of the vector fields $\omega^{\mu}$,
$\rho^{\mu}$, and $A^{\mu}$ are separated into the time-like ($\omega_0$, $\rho_0$
and $A_0$) and space-like [${\bm{\omega}}=(\omega^x, \omega^y, \omega^z)$,
${\bm{\rho}}=(\rho^x, \rho^y, \rho^z)$, and ${\bm{A}}=(A^x, A^y, A^z)$] components.
The equations similar to equations~(\ref{KGomega0}) and (\ref{KGomegav}) exist also for 
the $\rho$-meson, but the contribution of this meson to time-odd mean fields is marginal 
\cite{AA.10}.

   The magnetic potential $\bm{V} (\bm{r})$ in the Dirac equation as well as the currents
${\bm{j}}^{n,p}(\bm{r})$ in the Klein-Gordon equations break time-reversal symmetry in the 
intrinsic frame and lead to the appearance of time-odd mean fields. The effect is often called \textit{nuclear magnetism} (NM) in the CDFT framework~\cite{(Koe89)}. The currents ${\bm{j}}$ together with densities $\rho$ form covariant 
four-vector $j^{\mu}=\{\rho, \bm{j}\}$. As a consequence, these two quantities ($\rho$ 
and $\bm{j}$) do not transform independently under Lorentz transformation. This explains 
why the structure of the Klein-Gordon equations for time-like and space-like components 
of vector mesons is the same and why the same coupling constant stands in front of the densities and currents on the right-hand side of these equations.

  The impact of time-odd mean fields on the energies of single-particle states has been studied in~\cite{Rutz98a,AA.10} and on 
the alignment properties of the single-particle orbitals and
total moments of inertia in rotating nuclei by~\citeasnoun{VALR.05,TO-rot}
and references quoted therein. However, there
are still open problems in our understanding of time-odd mean 
fields in the CDFT. These are:

\begin{itemize}
  \item 
   Experimental absolute (moments of inertia) and relative
   (effective alignments)  properties of rotational bands 
provide important constraints on time-odd mean fields since
they substantially affect these observables~\cite{VALR.05,TO-rot}. At 
present, they have been confronted in a systematic way only for non-linear 
meson exchange functionals, and such a test shows that the time-odd fields 
are well described in this class of functionals. However, such a stringent 
test has not been carried out for the PC and density-dependent ME 
functionals, which are frequently used in model calculations.

\item
In the rigorous CDFT equations
(\ref{KGomega0}--\ref{KGomegav}),
the Gallagher-Moszkowski splittings in odd-odd and 
even-even nuclei can only originate from current-current
interactions of unpaired particles. As the phenomenology of Gallagher-Moszkowski (GM) splittings undoubtedly points to a spin-spin interaction at its origin, see~\citeasnoun{Gallagher58a} and  section~\ref{sect:odd-odd:GM}, this question certainly deserves further study. 
     
\item 
  Comparative analysis of the emergence and properties of 
time-odd mean fields in covariant and non-relativistic
DFTs is mostly absent. At present, it is carried
out only for a simplified form of the covariant point-coupling 
model which does not contain gradient terms simulating the
effects of finite range \cite{Sulaksano07a,Sulaksano11a}.
These studies derived forms that can be directly compared with the Skyrme EDF. They underline the relevance of constructing the non-relativistic limit in such a way that one recovers a Galilean-invariant theory in the end,
which, however, is not possible for some terms of
the point-coupling model discussed by \citeasnoun{Sulaksano11a}.
As stressed in section \ref{timeoddskyrme}, all time-odd terms with either currents 
or other densities containing gradients are coupled by Galilean invariance to time-even terms and cannot be chosen independently. The only spin terms that could be chosen independently in a Skyrme EDF are spin-spin interactions 
{that are built solely using the spin density $\bm{s}_t (\bm{r})$
in equation~\eqref{eq:Skyrme_to}.}
Undoubtedly, further studies are needed  for covariant functionals
employing either density-dependent or non-linear meson-nucleon 
couplings and the most general form of the point-coupling model to
shed additional light on the 
properties of time-odd fields and provide useful constraints. 

\end{itemize}

It is necessary to recognise that there are some limitations which exist at 
the Hartree level of the CDFT but which can be resolved at the Hartree-Fock level. For example, spin-spin interactions are absent in the Lagrangian at the Hartree level in part because
the pion contribution vanishes at this level.  As a consequence, the Landau parameters $g_\ell$ and $g_\ell'$ are 
zero for all $\ell$ in such models. In calculations of spin- and spin-isospin excitations 
in (Q)RPA which always include pion, this poses an awkward problem, as the residual interaction is essentially zero 
(modulo perhaps a small contribution from the spin-orbit interaction), which contradicts 
the phenomenology of giant resonances. Because of this, one has to add a residual interaction 
in the spin- and spin-isospin channels that have no time-even counterpart. The usual choice 
is a contact spin-spin residual interaction which depends on the pion properties
that is parametrised through Landau parameters 
~\cite{Paar04a,(Nik05)}  and adjusted to experimental data. However, 
the transition to the Hartree-Fock level eliminates the need for adjusting the Landau
parameters \cite{BMGM.87,LGM.08}: the experimental data on the Gamow-Teller and spin-dipole resonances are well described in 
the RHF+RPA approaches in a fully self-consistent way.  This is also the reason why 
relativistic studies of polarised matter are made with relativistic HF models~\cite{Niembro90a,Tachibana25a}.

\subsection{The competition of tensor interaction and particle-vibration coupling}

\begin{figure}[htb!]
\centering
\includegraphics[width=0.8\textwidth, angle=0]{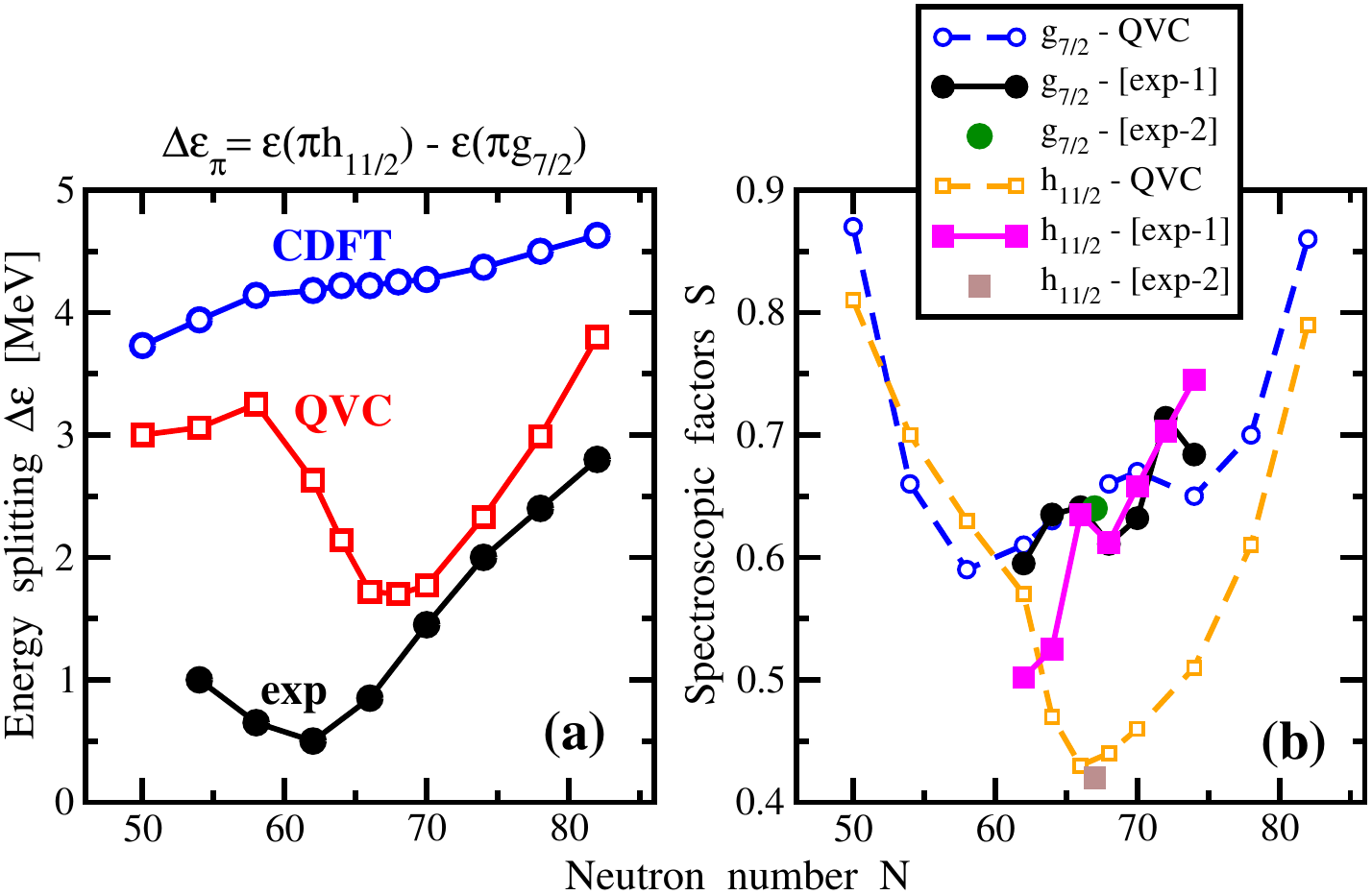}
\caption{(Panel (a)) The energy splitting $\Delta \varepsilon_{\pi}$ 
of the proton one-quasiparticle states with dominant
$h_{11/2}$ and $g_{7/2}$ single-particle content in the Sb isotopes. The results of the mean-field (CDFT) and beyond-mean-field (QVC) calculations with the NL3* functional
are compared with experimental data. (Panel (b)) 
Spectroscopic
factors $S$ of the one-quasiparticle states of the Sb isotopes  under study, compared with two sets of experimental data
('exp-1' is taken from~\citeasnoun{Z51-Sb}
and 'exp-2' from~\citeasnoun{(Gal85)}). 
Based on the results presented in~\citeasnoun{AL.15}. }
\label{tensor-spli}
\end{figure}

The relative energies of specific pairs of single-particle states emerging from different subshells are expected to be strongly affected by the tensor interactions~\cite{SS.14}. As a result, they were used in several papers to fit the strength of the tensor interaction at the mean-field level (see discussion in~\citeasnoun{AL.15} and
in section 3.2 of~\citeasnoun{SS.14}). The situation can be illustrated on the pairs  of the $\pi h_{11/2}$ and $\pi g_{9/2}$
states observed in the Sb $(Z=51)$ isotopes. Figure~\ref{tensor-spli}(a)
demonstrates that the experimental $\Delta \varepsilon_{\pi} = \varepsilon (\pi h_{11/2})- \varepsilon (\pi g_{9/2})$
splitting cannot be reproduced at the mean-field (CDFT) level: the same is true in non-relativistic DFTs (see 
discussion in section 3.2 of~\citeasnoun{SS.14}).
As illustrated in the references discussed and quoted in section 3.2 of~\citeasnoun{SS.14}, including the tensor interaction at the mean-field level can substantially reduce the difference between experimental and calculated $\Delta \varepsilon_{\pi}$ splittings. 

However, such an approach ignores the fact that observed states in odd-A Sb nuclei are not pure single-particle states (with spectroscopic factor $S=1.0$): strong vibrational admixtures
to the structure of these states are clearly seen in low experimental 
values of $S\approx 0.6$ (see figure~\ref{tensor-spli}(b)). 
The accounting of such vibrational admixtures within the relativistic quasiparticle-vibration coupling (QVC) model in~\citeasnoun{AL.15} allows to explain low experimental values of spectroscopic $S$ factors (see figure~\ref{tensor-spli}(b)) and reduce (substantially for $N\geq 64$)  
 the difference between calculated and experimental $\Delta \varepsilon_{\pi}$
 values  (see figure~\ref{tensor-spli}(a)) for the functionals which do not include tensor interaction.
 
   Thus one can conclude that both quasiparticle vibration coupling and effective tensor interaction act in the same direction. This leads to the following open problems:
   
\begin{itemize} 
\item 

The strengths of effective tensor interactions are defined exclusively at the mean-field level (see section 3.2 of~\citeasnoun{SS.14}). Including QVC in the fitting protocol (which has never been done before) will reduce these strengths. It remains to be seen whether such a refit of the functionals at the QVC level will allow us to obtain consistent improvement in the description of the single-particle properties (energy splittings
$\Delta \varepsilon_{\pi, \nu}$,
spectroscopic factors $S$) across the nuclear chart and define the strength of tensor interaction with a reasonable precision.

\item
  The competition of tensor interaction and particle-vibration coupling has not been investigated in deformed nuclei. However, general arguments suggest that their individual impact will be reduced. A dedicated study is needed to determine the magnitude of this reduction.

\end{itemize} 

  Pseudospin doublets are another example where particle-vibration coupling (PVC) substantially improves the description of splitting energies 
as compared with mean-field results
(see section IV.G in~\citeasnoun{LA.11}). Moreover, the observed
similarity of the experimental splitting energies of proton and neutron
pseudospin doublets with the same single-particle
structure in medium- and heavy-mass nuclei can only
be reproduced when the PVC is
taken into account
(see section IV.G in~\citeasnoun{LA.11}).

%%%%%%%%%%%%%%%%%%%%%%%%%%%%%%%%%%%%%%%%%%%%%%%%%%%%%%%%%%%%%%%%%%%%%%%%%%%%%%%%%%%%%%

\subsection{Adjustment of non-relativistic functionals to time-odd observables\label{sec:adjust}}

Adjusting novel density functionals to the dipole or octupole magnetic moments~\cite{(Dob25k),(Stu26a)} of odd nuclei is likely the most promising avenue for improving their properties in the time-odd sector of the mean field. This is not as easy as adjusting the time-even sector, which is done routinely. Indeed, it requires breaking not only spherical symmetry but also the time-reversal symmetry of the DFT calculation. In addition, even in the simplest physically interesting case of conserved axial symmetry, it requires aligning the odd particle's angular momentum along the symmetry axis. This latter requirement does not align well with existing DFT solvers, which usually use the $z$-axis to align the axial shape and the $y$-signature or $y$-simplex as s.p.\ symmetries. In such codes, one must thus work with a broken signature or simplex symmetry.

This is not an optimal setting for the large-scale adjustment of functional parameters, which requires a fast, stable method for obtaining self-consistent solutions. Therefore, in these projects, one would like to rewrite the codes to align shapes and angular momenta along the same axis. In principle, this does not seem to be a problem for triaxial codes, which have full flexibility in shape alignment. However, such codes are not only slower than axial codes but also routinely use the $z$-axis for angular-momentum restoration, which is required to determine spectroscopic magnetic moments and is again incompatible with their typical use of signature or simplex symmetry. Apart from the unpublished code {\sc hfbtemp}~\cite{(Kor26a)}, it appears that more code development is required to proceed.

The need to use deformed codes that break time-reversal symmetry to adjust the time-odd mean-field sector of novel functionals creates a mismatch between numerical efforts to reproduce time-even and time-odd observables. Bulk time-even observables, such as binding energies and radii, can be meaningfully adjusted using rapid codes in spherical symmetry. Therefore, it would be beneficial to develop hybrid algorithms that combine rapid adjustments of time-even observables with slow adjustments of the time-odd ones. To this end, one can use Tikhonov or ridge regression methods~\cite{(Hoe70a)} to drive the rapid adjustments toward the solutions preferred by the slow ones, alternating between the two. Alternatively, though probably less efficiently, one can mix gradients in the parameter space derived from both phases. Practical implementations of these ideas remain to be developed.

It also seems likely that emulators might be developed to replace evaluations of magnetic moments or other relevant observables at a fraction of the computational cost. Both deterministic emulators based on the reduced basis method - demonstrated already for the mean-field problem in spherical symmetry~\cite{bonilla2022a,giuliani2023a} and the RPA problem even for deformed states~\cite{hinohara2026,vancayseele2026} - or more physics-agnostic machine learning approaches~\cite{jin2025,lei2026} offer possible avenues of exploration.

A final possibility is simply accelerating the slow tools, either by implementing more advanced algorithms for the non-linear mean-field problem or by leveraging novel computing architectures. To the best of our knowledge, no currently existing solver that can access magnetic moments is capable of leveraging the GPU's and/or APU's that will surely dominate the next generation of HPC systems.

\newpage
\section{Self-interactions and singularities\label{singularities}}

One aspect of the discussion in this section concerns the configuration mixing of $K$ substates near sphericity; see section~\ref{K-mixing-spherical}.

\subsection{Minimisation after projection of deformed and/or cranked states\label{sec:MAP}}

In even-even nuclei, minimisation after projection (MAP)~\cite{(Sab07),(Rob11a)} has been an interesting approximate alternative to the full (and complex) VAP method~\cite{(Rin80b),(Sch19b)}. In particular, it improved the properties of quadrupole and octupole deformations. In this approach, self-consistent solutions are first obtained by constraining a set of deformations, and then energies of the symmetry-restored states are minimised within this set.

As shown in parallel by~\citeasnoun{(Zdu07c)}, the method does not seem to work correctly for cranked states projected on good angular momentum and minimised over the cranking frequency. Here, we test its deformation properties in odd nuclei.
    
	\begin{figure}[h]
		\centering
			\includegraphics[width=0.98\textwidth]{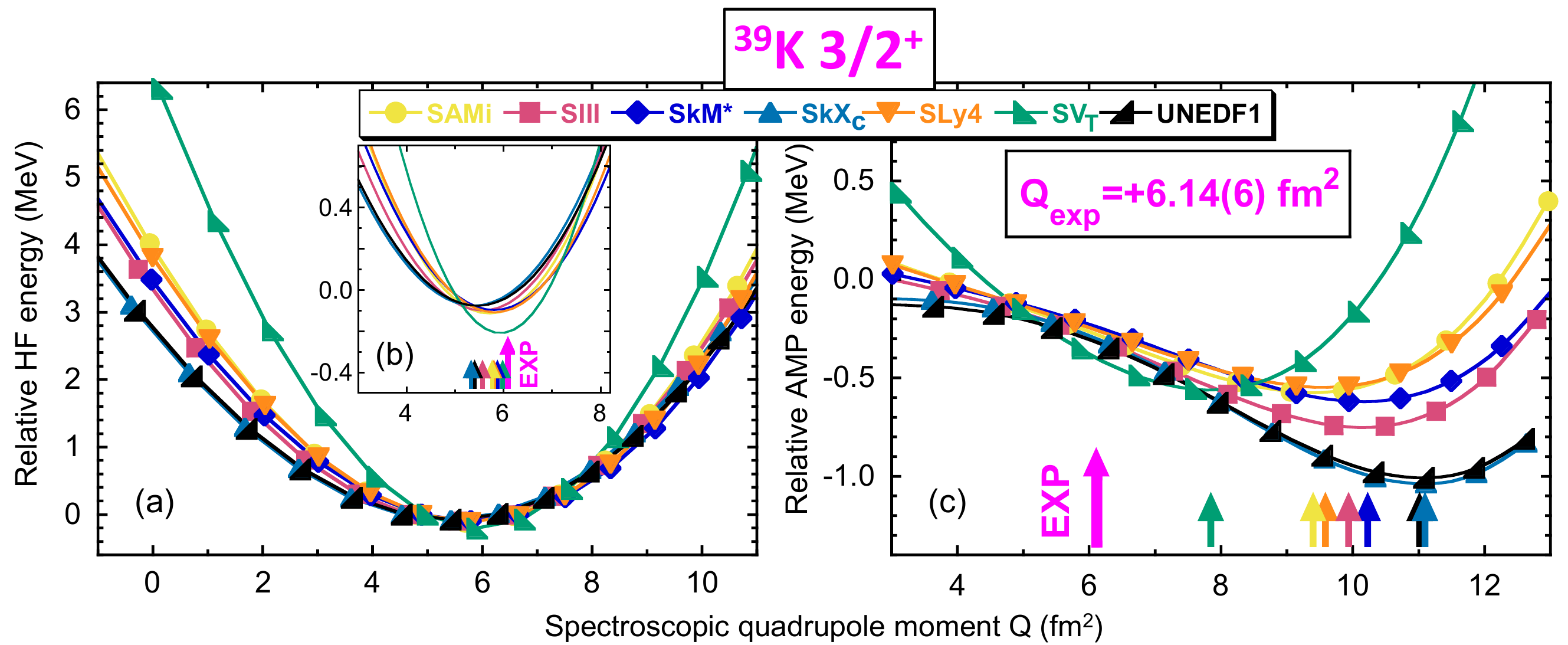}        
		\caption{Relative HF energies, panels (a) and (b), compared with those of the HF+AMP, panel (c), determined for the $3/2^+$ ground state of $^{39}$K as functions of the spectroscopic quadrupole moment $Q$; see text.
		\label{MAP}}
	\end{figure}
In figure~\ref{MAP}, we show the results of the HF calculations, panels (a) and (b), compared with those of the HF+AMP, panel (c), performed for the $3/2^+$ ground state of $^{39}$K, constrained to a sequence of the total intrinsic quadrupole moments $Q^{\text{tot}}_{20}=Q^{\text{n}}_{20}+Q^{\text{p}}_{20}$ and plotted in function of the PAV spectroscopic quadrupole moments $Q$. All panels present the energies relative to the HF energies constrained to $Q^{\text{tot}}_{20}=0$, determined for seven Skyrme functionals:
SAMi~\cite{(Roc12c)},
SIII~\cite{(Bei75b)},
SkM*~\cite{(Bar82c)},
SkX$_{\text{c}}$~\cite{(Bro98c)},
SLy4~\cite{(Cha98a)},
SV$_{\text{T}}$~\cite{(Bei75b),(Sat14g)},
UNEDF1~\cite{(Kor10c)}.

As shown in figure~\ref{MAP}(b), all Skyrme functionals consistently predict the minima of the HF energies (short arrows) clustered near the experimental value of $Q_{\text{exp}}$=6.14(6)\,fm$^2$ by~\citeasnoun{(Pyy18)}, shown by the long arrow with its width indicating the experimental uncertainty. Conversely, the energies of the $3/2^+$ symmetry-restored AMP states show minima ranging from $Q_{\text{AMP}}$=9.4\,fm$^2$ to 11\,fm$^2$ (density-dependent functionals) or at $Q_{\text{AMP}}$=7.9\,m$^2$ (density-independent functional SV$_{\text{T}}$), which are far from the experimental values. In this case, we observed no singularities from self-interactions; however, self-interactions are most likely responsible for the markedly different results obtained with the density-dependent and density-independent functionals.  

Further investigation into the relationship between the PAV method and the VAP or MAP methods for odd nuclei appears necessary. 

\subsection{Impact of self-interactions on rotational energies\label{self-interaction:rot}}

	\begin{figure}[h]
		\centering
		\includegraphics[width=0.90\textwidth]{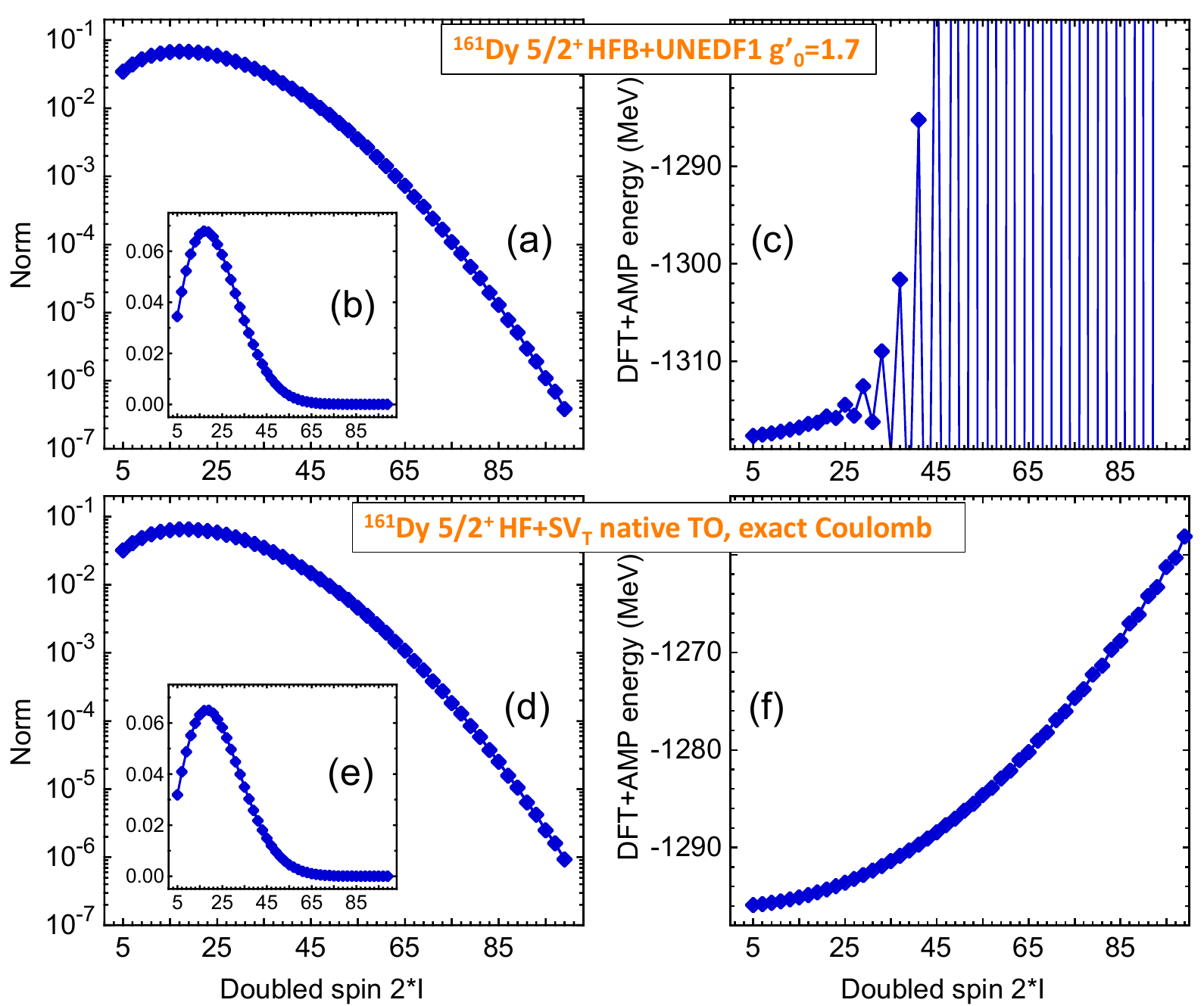}
		\caption{Properties of the $^{161}$Dy, $K^\pi=5/2^+$ rotational band calculated for functionals UNEDF1~\protect\cite{(Kor10c)}, panels (a)--(c), and SV$_{\text{T}}$~\protect\cite{(Bei75b),(Sat14g)}, panels (d)--(f), see text}
		\label{slide7}
	\end{figure}
To illustrate the disastrous effects of self-interaction and/or self-pairing on rotational-band energies, we use the example of the $^{161}$Dy $K^\pi=5/2^+$ rotational band calculated by~\citeasnoun{(Dob26g)}. In figure~\ref{slide7}, we show the calculated norms $N(I)=\langle\Phi|\hat{P}^I_{KK}|\Phi\rangle$ and energies $E(I)=\langle\Phi|\hat{H}\hat{P}^I_{KK}|\Phi\rangle/N(I)$ for the HFB AMP states, with $|\Phi\rangle$ taken as the intrinsic deformed state for the blocked [642]5/2 quasiparticle with the angular momentum aligned along the axial-deformation axis. Figure~\ref{slide7}(c) shows the HFB AMP energies obtained for the UNEDF1~\cite{(Kor10c)} functional, which exhibits self-interactions and/or self-pairing due to several possible effects:
\begin{itemize}
\item
mismatch between the time-even (TE) and time-odd (TO) mean fields,
\item
mismatch between the mean-field and pairing terms
\item
density dependence of the mean-field and pairing terms,
\item
density dependence of the Slater approximation of the Coulomb exchange term.
\end{itemize}

We observe that beyond $I=25/2$, the energy values clearly diverge, while the norms are precisely evaluated up to a very high angular momentum of $I=95/2$. As with the norms, the one-body electromagnetic moments, determined by~\citeasnoun{(Dob26g)}, are unaffected by self-interactions and self-pairing.

For comparison, the HF AMP energies obtained for the SV$_{\text{T}}$~\cite{(Bei75b),(Sat14g)} functional, with the Coulomb direct and exchange terms evaluated exactly, as shown in figure~\ref{slide7}(f), exhibit no divergence. This is because, by definition, the SV$_{\text{T}}$ functional corresponds to the exact average of a two-body functional generator and is thus unaffected by self-interactions.    

\subsection{Impact of self-interactions on single-particle energies\label{impact_on_spe}}
\label{self-interaction:spe}

\begin{figure}[h]
\hfill\includegraphics[width=0.4\textwidth]{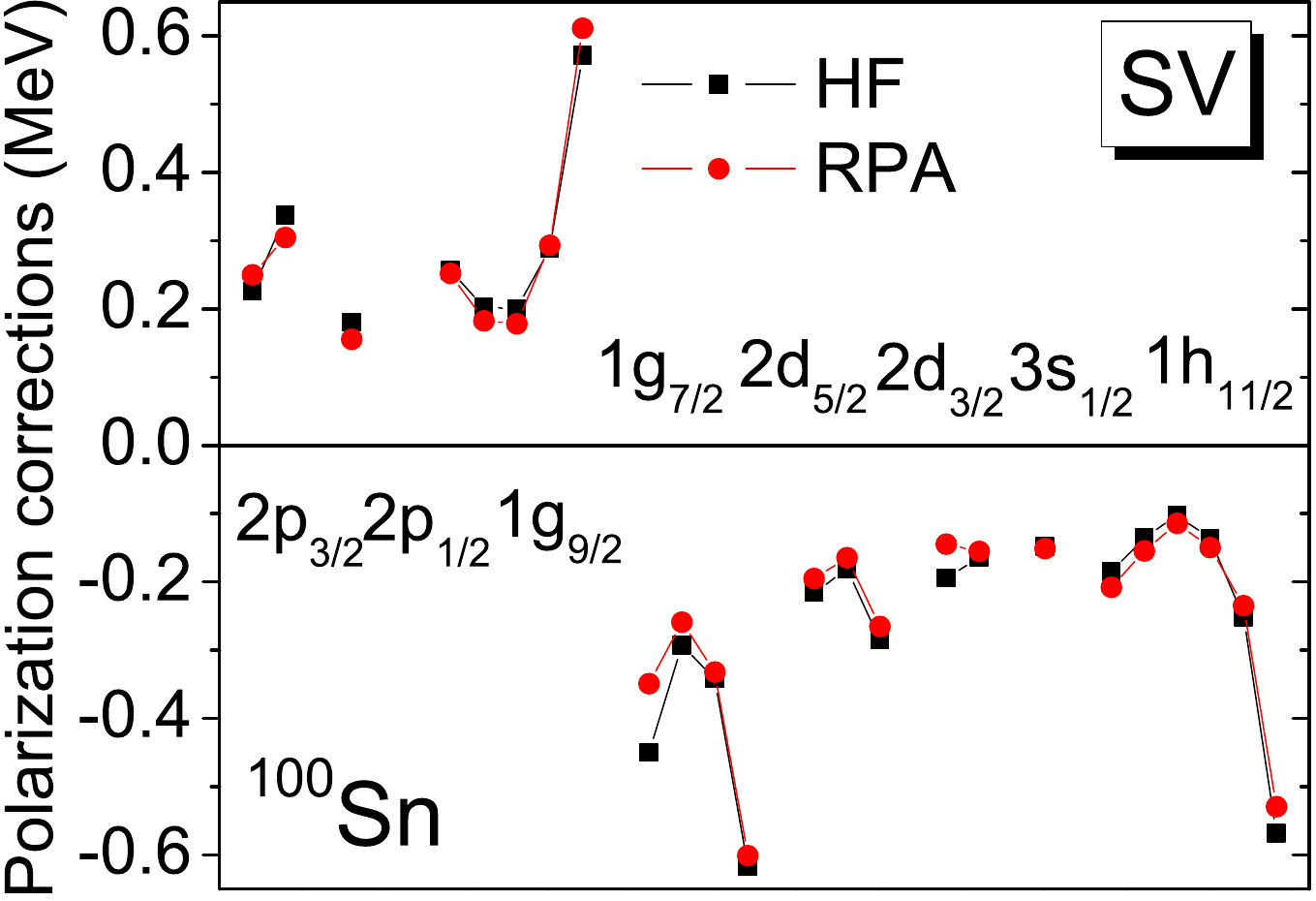}
    ~~\includegraphics[width=0.4\textwidth]{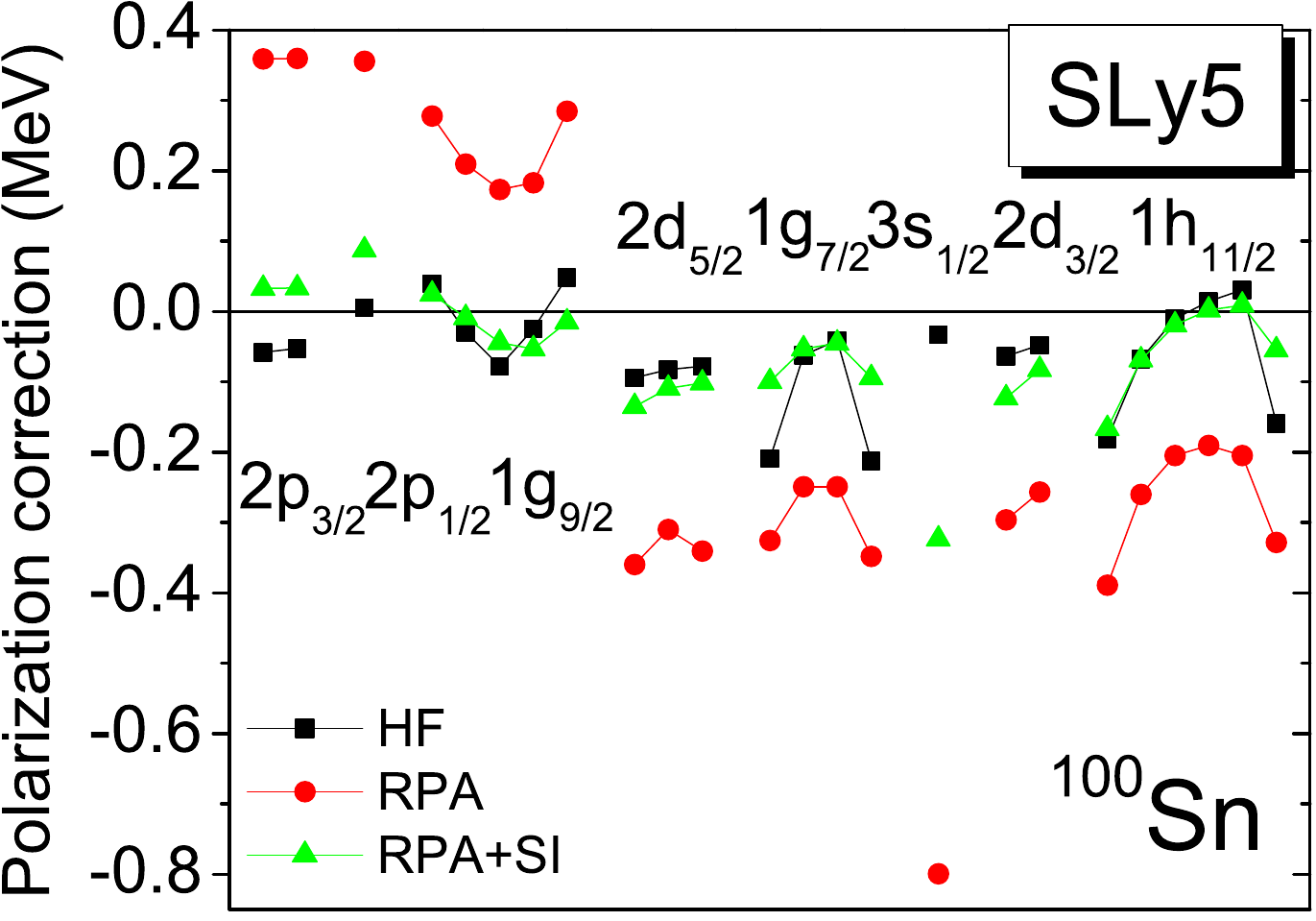}
		\caption{Calculated HF s.p.\ energies in $^{100}$Sn (squares) compared with those obtained using the RPA method with (triangles) and without (circles) self-interaction contributions. The left and right panels show results obtained with the SV$_{\text{T}}$~\protect\cite{(Bei75b),(Sat14g)} and SLy5~\protect\cite{(Cha98a)} Skyrme functionals. Reprinted with permission from~\protect\citeasnoun{(Tar14a)}. Copyright (2014) by the American Physical Society.\label{slide9}}
	\end{figure}

In nuclear physics, a body of studies determines the particle-vibration corrections (PVCs)~\cite{(Dic05)} to HF s.p.\ energies for Skyrme functionals; see~\citeasnoun{(Tar14a)} and the references cited therein. As shown by~\citeasnoun{(Bla86b)} and re-derived by~\citeasnoun{(Tar14a)}, under several simplifying assumptions and for density-independent functionals, the self-consistent HF energies of odd nuclei indeed include the RPA PVCs. Numerical examples in~\citeasnoun{(Tar14a)} illustrate this; see the left panel of figure~\ref{slide9}.

However, for standard density-dependent functionals, which include self-interaction terms, the HF results agree with RPA PVC only when these terms are included in the latter; see the right panel of figure~\ref{slide9}. Without them, the HF and correct RPA PVC results disagree catastrophically. This shows that the HF masses of odd nuclei are driven predominantly by self-interaction terms, are thus incompatible with experiment, and should not be adjusted to experimental data.

We conclude that meaningful constraints on the time-odd mean fields can be derived from the masses of odd nuclei only for future self-interaction-free functionals. Otherwise, this task would require adjustments to s.p.\ energies corrected by the RPA-PVC, which is difficult, especially in deformed nuclei. In addition, the masses of odd nuclei are affected not only by time-odd mean fields but predominantly by pairing correlations, which adds another layer of difficulty.

\clearpage
\section{Collective models\label{sec:collective}}

\hide{EDF calculations for collective
motion in even-even nuclei have
been extensively performed and,
in most cases, successful in describing
observed systematics.
Major questions arise: whether these models
have equal levels of accuracy in
computing collective properties
of odd nuclei, and what would be
cures for conceptual and technical
difficulties arising specifically
in odd nuclear systems, cf.~section~\ref{Theory}.
To construct certain collective models,
for instance, one should choose
appropriate collective degrees of
freedom to obtain potential energy
surfaces, such as triaxial, pairing,
etc., which are particularly relevant
to the modelling of fissions.
Traditional microscopic approaches
to collective motions in odd nuclei
include particle-rotor coupling,
cranking, etc.
Calculations of low-lying properties
in odd nuclei are made
by means of the beyond mean-field
method, which requires taking into account
symmetry restorations (angular momentum,
particle number, time-reversal),
blocking effects, and
configuration mixing.
Approximate methods that make the
calculations for odd nuclei
more feasible have been
developed, e.g., in terms of the IBFM,
while some other promising approaches,
such as those based on the quasiparticle RPA (QRPA) method,
are also available.}

\subsection{Schr\"odinger Collective-Intrinsic Model to describe odd 
nuclei in nuclear structure and fission}

Recently developed one-dimensional implementation of the Schr\"odinger Collective-Intrinsic Model (SCIM) 
in the context of even-even fissioning system studies has opened an alternative way of treating 
odd nuclei~\citeasnoun{TPaul,carpentier2024,trilogy1,trilogy2,trilogy3,bernard1} and excited states, in a more general way.
Indeed, in this model, the main idea is to couple single-particle and collective degrees of 
freedom by enlarging the GCM wave function ansatz to intrinsic excitations:
\begin{eqnarray}\label{eq:SCIM1}
\ket{\Psi_{\rm SCIM}} = \int dq^n f_0(q) \ket{\Phi_0(q)} + \sum_{i=1}^{N}
 \int dq^n f_i(q) \ket{\Phi_i(q)}
\end{eqnarray}
where $q$ represents the set of the collective coordinates, $\{\ket{\Phi_0(q)}, ~f_0(q) \}$ the set of the HFB lowest energy 
solutions and $\{\ket{\Phi_i(q)}, ~f_i(q) \}$ the set of intrinsic excitations.

Then, we transform the associated non-local Hill-Wheeler equation, obtained from a variational principle, using the Symmetric Ordered Product of Operators techniques and truncate it at second order. This last approximation leads to a local Schr\"odinger-like equation 
which defines a Hamiltonian that incorporates non-locality and naturally introduces couplings between collective and intrinsic degrees of freedom.
In particular, a first-order additional term appears, introducing first derivatives with respect to the collective degrees of freedom.
In the context of fission, two main difficulties have been encountered in applying the SCIM formalism: (i) discontinuities in the state between two adjacent HFB states already at the adiabatic level and (ii) level repulsions. Those induce unsolvable discontinuities when excited states are built as two-quasiparticle states, as originally proposed by~\citeasnoun{bernard1}.
The solution proposed by~\citeasnoun{carpentier2024,TPaul,trilogy1,trilogy2,trilogy3} was to perform HFB calculations with overlap constraints associated with both continuity and orthogonality. In particular, we used deflation techniques to generate excited states as HFB vacua, thereby solving the problem of level-repulsion crossings. The associated protocols (\enquote{Link}, \enquote{Drop}, and \enquote{Continuous Deflation})
were independent of the symmetries imposed on the system and defined a new collective variable.

Even if the first application of the SCIM has been done for even-even fissioning systems, it is clear that this formalism is able to
treat odd nuclei from both structure and fission perspectives. In particular, static calculations in the first or second well will allow spectroscopic studies (spectra, magnetic moments, half-lives of fission isomers, etc.). Dynamical calculations (based on the collective-intrinsic Hamiltonian) that require potential energy surfaces spanning from the first prolate well to the scission area and beyond will make it possible to describe fission of odd systems.
The main remaining challenges and difficulties will be related to the
introduction of time-reversal symmetry breaking, the two-dimensional extension of the SCIM with two new collective coordinates
(which ideally will allow taking into account axiality and parity breaking) and the projection on the total angular momentum $J$.

\subsection{EDF-mapped interacting boson-fermion model for describing spectroscopy of odd nuclei}

This section gives basic formulations
and open problems concerning a
method developed in the last decade
\cite{nomura2016odd,homma2025},
for computing low-energy spectra
and transition properties in odd nuclei
using the interacting boson-fermion model (IBFM)
based on the nuclear EDF.
This approach could be an alternative promising
EDF-based method allowing for detailed
spectroscopic calculations in medium-heavy
and heavy nuclei, and has by now been
employed in numerous nuclear structure,
$\beta$ decay, and $\beta\beta$ decay
studies.

The IBFM consists of the even-even
boson core, which is described in terms
of the interacting boson model (IBM),
and an unpaired nucleon degree of
freedom.
The coupled boson-fermion Hamiltonian
is given in general as
\begin{eqnarray}
\label{eq:ham}
 \hat H = \hat H_{\rm B} + \hat H_{\rm F}
+\hat V_{\rm BF} \; .
\end{eqnarray}
$\hat H_{\rm B}$ is the even-even
core Hamiltonian.
$\hat H_{\rm F}$ is
the single-fermion Hamiltonian
$\hat H_{\rm F} = - \sum_{j}\sqrt{2j+1} \epsilon_j
(a_j^{\dagger}\tilde a_j)^{(0)}$,
and $\hat V_{\rm BF}$
denotes the boson-fermion interaction.
The assumption of the IBM is
to approximate collective monopole
and quadrupole pairs of
valence nucleons to effective
$s$ and $d$ bosons.
The IBM Hamiltonian giving rise
to energies and wave functions for
excited states is expressed
by interaction terms formed by
$s$ and $d$ bosons with relevant strength
parameters.
The method of deriving the
IBM strength parameters for even-even nuclei
from the EDF-SCMF calculation
was developed,
and has been extensively used
for nuclear structure calculations
(see the review by~\citeasnoun{nomura2025rev}
and references therein).
The method consists of mapping
the SCMF energy surface
$E_{\mathrm{EDF}}(\beta,\gamma)$
onto the equivalent energy surface in
the boson system $E_{\mathrm{B}}(\beta,\gamma)$
so that
$E_{\mathrm{EDF}}(\beta,\gamma) \approx E_{\mathrm{B}}(\beta,\gamma)$
should hold.
The bosonic energy surface is
obtained as the expectation value
of the bosonic Hamiltonian
in the condensate state of $s$
and $d$ bosons,
\begin{eqnarray}
\label{eq:coherent1}\hspace*{-10mm}
 |{\phi_{\mathrm{B}}}\rangle
=(1+\beta^2)^{-1/2}\left[
s^{\dagger}+\beta\cos{\gamma}d^{\dagger}_{0}
+\frac{1}{\sqrt{2}}\beta\sin{\gamma}
(d^{\dagger}_{+2}+d^{\dagger}_{-2})
\right]^{N_{\mathrm{B}}}
|0\rangle
\; ,
\end{eqnarray}
where the amplitudes
$\beta$ and $\gamma$ represent
quadrupole deformations,
$N_{\mathrm{B}}$ is the number
of bosons, and the state $|0\rangle$ is a boson vacuum representing the inert core.

Essential boson-fermion interaction
$\hat V_{\rm BF}$ for the odd nuclei
is of the form~\cite{IBFM},
\begin{eqnarray}
\label{eq:vbf}
 \hat V_{\rm BF}
&=&\sum_{jj'}\Gamma_{jj'}
(a^{\dagger}_{j}\times\tilde{a}_{j'})^{(2)}
\cdot\hat{Q}
\nonumber \\
&+&\sum_{jj'j''}\Lambda_{jj'j''}
:\left[(a_{j}^\dagger\times\tilde d)^{(j'')}
\times
(d^\dagger\times\tilde a_{j'})^{(j'')}\right]^{(0)}: \\
&+&\sum_{j}A_{j}\hat{n}_{d}\hat{n}_{j} \; ,
\nonumber
\end{eqnarray}
where the first, second, and
third terms are referred to as
quadrupole dynamical (direct),
exchange, and monopole terms,
respectively.
$\hat Q$ is the quadrupole operator
in the boson system.
The coefficients for the three terms
in equation~(\ref{eq:vbf}) are conveniently
formulated by using the generalised
seniority scheme, and take the following
forms~\cite{IBFM}:
\begin{eqnarray}
\Gamma_{jj'}
&=&\Gamma_0\gamma_{jj'}^{(2)}
\;,\nonumber \\
\Lambda_{jj'j''}
&=&-2\Lambda_0\sqrt{\frac{5}{2j''+1}}
\beta_{jj''}\beta_{j'j''}
\;,  \\
A_{j}&=&-A_0\sqrt{2j+1}
\; , \nonumber
\end{eqnarray}
with $\Gamma_0$, $\Lambda_0$, and $A_0$
denoting strength parameters.
Here the coefficients
$\gamma_{jj'}=(u_{j}u_{j'}-v_{j}v_{j'})q_{jj'}$
and
$\beta_{jj'}=(u_{j}v_{j'}+v_{j}u_{j'})q_{jj'}$
are specified by the occupation $v_j$
and unoccupation $u_j$ amplitudes.
$q_{jj'}$ stands for the matrix element
of the fermion quadrupole
operator in the single-particle
basis.

The problem of the conventional
IBFM calculations
was that a number of parameters
involved were directly fit
to the experimental low-energy spectra.
The method of~\citeasnoun{nomura2016odd}
allowed fixing most of these parameters
from the EDF calculations:
The $v_j$ and $u_j$ values
are determined by the SCMF
calculations for odd nuclei,
constrained to zero
deformation;
Blocking is not carried out in
this framework, but with the particle
numbers constrained to odd numbers;
The microscopically calculated
$v_j$ and $u_j$ values
are plugged into the above formulas,
and the spherical single-particle
energy $\epsilon_j$ is another
input from the EDF calculation
necessary for $\hat H_{\mathrm{F}}$.
However, in the method of~\citeasnoun{nomura2016odd},
the remaining coupling constants
$A_0$, $\Gamma_0$, and $\Lambda_0$
for the boson-fermion interaction
still had to be determined
in a phenomenological manner
so as to reproduce the low-energy levels
for each odd nucleus.

The use of the seniority argument
is assumed to be valid for nearly
spherical nuclei and hence may not
be appropriate for deformed odd nuclei.
From that point of view,
significant progress was made quite recently 
with the EDF-mapped IBFM,
so that all the parameters should
be determined fully microscopically
by the EDF calculation
\cite{homma2025}.
The principal idea is to use the
geometry of the boson-fermion system.
By taking an expectation value of
the Hamiltonian (\ref{eq:ham})
in the boson coherent state (\ref{eq:coherent1})
and by integrating out the boson
part, the deformed single-particle
Hamiltonian is obtained as follows,
\begin{eqnarray}\label{eq:ham2}\hspace*{-25mm} 
\hat H_{\mathrm{IBFM}}(\beta,\gamma)
= E_{\mathrm{B}}(\beta,\gamma)
+ \sum_{j_1,j_2,m_1,m_2}
g_{j_1m_1,j_2m_2}(\beta,\gamma)
\frac{a^{\dagger}_{j_1m_1}a_{j_2m_2}+
a^{\dagger}_{j_2m_2}a_{j_1m_1}}
{1+\delta_{j_1j_2}\delta_{m_1m_2}}
\; .
\end{eqnarray}
The above Hamiltonian is diagonalised
in the spherical basis $|{jm}\rangle$.
In the case of single-$j$ problems,
and assuming further the axial symmetry,
the Hamiltonian (\ref{eq:ham2})
only has diagonal elements,
which are the energy curves
for different projections $K$:
\begin{eqnarray}
\label{eq:pes2}
 E_{\mathrm{IBFM},K}(\beta)
=E_{\mathrm{B}}(\beta)+\lambda_{K}(\beta) \; .
\end{eqnarray}
The entry $\lambda_{K}(\beta)$
was shown to be analogous to Nilsson orbits.

\begin{figure}
\begin{center}
\includegraphics[width=.4\linewidth]{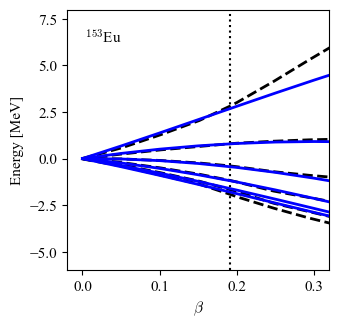}
\caption{
Deformed single-particle spectra
$\epsilon_{K}(\beta)$
for $^{153}$Eu calculated by using the HFBTHO
with the Skyrme-SkM* force (dashed lines),
resulting from the proton $h_{11/2}$
spherical orbit,
and the IBFM counterparts
$\lambda_K(\beta)$ (solid lines).
Different single-particle levels
are labelled by asymptotic quantum
numbers.
Figure based on~\cite{homma2025}.
The vertical dotted line stands for
the minimum in the energy curve
for the odd nucleus.
}
\label{fig:spe}
\end{center}
\end{figure}

\begin{figure}
\begin{center}
\includegraphics[width=.49\linewidth]{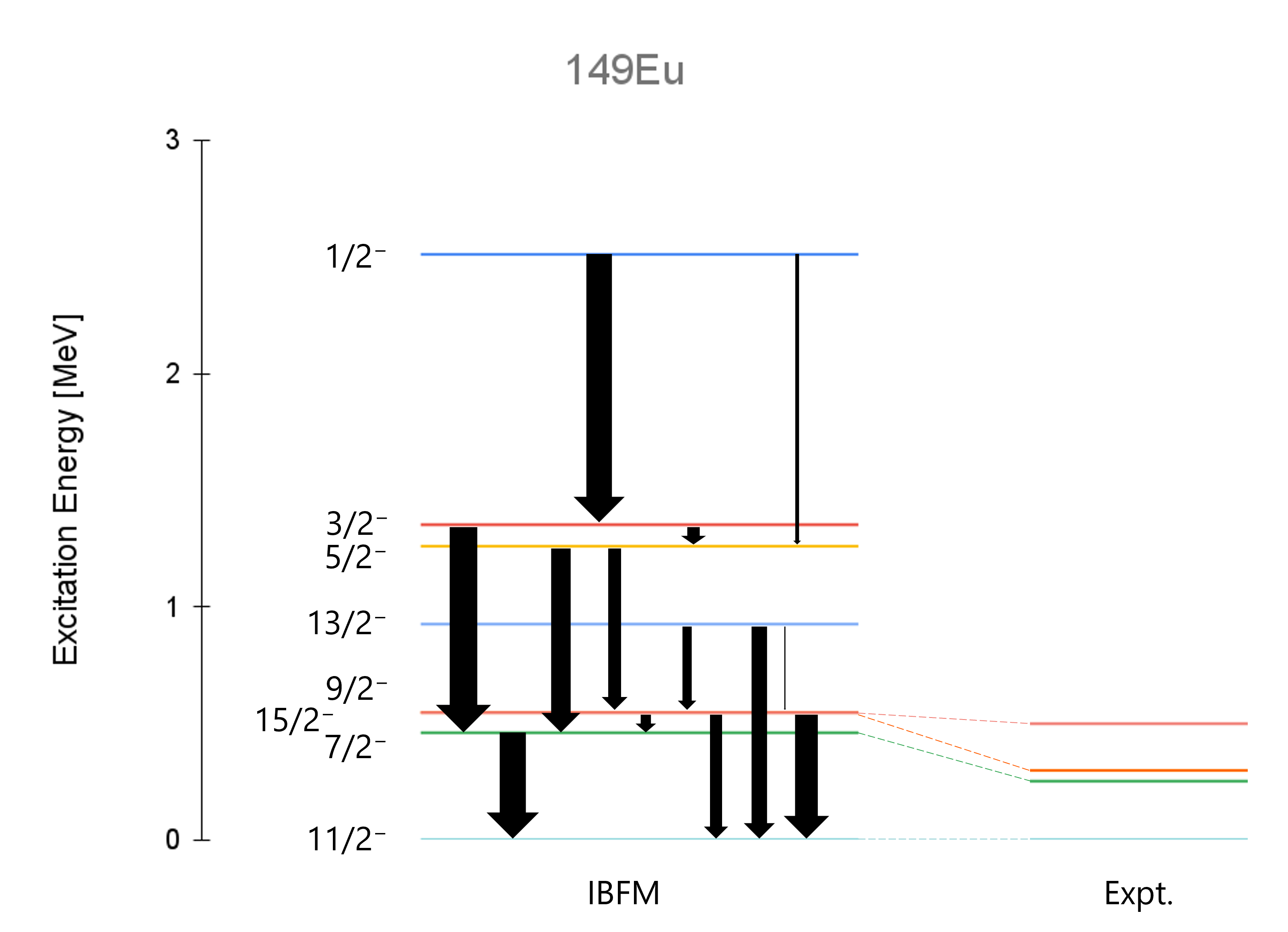}
\includegraphics[width=.49\linewidth]{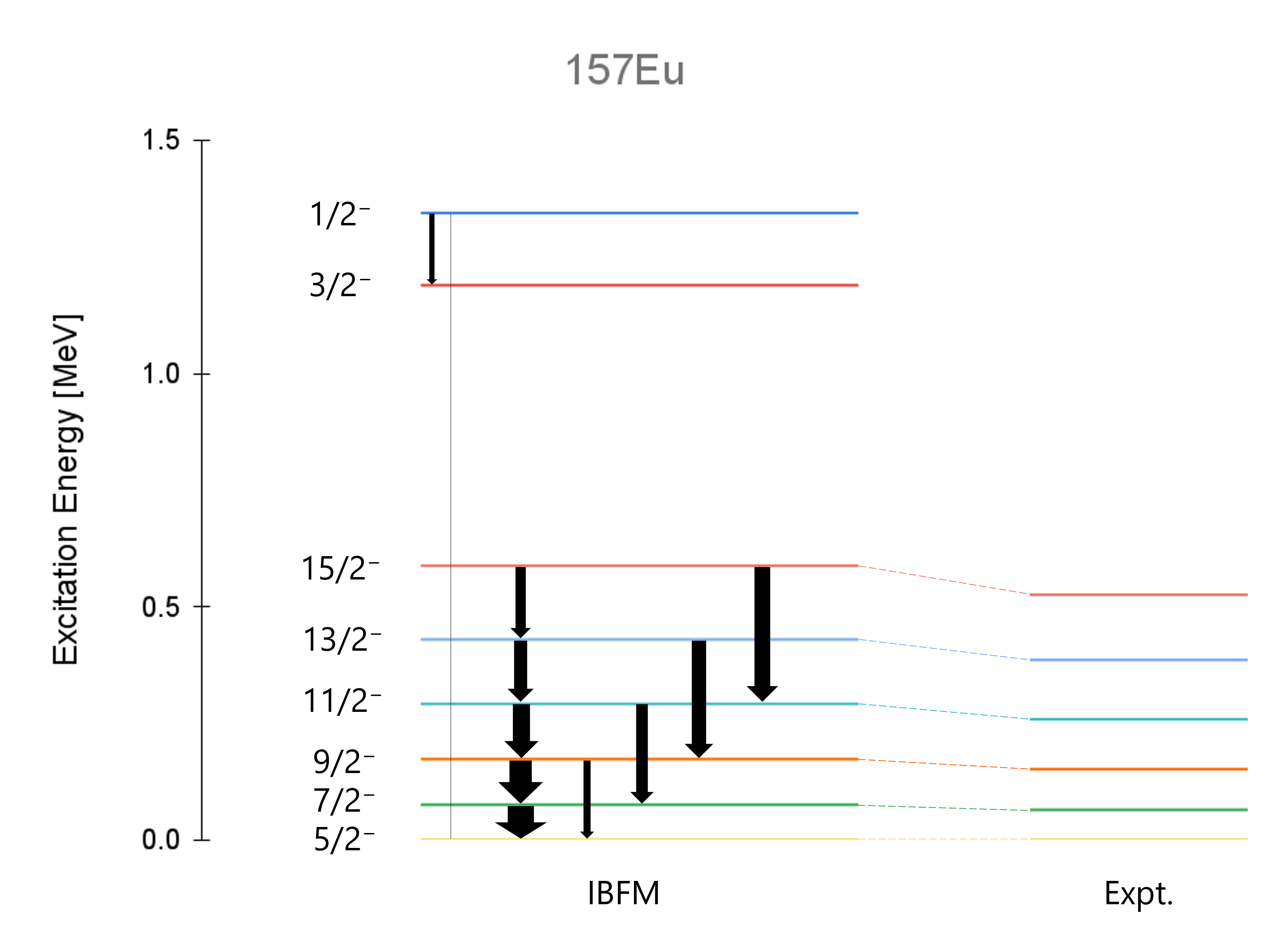}
\caption{
Calculated excitation energies of
low-lying negative-parity states for
$^{149}$Eu and $^{157}$Eu in the
EDF-mapped IBFM.
Arrows represent $E2$ transition
strengths, and the thickness of the
arrow is proportional to
the $B(E2)$ value.
The experimental data are adopted
from NNDC database~\cite{data}.
}
\label{fig:eu-level}
\end{center}
\end{figure}

In the new method~\cite{homma2025}
the parameters for the bosonic
core Hamiltonian are determined
by matching
$E_{\mathrm{B}}(\beta)$ to the
corresponding EDF energy surface
$E_{\mathrm{EDF}}(\beta)$,
and the three boson-fermion strength
parameters are then determined so that
the behaviour of the deformed
single-particle levels
$\lambda_{K}(\beta)$ be equal to
that of the EDF counterpart for
an odd-$A$ nucleus
denoted $\epsilon_{K}(\beta)$.
Figure~\ref {fig:spe} shows an example.
The $\epsilon_{K}(\beta)$ was computed 
with the HFBTHO code (v4.00),
in which the Skyrme SkM* force
was employed, and blocking effects were
taken into account at each $\beta$
and for all possible single-particle
orbits.
All the model parameters for the IBFM
were thus obtained only from the
EDF inputs. The major consequences were 
the ability to reproduce
low-energy negative-parity levels in
odd-$Z$ Eu isotopes exhibiting
structural evolution from
near vibrational to rotational
limits without any adjustment to the
experimental data.
Figure~\ref{fig:eu-level}
depicts the calculated
negative-parity energy levels
and preliminary results for
the $B(E2)$ transitions for the nearly
spherical $^{149}$Eu and
axially deformed $^{157}$Eu nuclei.

%-----------------------------------------------------------
%
%
%
%-----------------------------------------------------------

%-----------------------------------------------------------
%
%       Eu spectra
%
%-----------------------------------------------------------

There are several unsolved problems and future
directions of the EDF-mapped IBFM method.
\begin{itemize}
\item
A straightforward extension of the method
of~\citeasnoun{homma2025}
is to include multi-$j$ orbits,
which will allow us to study states
of both parities and predict more
complicated electromagnetic transition
properties.
In this case, the Hamiltonian in a
matrix form (\ref{eq:pes2}) should be diagonalized.
A problem is that for the multi-$j$
shells the quantities
$g_{j_1m_1,j_2m_2}(\beta,\gamma)$
of (\ref{eq:pes2})
are not given in a compact form
as in the case of the single-$j$ shell.
While the main focus is placed on
the properties of deformed odd nuclei,
one may have to go back to the formalism in
the generalised seniority scheme
and consider the $v_j$ and $u_j$
amplitudes for different orbits $j$,
which quantities one did not need to
consider for the single-$j$ problem.
The validity of employing
the seniority consideration
in this case should be examined.

 \item
The like-neutron bosons should be
distinguished from the like-proton bosons
in terms of the neutron-proton version
of the IBFM (IBFM-2).
The distinction will be essential especially
when one studies collective states in which
neutron and proton contributions should be treated
on the same footing, e.g., mixed symmetry states
including scissors mode and enhanced
$B(M1;1^+ \to 0^+_{gs})$,
and nuclear decay processes such as
$\beta$ decay and $\beta\beta$ decay,
for which the two degrees of freedom
should necessarily be considered.
The work along this line is progress
in the case of the proton $h_{11/2}$ orbit.
A preliminary result for the IBFM-2
deformed single-particle levels
in comparison with the Skyrme EDF ones
for $^{153}$Eu has been obtained with
the deformed single-particle levels
similar to those shown in figure~\ref{fig:spe}.
The derived boson-fermion strength
parameters for the IBFM-2 are,
however, anomalously large
as compared with the empirical values.
The reason for this is that
the intrinsic-state formulation
of the boson-fermion interactions
in the IBFM-2 could be quite different
from that in the case of the
IBFM-1.
The next step is to interpret this
result and reformulate the IBFM-2
boson-fermion interactions.

\item
The present version of the framework
assumes axial symmetry and
only the quadrupole shape is considered.
These restrictions should be lifted.
To include the triaxial (quadrupole)
degree of freedom,
referring to the Nilsson-like
single-particle spectra alone may not be
sufficient, and therefore
one can generalise
the matrix (\ref{eq:pes2})
to cover both $\beta$ and $\gamma$
deformations, as was recently done by
\citeasnoun{MAYABARBECHO2025},
to consider the energy surface
in the $(\beta,\gamma)$ space.
Among the additional collective
coordinates that are relevant to
low-lying states is the octupole
deformation.
As far as the axial symmetry is
assumed, an extension to include
the octupole degree of freedom could
be done by associating
$\lambda_{K}(\beta_{20},\beta_{30})$
with the HFB energy curve along
$\beta_{20}$ deformation and with
that along $\beta_{30}$
separately.
Whether this procedure works for
realistic calculations and whether
it is feasible, is still to be examined.

\item
In many of the heavy nuclei
considerable degrees of shape mixing
and coexistence are present,
which are inferred from the
softness in certain deformations
and appearance of local minima
in the potential energy surfaces.
The phenomenon of shape coexistence in
even-even systems has
been studied by numerous mean-field
and beyond approaches,
but that of the odd nuclei
would be quite complicated.
The present framework can be
extended to include effects of
configuration mixing of normal and
intruder states,
coming from the next major shell.
This version of the IBFM
is often referred to as IBFM-CM,
in which case two independent
IBFM Hamiltonians differing
in boson numbers should be
introduced, hence more parameters.
Developing an efficient way of determining
these parameters from the
EDF calculations will be
an interesting future study.

\item
The IBFM for odd nuclei could also
be extended to treat more challenging
cases of odd-odd nuclei.
For the calculations for odd-odd nuclei
an unpaired neutron and an unpaired proton
degrees of freedom, their mutual coupling,
and couplings to the even-even core
would have to be
explicitly considered.
The method of~\citeasnoun{nomura2016odd},
which is based on the seniority scheme,
was extended to the odd-odd cases
\cite{nomura2019dodd}.
However, the residual interaction
between the unpaired neutron and proton
as well as the boson-fermion interactions
were phenomenological parameters
that were determined to fit the
energy spectrum of each odd-odd nucleus.
The way to solve this is to
extend the new method of~\citeasnoun{homma2025}.
This would be feasible to
derive the coupling constants for the
boson-fermion interactions,
but how to determine from the EDF
calculations the
residual neutron-proton interaction
is an open question.

\end{itemize}
These are challenging
problems that require extensions
in many aspects and verification of
the EDF-based IBFM method,
but this approach will also potentially
allow for a systematic EDF description
of detailed low-energy structure
in odd nuclei in heavy mass regions.

\subsection{Pairing as a relevant degree of freedom in fission}

\begin{figure}[ht]
\hfill\includegraphics[width=0.95\textwidth]{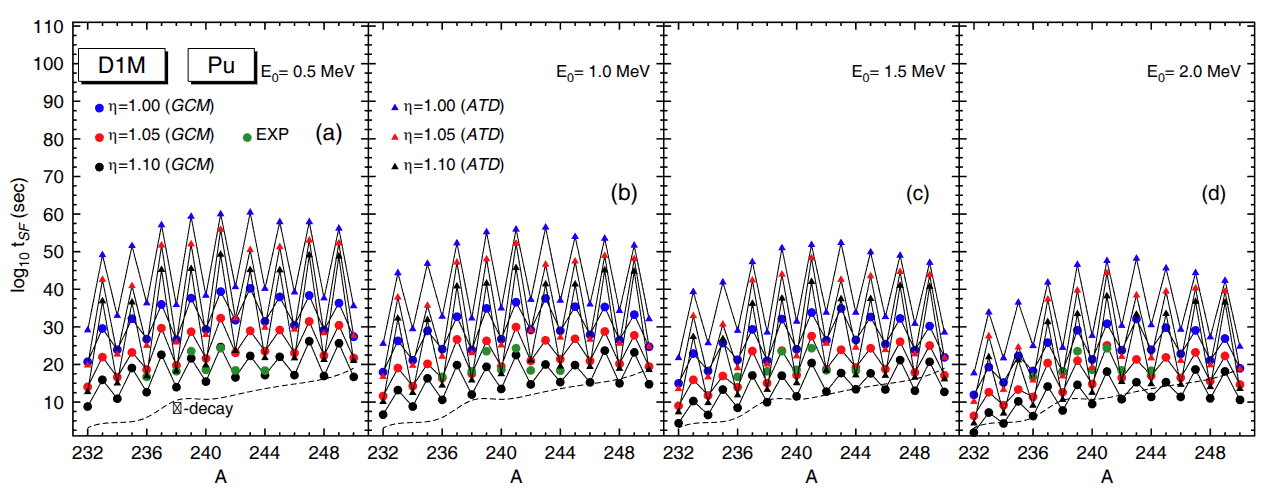}
\caption{Spontaneous fission half-lives, predicted within the GCM and ATDHFB schemes, for the isotopes $^{232-250}$Pu are depicted as functions of the mass number A. Results have been obtained with the normal ($\eta$ = 1.00) and modified
($\eta$= 1.05 and 1.10) Gogny-D1M EDF. Calculations have been carried out with $E_0$ = 0.5 (panel (a)), 1.0 (panel (b)), 1.5 (panel
(c)) and 2.0\,MeV (panel (d)). Results for even-even Pu isotopes are taken from~\citeasnoun{(Rod14a)}. The experimental $t_{SF}$ values are taken from~\protect\citeasnoun{(Hol00)}. In addition, we plot $\alpha$-decay half-lives with short dashed lines (from~\citeasnoun{RodriguezGuzman2017}). See
text for details. }
\label{PuOdd}
\end{figure}

From a dynamic perspective, nuclear fission is nothing more than a tunnelling process through a barrier. The barrier arises from the potential energy of the fission process, parametrised as a function of a relevant fission degree of freedom (traditionally the quadrupole moment). In the traditional tunnelling process analysed in standard quantum mechanics textbooks, only the parameters characterising the barrier's geometric properties matter, since the particle's mass remains constant in time and space. In fission, however, one deals with collective coordinates with an associated collective inertia that depends on the specific value of the collective coordinate driving the fission process. In addition, as it is well established in the literature, collective inertia depends upon the inverse of the square of the pairing gap present in the system, making it very sensitive to the evolution of the pairing degree of freedom as a function of the fission driving coordinate as well as to the pairing contents of the interaction. As a consequence, the tunnelling probability, linked to the fission lifetime, depends strongly not only on the barrier properties (essentially height and width) but also on the amount of pairing correlations in the configurations relevant to fission. This correlation between pairing and fission properties is also evident in traditional approaches to describing fission-fragment mass distributions. In odd systems (as well as in odd-odd ones), the blocking procedure drastically reduces the amount of pairing correlations present in the system, driving the dynamics from a static pairing regime to a dynamic one where correlations beyond the mean field associated with particle number restoration and fluctuations in the pairing gap become essential for a proper understanding of pairing correlations. 

In the traditional mean-field approach to fission in odd nuclei, one often assumes that axial symmetry is preserved and that the $K$ quantum number of the ground state remains constant throughout the dynamic evolution towards fission~\cite{Perez2005,PerezMartin2009}. It is also common to replace the exact blocking procedure by the less demanding Equal Filling Approximation (EFA). Finally, it is customary to treat the dynamics using collective inertia derived from several approximations of the ATDHFB or GCM theories. This step is not fully justified in the ATDHFB case, as the dynamics of the odd system is not adiabatic, cf.~section~\ref{atdhfb}. For each fission configuration, the blocking procedure substantially quenches pairing correlations, increasing the collective inertia. Consequently, the collective action in the WKB approximation increases as compared to neighbouring even-even systems and the spontaneous fission half-life $t_{SF}$ increases accordingly. The main consequence is that the odd-even staggering in $t_{SF}$ observed experimentally is predicted to be a few orders of magnitude larger in the calculations. To analyse the relevant degrees of freedom, we performed calculations that artificially increase the pairing strength and the $E_0$ parameter in the action calculation. Increasing the pairing strength reduces inertia, and increasing $E_0$ effectively reduces the barrier height. Increasing the pairing strength by 10\% and $E_0$ from the standard value of 0.5 MeV to 2 MeV substantially improves agreement with experimental data, but the odd-even staggering remains larger than expected. The explanation is simple: dynamic pairing is far more relevant in odd systems than in even-even ones, as it reduces the collective inertia. Figure~\ref {PuOdd} gives an example calculation.

Given the importance of pairing in fission and the pairing-quenching mechanism inherent to quasiparticle blocking, fission studies must incorporate dynamical pairing in the theoretical description. This can be done by incorporating variation after particle number projection as the method of choice for determining the intrinsic mean-field configuration~\cite{(Giu13b),(Ber19g)} or by including the pairing gap as a dynamic degree of freedom. A consistent theory of collective inertia based on PNP states must then follow. While the first step is feasible over short timescales, the latter requires significant advances in describing collective inertia~\cite{(Giu18a)}. An alternative method is to consider pairing as a relevant degree of freedom in fission~\cite{(Giu14),(Sad14a),zdeb25} and explore the least action principle~\cite{rodriguez2018} including this degree of freedom instead of the least energy, as is customary in most calculations. 

\subsection{Particle-c{\oe}ur coupling in nuclear DFT}

Here we discuss the recently introduced method~\cite{(Ver22b)} for analysing the nuclear electromagnetic moments of odd nuclei in terms of contributions characterising the odd nucleon and the even-particle-number subsystem. \citeasnoun{(Bon26)} introduced a new term, "c{\oe}ur", for the even-particle-number subsystem, distinguishing it from the traditional term "core", which is primarily used in shell-model and valence-space approaches. This terminological distinction aims to clarify the ongoing confusion surrounding the meaning of core polarisation in odd systems.

To cite from~\citeasnoun{(Dob25k)}, {\it traditionally, the odd-particle-number states are described in the language of the so-called
particle–core coupling approach~\cite{(Suh07)}, in which the states of its even-particle-number subsystem
(the core) are coupled with the states of the odd single particle. In this approach, the core polarisation is understood as taking into account not only the ground state but also a suitable linear
combination of the excited core states.
In nuclear DFT, the approach differs because of the self-consistent angular-momentum and
shape polarisation exerted by the blocked quasiparticle on the even-particle-number subsystem, which means that the latter is always distinct from the genuine even system. Such polarisation is not described solely by the spectrum of eigenstates of the even
system. Instead, it is nonperturbatively embedded into a single, modified state, which is different
for every blocked quasiparticle.}

The new approach involves independently restoring symmetry for the odd-particle and c{\oe}ur wave functions, then calculating the diagonal and transition matrix elements of the electric and magnetic operators in both subsystems. These are subsequently recombined to derive those of the odd-particle state. This approach allows scrutiny of the role of the polarised c{\oe}ur contribution. In the future, this type of analysis may become the method of choice for identifying the salient features of electromagnetic moments in odd nuclei.

\subsection{Transition rates and spectroscopic moments along collective rotational bands built on blocked states}

There is a vast body of experimental data on the intra- and inter-band B(M1) and B(E2) transition probabilities within rotational bands of odd nuclei. Within the newly defined methodology for determining nuclear-DFT electromagnetic moments~\cite{(Sas22c),(Bon23c),(Wib25d),(Dob26g)}, these data are now available for systematic comparison with calculations and for evaluating and improving their theoretical description. This may constitute an interesting area of future research~\cite{(But26a)}.

	\begin{figure}[ht]
\hfill\includegraphics[width=0.99\textwidth]{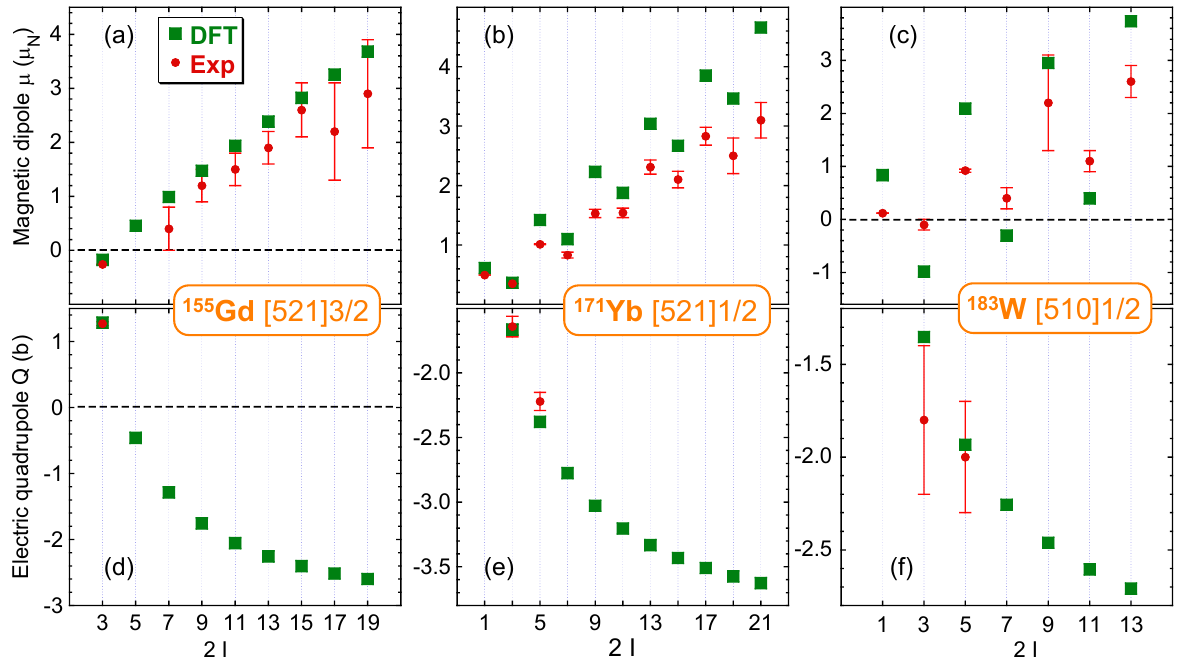}
		\caption{The DFT magnetic dipole $\mu$ (upper panels) and electric quadrupole $Q$ (lower panels) spectroscopic moments compared with experimental data in $^{155}$Gd~\cite{(Stu98)} (left panels), $^{171}$Yb~\cite{(Stu00)} (middle panels), and $^{183}$W~\cite{(Lam92)} (right panels).}
		\label{slide1}
	\end{figure}
Similarly, in all calculated bands, the magnetic dipole $\mu$ and electric quadrupole $Q$ spectroscopic moments of high-spin states are available for analysis~\cite{(But26a)}. Although experimental data are available in only a few cases~\cite{(Lam92),(Stu98),(Stu00)}, the consistency between calculated moments and transition rates warrants attention.

In figure~\ref{slide1}, panels (a)--(c) show the three known cases of experimental spectroscopic magnetic moments compared with DFT calculations performed by~\citeasnoun{(Dob26g)}, while panels (d)--(f) present the corresponding comparison of electric quadrupole moments known only at low spins. It is striking that symmetry restoration alone, without any external parameter, generates the signature staggering in the $K=1/2$ bands in $^{171}$Yb and $^{183}$W, which is usually interpreted in terms of Coriolis coupling. Indeed, the Coriolis coupling is a first-order approximation for describing rotation in the intrinsic frame. Consistent with this interpretation, the $K=3/2$ band in $^{155}$Gd shows no signature staggering. The calculated signature-staggering strength does not match the data well; however, it reproduces the low-spin values. This may indicate the need to address the problem within the dynamical cranking approach, that is, by projecting from rotating states~\cite{(Zdu07c)} rather than from the non-rotating solutions shown in figure~\ref{slide1}. This points to another interesting avenue for future research on odd nuclei.

\subsection{Pseudospin}
	\begin{figure}[ht]
\hfill\includegraphics[width=0.95\textwidth]{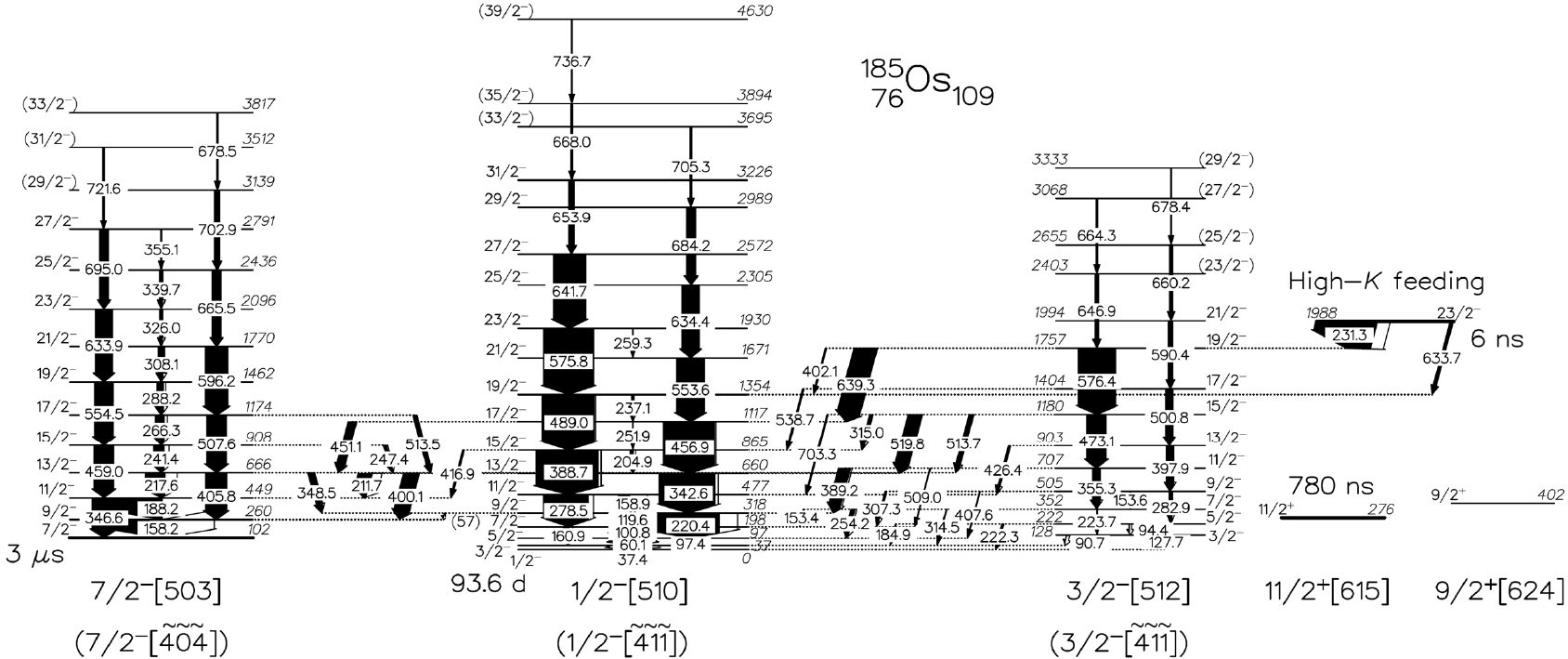}
		\caption{Rotational bands measured in $^{185}$Os. Reprinted with permission from~\protect\citeasnoun{(Whe04)}, figure~1.}
		\label{185OsEXP}
	\end{figure}
	\begin{figure}[ht]
\hfill\includegraphics[width=0.60\textwidth]{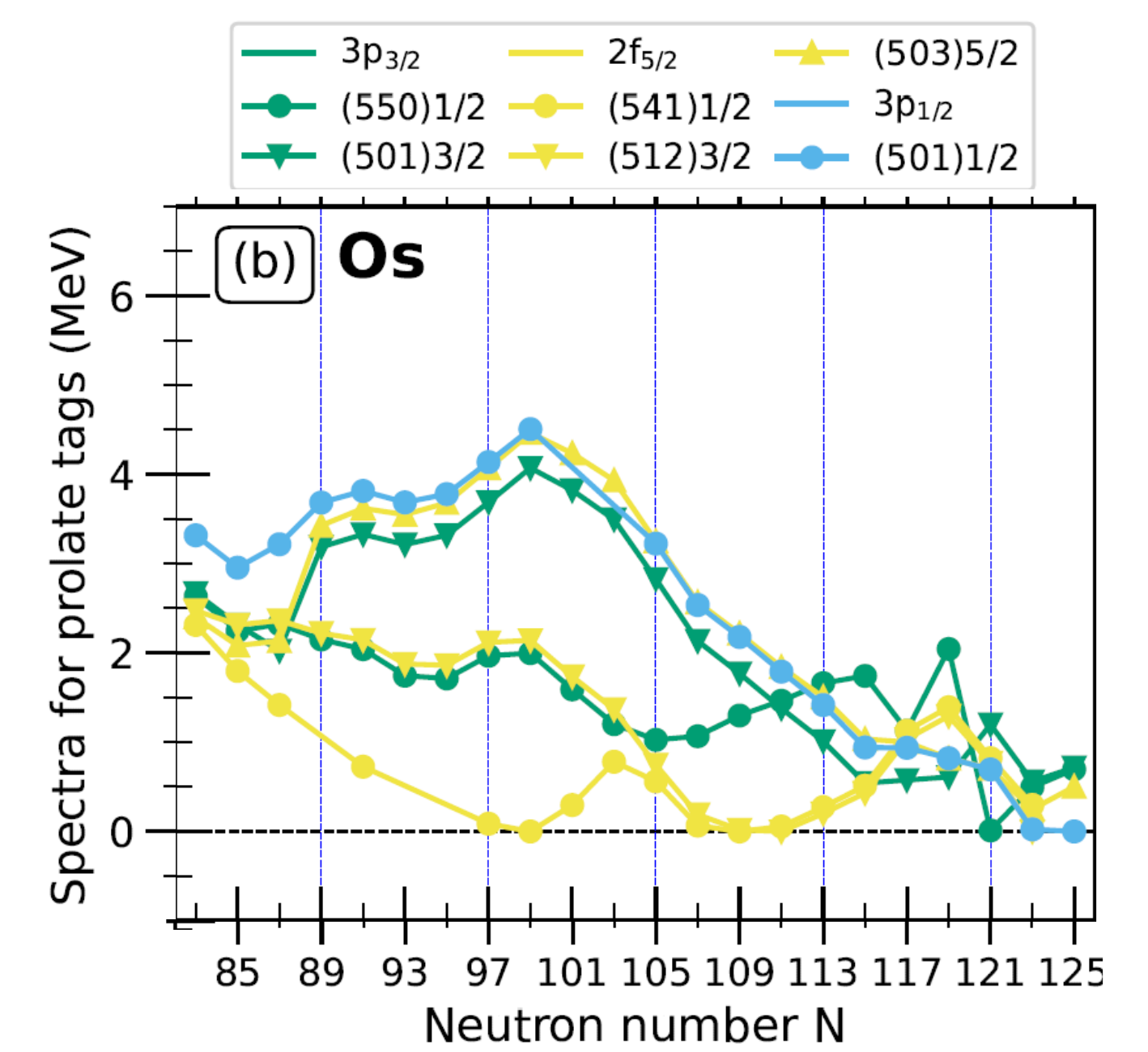}\hspace*{15mm}
		\caption{The nuclear-DFT excitation energies of selected quasiparticle configurations determined in the odd-$N$ osmium isotopes. Adapted with permission from~\protect\citeasnoun{(Dob26g)}, figure~46(b) of the Supplemental Material.}
		\label{185OsDFT}
	\end{figure}

Pseudospin symmetry,~\cite{(Ari69a),(Hec69a)}, and~\cite{(Stu02a),(Bor03a)} and references cited therein, relates to an approximate but noticeable degeneracy of pairs of rotational bands in odd nuclei. Its origins are thought to stem from the degeneracies of specific deformed single-particle states. For example, as shown in figure~\ref{185OsEXP}, transition energies in two $^{185}$Os bands, identified with two Nilsson configurations $[510]1/2^-$ and $[512]3/2^-$, seem to be nearly identical. In the pseudospin scheme, these two configurations are identified with two pseudo-Nilsson labels $[\tilde{4}\tilde{1}\tilde{1}]1/2^-$ and $[\tilde{4}\tilde{1}\tilde{1}]3/2^-$ that correspond to the same harmonic-oscillator quantum numbers, and antiparallel and parallel pseudospins, respectively, suggesting that the energy does not depend on pseudospin. Other examples of this type are common in experimental data.

Recent nuclear-DFT calculations of excited quasiparticle configurations in odd-$N$ open-shell nuclei~\cite{(Dob26g)} reveal multiple clearly degenerate states across long isotope chains. For example, as depicted in figure~\ref{185OsDFT}, low-lying $1/2^-$ and $3/2^-$ configurations have degenerate excitation energies not only in $^{185}$Os ($N=109$), but also across nearly the entire major shell, from $N=85$ to $N=123$. Note that between $N=85$ and $N=103$, the calculated self-consistent Nilsson label of $[510]1/2^-$ corresponds to the tag Nilsson label of $(550)1/2^-$ and then to $(541)1/2^-$, see discussion by~\citeasnoun{(Dob26g)}. At the same time, the calculated self-consistent $[512]3/2^-$ and tag $(512)3/2^-$ Nilsson labels are identical.

Three higher states corresponding to the tag Nilsson labels $[501]1/2^-$, $[501]3/2^-$, and $[503]5/2^-$ also closely follow one another. This near-degeneracy does not fit the pseudospin scheme but relates to the pseudo-SU(3) symmetry introduced by~\citeasnoun{(Ari69a)}. Other examples of two- or three-fold degeneracies are common in the results of~\citeasnoun {(Dob26g)}. Overall, the self-consistent nuclear DFT reveals degeneracies in excitation energies that are probably difficult to explain by relativistic effects or fine-tuning of the spin-orbit strength reviewed by~\citeasnoun{(Stu02a)}. These puzzling results warrant further investigation.

\subsection{QRPA for odd nuclei\label{QRPA}}

First, we recall the consequence of the blocking procedure in the HFB framework, as a starting point for the QRPA equations.
Let us define the HFB state of an odd system obtained by the standard blocking procedure as $\vert{b}\rangle = \beta^\dagger_b \vert0_{qp}\rangle$, 
with the blocked quasiparticle identified with index $b$, as one quasiparticle excitation on the true quasiparticle vacuum $\vert0_{qp}\rangle$. 
The latter solves the HFB equation to minimise the binding energy of the odd-nucleus ground state $\vert b \rangle$. It is possible to define a new set of quasiparticle operators, $\{\eta, \eta^+\}$, such that
\begin{equation}
\begin{split}
\eta_i &= \beta_i\,, \text{ for } i\neq b, \\
\eta^\dagger_i &= \beta^\dagger_i\,, \text{ for } i\neq b, \\
\eta_b &= \beta^\dagger_b,\\
\eta^\dagger_b &= \beta_b \label{beta_eta},
\end{split}
\end{equation}
as to satisfy the useful relation $\eta_i \vert {b} \rangle =0$ for all $i$.
Within this convention, all QRPA equations valid for even-even systems can be easily extended to odd systems. 

For the matrix QRPA approach, the above definition of the $\eta$ quasiparticle operators induces a two-step protocol.
As a first step, the blocked quasiparticle is simply frozen; the single nucleon becomes a spectator and does not contribute to the two-quasiparticle excitation basis. This level of treatment for odd nuclei is simple, easy to implement, and has often been used successfully (see, for example,~\cite{PhysRevC.94.044325,PhysRevC.102.064309,PASQUALATO2023138148}).
Nevertheless, its predictive power is limited, as it may fail to reproduce the excitation energy when the blocked or frozen quasiparticle should contribute to the collective excitations, as shown in Fig. 13 of \cite{PhysRevC.97.054332}.
In the charge-exchange QRPA framework, this first-order treatment is expected not to reproduce the Fermi and Gamow-Teller sum rules. 

The second step is to include the two-quasiparticle excitation involving the blocked one. This treatment modifies the QRPA matrix element depending on the creation- or annihilation-character of the quasiparticle operators used to generate the single-particle excitation basis, namely the two-quasiparticle basis. 
With respect to the reference HFB state, the two excitation types can be expressed as three-quasiparticle ($\beta^+_i\beta^+_j\beta_b$ for $i,j\neq b $) and one-quasiparticle excitations ($\eta^+_i$ for $i\neq b$) on the HFB vacuum.  
In the framework discussed here, the QRPA solutions still represent bosons acting on the quasiparticle vacuum, coupled to a one-fermion excitation associated with the single nucleon. Thus, all restoration formulas, including those applied to transition probabilities, must account for the non-integer value of the ground-state spin; see, for example, the formula given in~\cite{PhysRevC.102.064309}.

The finite-amplitude method (FAM) \cite{Nakatsukasa2007} offers a computationally lighter approach to solving the linear-response problem than traditional matrix QRPA via iterative means. So far, FAM has been applied mostly to even-even nuclei, although it has been used for beta decays in odd nuclei within the equal-filling approximation~\cite{Shafer2016,Ney2020}.
FAM equations in the general case can be derived by starting from the time-dependent HFB (TDHFB) equation
\begin{equation}
i\dot{R}(t) = \left[H(R(t))+F(t),R(t)\right] \,,
\end{equation}
where $R(t)$ is the time-dependent generalized density matrix, $H(R(t))$ is HFB matrix and $F(t)$ is an external perturbing field driving the transition. We can now assume that the time-dependent part of $R(t)$ consists only of small-amplitude oscillations, as
\begin{equation}
R(t) \approx R_0 + \eta \delta R(t) \,,
\end{equation}
where $\eta$ is a small parameter, $R_0$ is generalized density matrix of static HFB solution, and $\delta R(t)$ is oscillating time-dependent part. Similarly, oscillations in the time-dependent
Hamiltonian $H(R(t))$ can be assumed to be small. With these approximations, by utilising the TDHFB equation, the FAM equations in the general case can be derived as
\begin{eqnarray}
(E_i+E_j -\omega)X_{ij}(\omega) &=& -(1-f_i-f_j)(H^{20}(\omega)+F^{20})_{ij} \,,\\
(E_i+E_j +\omega)Y_{ij}(\omega) &=& -(1-f_i-f_j)(H^{02}(\omega)+F^{02})_{ij} \,,\\
(E_i-E_j-\omega) P_{ij}(\omega) &=& (f_i-f_j)(H^{11}(\omega)+F^{11})_{ij} \,,\\
(E_i-E_j+\omega) Q_{ij}(\omega) &=& (f_i-f_j)(H^{11}(\omega)+F^{11})^T_{ij} \,.
\end{eqnarray}
Here, $X_{ij}$, $Y_{ij}$, $P_{ij}$, and $Q_{ij}$ are the FAM amplitudes, depending on the excitation energy $\omega$. Furthermore, $E_i$ denotes the quasiparticle energy and 
\begin{equation}
f_i= \left\{\begin{array}{ll}
   1 &\textrm{for blocked q.p. state } i \\
   0 &\textrm{otherwise }
\end{array}\right. \,. \label{eq:fFAM}
\end{equation}
is connected to quasiparticle blocking. Introducing an odd particle via quasiparticle blocking now yields additional nonzero amplitudes $P_{ij}$ and $Q_{ij}$, which are absent in the even-even HFB case.

The external field $F$ dictates the type of transition described by the FAM equations. For charge conserving transitions (e.g. electromagnetic transitions) the FAM amplitudes are diagonal in particle type, meaning that both indices $i$ and $j$ are of the same particle type. However, the nucleon-nucleon interaction in the oscillating part of the Hamiltonian, $\delta H(\omega)$, still couples proton and neutron fields. For beta decays, one of the indices in FAM amplitudes is a proton index; the other is a neutron index.

As far as we are aware, FAM calculations for odd nuclei have relied so far on the equal filling approximation where factor $f_i$ in equation~(\ref{eq:fFAM}) has been set to $f_i=f_{\bar{i}}=1/2$ for blocked quasiparticle state $i$ and its conjugate state $\bar{i}$, leading the underlying HFB solution to lack spin polarisation effects. As was demonstrated in~\cite{(Sas22c)}, nuclear magnetic moments are often very sensitive to these polarisation effects. Therefore, the nuclear response to spin-dependent operators is likely to become sensitive to spin-polarisation effects, calling into question the validity of the equal-filling approximation.

To describe excitations in odd nuclei beyond the mean-field approximation, it is desirable to account for quasiparticle-vibration coupling (QVC) effects within the QRPA.  This is already known to be a 
critical ingredient for calculations in medium and heavy even-even nuclei~\cite{Litvinova2025} and, as matter of fact, also for odd nuclei \cite{LA.11,CCSB.14,AL.15}.
In addition, the QVC will be needed to address well-known effects in odd nuclei, such as multiplet splitting~\cite{Trache1989} and the fragmentation of single-particle strength~\cite{LA.11,Mizuyama2012,AL.15}.  Methods based on Green functions in particular have been used to formulate microscopic versions of the QVC (e.g.,~\cite{Kamerdzhiev2001,Litvinova2006}).  However, these approaches are computationally demanding and have necessarily been limited in their implementation.  The impact of limitations imposed in calculations for even nuclei will have to be systematically investigated for odd nuclei, so that they can be judiciously applied.  These limitations include: 1) truncation and optimization of the model space~\cite{Tselyaev2018}, especially when dealing with deformed nuclei (this includes the potential need to consider multi-phonon configurations~\cite{Litvinova2025}), 2) whether or not to account for ground-state correlations~\cite{Kamerdzhiev2001,Litvinova2025}, 3) double-counting from QVC implicitly contained in the effective interaction and whether it is best handled by refitting parameters or a subtraction method~\cite{Tselyaev2013,Baldo2015}, 4) whether it is sufficient to use a perturbative treatment of the QVC in powers of the interaction, or if some other approach (e.g., approximations based on cluster decompositions~\cite{Litvinova2022}) should be used instead. The question of ensuring QVC corrections are compatible with any newly developed nuclear EDFs will also need to be addressed~\cite{Tarpanov2014} (see also sections \ref{impact_on_spe} and \ref{qp_energies}).

\newpage
\section{Observables and pseudo-observables\label{sec:observables}}

\subsection{Single-particle observables in covariant density
functional theory: confronting theory with experiment}

\subsubsection{Energies of one-(quasi)particle configurations in odd nuclei:\label{qp_energies}}
The detailed and systematic investigation of relative energies of one (quasi)particle configurations in 
 spherical odd nuclei has been performed at the mean-field level and in (quasi)particle-vibration coupling (QVC) PVC models in~\citeasnoun{LA.11,AL.15}. It was shown that including coupling between single-particle motion and vibration substantially improves the accuracy of describing experimental relative energies of one-(quasi)particle configurations compared with the mean-field level. In addition, it brings calculated spectroscopic factors $S$ into much closer agreement with experiment. In contrast, although being more extensive and systematic, the studies of relative energies of one-quasiparticle configurations in deformed nuclei have been carried out almost exclusively at the mean-field level in~\citeasnoun{AS.11,DABRS.15}. Only in~\citeasnoun{ZBNLRS.22} is the QVC coupling taken into account in calculations of two deformed odd nuclei. Note that these studies have been carried out only with the non-linear
 meson exchange (NLME) CEDFs. The analysis of these results reveals the following open problems:

\begin{itemize} 
\item 
Including PVC (QVC) substantially improves the spectroscopic quality of the description of experimental data. 
However, the underlying functionals are also inadequate for describing single-particle energies, and accounting for beyond-mean-field effects alone will not fully correct these deficiencies. This calls for the
development of the fitting protocols at the beyond-mean-field level and a search for missing terms of the 
functionals sensitive to single-particle properties.

\item 
   The generalisation of PVC/QVC codes to different classes of functionals (such as density-dependent meson exchange and point coupling)
    is
   required to define the impact of the form of the functional on particle-vibration coupling. Further 
   development of the QVC code for deformed nuclei and its systematic confrontation with experimental
   data in different regions is required for a better understanding of the QVC effects in deformed 
   odd nuclei.

\end{itemize}
 
\subsubsection{Alignment properties of the single-particle orbitals in rotating nuclei:}

The alignments $\left<j_x\right>_i$ of the single-particle angular momenta
of different orbitals  along the axis of rotation ($x$-axis) provide state-dependent 
information on the impact of time-odd mean fields on single-particle angular 
momenta~\cite{AR.00}.  They are not directly accessible in experiments, but some 
information on them can be extracted via the analysis of relative effective alignments 
$i^{B,A}_{eff} (\omega) = I_B(\omega) - I_A(\omega)$ defined as the differences 
between spins $I$ of the bands $A$ and $B$ at the same rotational frequency 
$\omega$ with the band $A$ being a reference band~\cite{(Rag93)}. These quantities
depend both on the $\left<j_x\right>_i$ features and on polarization effects. At 
present, a systematic comparison of calculated and experimental relative alignments 
has been carried out only for high-spin normal- and superdeformed rotational bands in 
which the pairing can be neglected in the first approximation and only for the NLME 
functionals (see section 8.3 of~\citeasnoun{VALR.05} and references quoted therein
and~\cite{AF.05}). It proves that the  alignments  $\left<j_x\right>_i$ and thus 
state-dependence  of the impact of time-odd mean fields are reasonably well 
described in the CDFT. 

However, further studies are required to better understand these physical observables.
They need to address the dependence of single-particle alignments on the class of
the functional, the impact of exact treatment of pairing on these observables in
the high spin/weak pairing regime, the impact of pairing on these observables at low
and medium spins, and finally the impact of particle-vibration coupling on these
observables.

\subsection{Observables sensitive to properties of odd nuclei}

Among the observables reachable by DFT approaches, some are more sensitive than others to the fermionic nature of odd-numbered nuclei and symmetry breaking (and symmetry restoration) needed to describe odd and odd-odd nuclei even in their ground states. Here, we will discuss two observables that are widely documented and/or easily accessible by experimentation and for which the different hypotheses of theoretical models make their predictions more or less precise. Thus, in the following two subsections we will discuss the particular cases of binding energies and magnetic moments.

\subsubsection{Binding energies versus single particle energies:}

\begin{figure}[h]
		\centering
\includegraphics[width=0.70\textwidth]{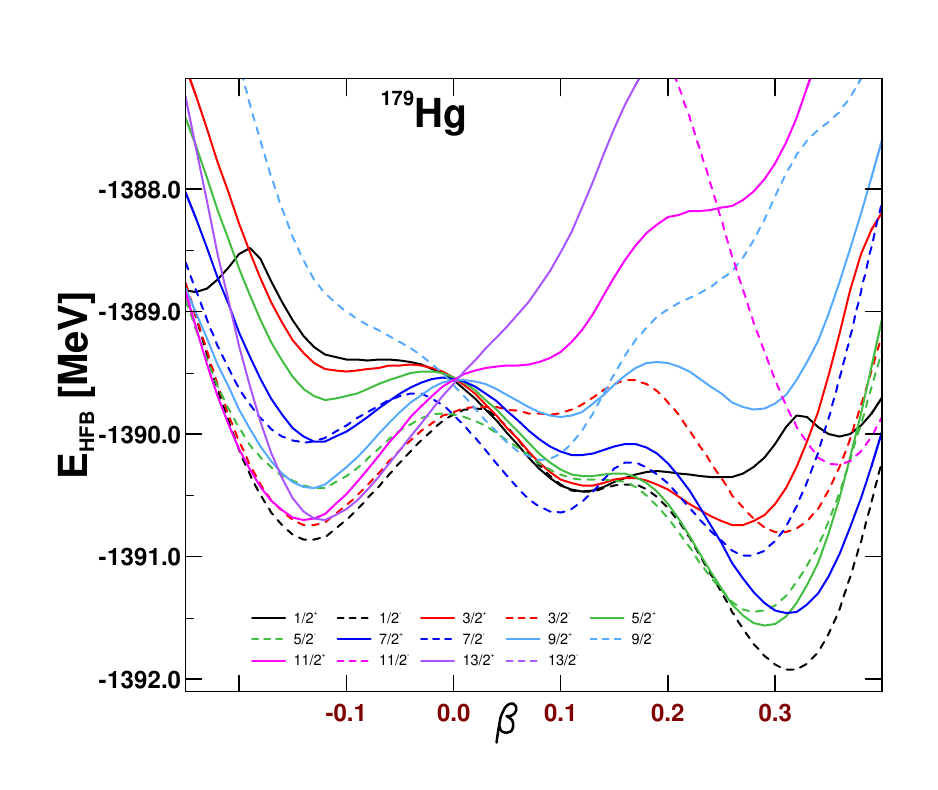}
		\caption{Caption: PES for different blocking in $^{179}$Hg}
		\label{fig:Hg}
	\end{figure}

In addition to the Gallagher-Moszkowski rules for odd-odd nuclei discussed in section~\ref{sect:odd-odd:GM}, we now consider the odd-even energy shift. Its theoretical prediction depends on the complexity and completeness of the model used to describe the ground states.
For quasi-rigid and spherical forms, this odd-even difference constrains effective interactions. For example, the binding energies of the odd and even isotopes of tin are commonly used to adjust the pairing-channel parameters in Gogny interactions; see, for example, \citeasnoun{PhysRevC.21.1568}. But when the theory deals with deformed systems or isotopic/isotonic chains far from shell closures, an arbitrary choice of the orbital occupied, partially or not, by the single nucleon affects the binding-energy calculation. Theories often fail to predict spins and parities of the ground states
in odd-$A$ nuclei \cite{BQM.07,AS.11},
leading to poor predictions 
of global properties such as binding energy and pairing indicators.  We develop this point in section ~\ref{sec:adjust}.
As an example of HFB/exp mismatch, figure~\ref{fig:Hg} shows binding energies versus the axial deformation parameter $\beta$ for different quasiparticle blocking; each is identified by its $K^\Pi$ value and selected as the lowest in energy. 
In this example, HFB+D1M predicts the wrong spin and parity of the Ground state of $^{179}$Hg.
In NNDC, the ground state is a $7/2^-$, but the HFB predicts a $1/2^-$ state with a large prolate shape.

\subsubsection{Magnetic moments:}

\begin{figure}[h]
		\centering
\includegraphics[width=0.70\textwidth]{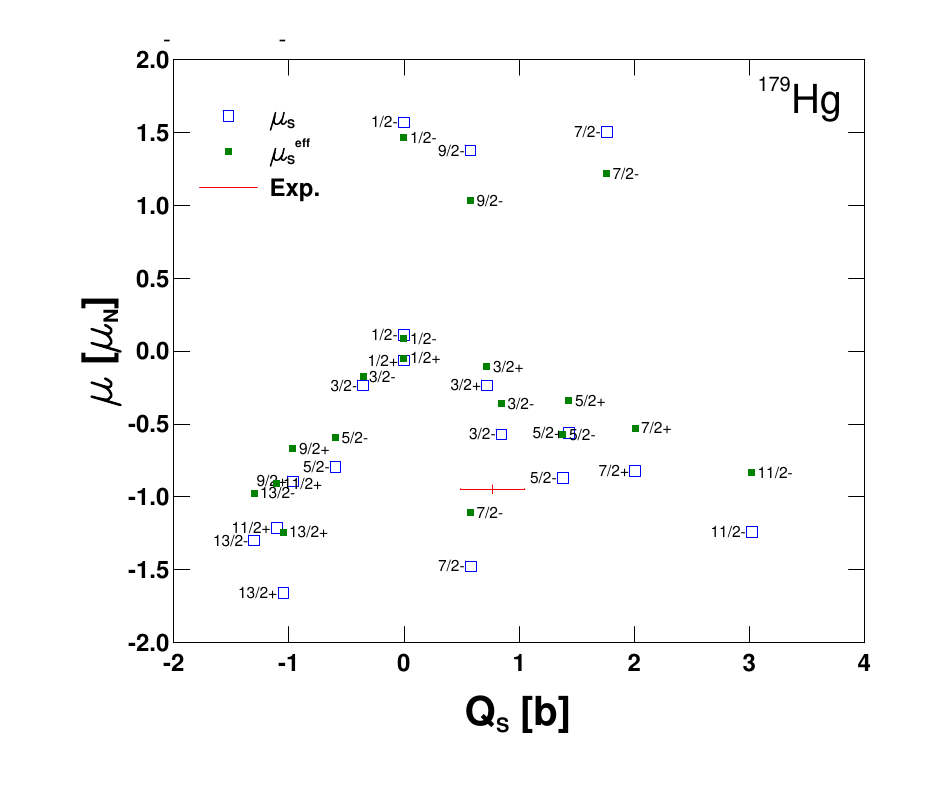}
		\caption{Spectroscopic magnetic moment obtained from different blocking in $^{179}$Hg, each value is related to a minimum in the PESs of figure~\ref{fig:Hg}}
		\label{fig:mntHg}
	\end{figure}

Up to recent works, reviewed by \citeasnoun{(Dob26g)}, magnetic moments have been poorly studied from a theoretical point of view because they are zero by definition for the fundamental states of even-even nuclei. The spectroscopic magnetic moment is of particular interest since it contains detailed structural information for odd and odd-odd nuclei, in contrast to the spectroscopic quadrupole moment $Q$, which is sero by definition for spin values of $J=0$ or 1/2).
  To illustrate this point, figure~\ref{fig:mntHg} shows the spectroscopic moment calculated for every PES minimum plotted in figure~\ref{fig:Hg}.
  This quantity appears to be very sensitive to the underlying structure. It can be used to identify the qp configuration by comparison with experimental data. To bridge the gap between theoretical and experimental data, particularly the impact of the quasi-equal filling approximation\footnote{In the quasi-equal filling technique, HFB equations are solved with the conservation of time-reversal symmetry and the blocking is treated as an equal-filling approximation. In a second step, the $U$ and $V$ Bogoliubov eigenvectors of the only selected blocked qp  ``up or down'' are exchanged before any calculation of average values or overlap}, some phenomenological corrections can be made using effective operators.
  It is well known that magnetic moments are strongly affected by time-odd mean fields (or nuclear magnetism) in the CDFT framework (see Refs.\ \cite{HR.88,YCM.06}].
  We stress that the HFB+D1M values can be improved by breaking time-reversal symmetry, which allows us to account for core polarisation \cite{(Dob26g)} and avoid the effective coefficients introduced in \cite{PhysRevC.104.024328}. For this reason, a new version of the HFB3 code \cite{dubray2025} will soon be available; in this version, time-reversal symmetry can be broken, and blocking techniques are implemented, allowing selection of the configuration $K$ value \cite{Newsome}.

\section{Odd-odd nuclei\label{sec:oddodd}}

\subsection{Masses of odd-odd nuclei}

\begin{figure}[h!]
\includegraphics[width=1.0\textwidth]{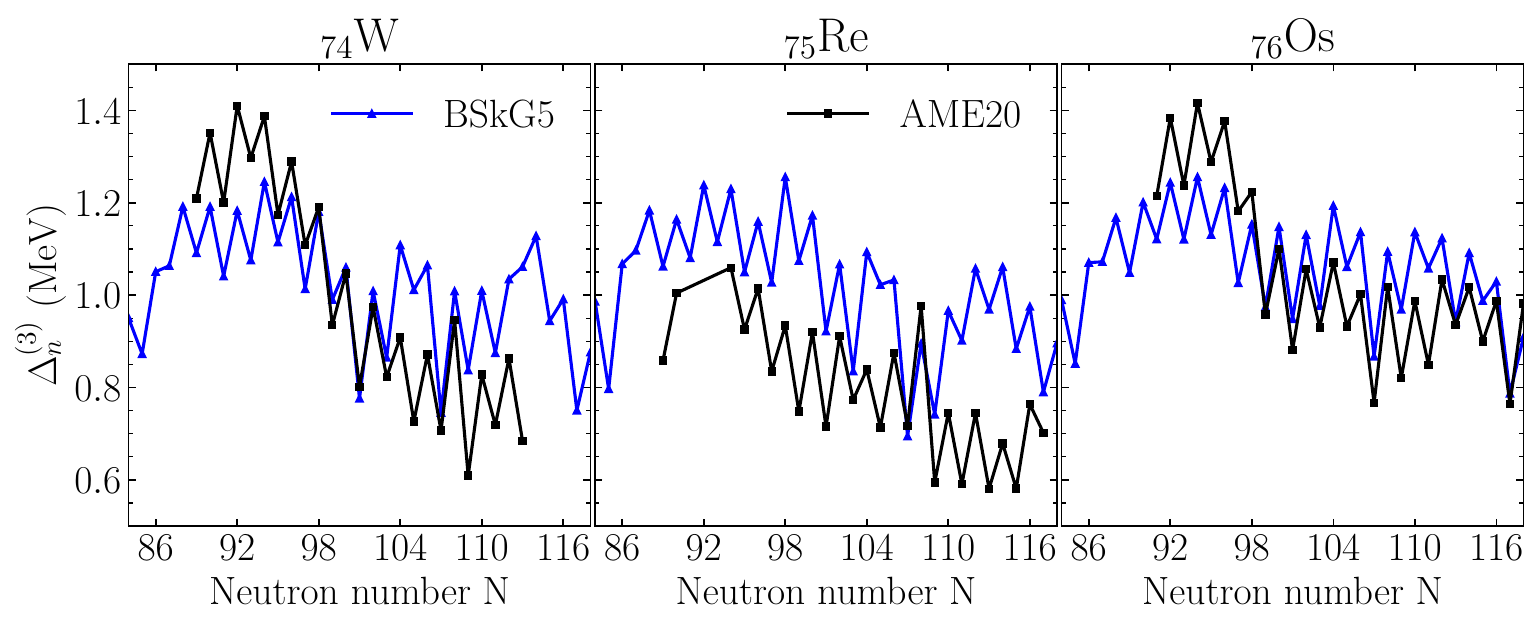}
\caption{Comparison of calculated (BSkG5) and experimental $\Delta^{(3)}_n$ (AME20 ~\cite{(Hua21)})
for the W ($Z=74$), Re ($Z=75$) and Os ($Z=76$) chains.}
\label{fig:odd-odd:masses:}
\end{figure}

In nuclear EDF models, there are contributions to the masses of odd-odd nuclei that are systematically omitted. This effect becomes apparent in mass differences. For example, comparing predictions from the recent BSkG1 mass model with new experimental data in the neutron-rich $Z \simeq 44$ region~\cite{Hukkanen23a} reveals this issue. 
There is a systematic offset in the
calculated $\Delta^{(3)}_n(N,Z)$ of Rh ($Z=45$) when comparing with Ru 
($Z=44$) and Pd ($Z=46$) chains. Adding 250 keV to the binding energies of the
odd-odd Rh isotopes by hand 
(which for this mass region corresponds to the size of the phenomenological 
``average neutron-proton interaction energy'' term $\delta_{np}$ in the FRDM~\cite{Moller16a}) 
makes the $\Delta^{(3)}_n(N,Z)$ smaller by 250 keV for \textit{all} nuclei 
in an isotopic chain with odd $Z$, thereby correcting the discrepancy. We show another example in Fig.~\ref{fig:odd-odd:masses:}: BSkG5 misses the systematic offset in the experimental $\Delta^{(3)}_n$ values of the odd-$Z$ Re isotopes compared to their even-Z W and Os neighbours~\cite{gramsBSkG5}.

Whether this effect could emerge within an EDF model remains open~\cite{Ryssens23i}.
Several mechanisms have been proposed for the possible microscopic origin(s) of such an effect 
that is not captured by any standard EDF modelling:
\begin{itemize}
\item
Missing spin-aligned $S=1$, $T=0$ proton-neutron pairing correlations. It 
has been argued, however, that their inclusion in the EDF modelling is 
unlikely to systematically produce additional binding throughout the nuclear 
chart at the mean-field level~\cite{(Fri07)}, such that a treatment beyond 
the mean field might be necessary. The absence of such pairing correlations 
is also an issue for the modelling of the $\beta$ decay of even-even nuclei 
to odd-odd ones in charge-exchange QRPA, where for the successful description 
of data they have to be added in an ad hoc phenomenological way~\cite{Engel99a}. 
However, the analysis of available experimental data does not provide any
strong fingerprints for the existence of isoscalar $T=0$ proton-neutron pairing
\cite{A.13,FM.14} and successful description of $\beta$-decay does not always
require such type of pairing \cite{MPK.03,LRW.20}.

\item
Missing explicit beyond-mean-field correlations from symmetry restoration or the 
mixing of specific configurations. Having a blocked proton and neutron near the chemical 
potential, pairing correlations of both species are in general quite suppressed in odd-odd nuclei, 
such that dynamic pairing arising in beyond-mean-field calculations may play a systematically
more important role than for even-even systems. 
\item
Missing or incorrectly parameterised time-odd mean fields in the EDF, which is also  
an issue for the modelling of Gallagher-Moszkowski splittings of 2qp states in odd-odd nuclei 
as discussed in section~\ref{sect:odd-odd:GM}.
\item 
Self-interactions and self-pairing effects, section~\ref{self-interaction:spe}, that adversely affect the energies of proton and neutron configurations, and which have never been studied in odd-odd nuclei.
\end{itemize}
Some hints about the systematic deficiencies of EDF calculations for odd-odd nuclei 
and the possible relevance of the beyond-mean-field treatment of $T=0$ and $T=1$ 
pairing correlations can be taken from the study of variational approximations to valence-space 
shell model calculations reported by \citeasnoun{Sanchez2021a}. These suggest that 
neutron-proton pairing is indeed relevant, and also that advanced beyond-mean-field modelling 
including variation after projection on particle number might be needed.

\subsection{Gallagher-Moszkowski splitting\label{sect:odd-odd:GM}}

Limiting the discussion to well-deformed \textit{axial} odd-odd nuclei, there are two 
distinct possibilities to construct two-quasiparticle (2qp) configurations that 
can be identified with the band heads of two distinct strongly coupled bands. 
Using the  asymptotic quantum numbers of the axial Nilsson model for 
illustration, the projection of total angular momentum of the blocked 
proton $\Omega_p$ and blocked neutron $\Omega_n$ on the symmetry axis 
can either be coupled  to be parallel $K_> = \Omega_n + \Omega_p$, or 
to be antiparallel $K_< = |\Omega_n - \Omega_p|$. Doing so, $K$ is 
associated with the observable angular momentum $I$ of the band head. 
Phenomenologically, the energy of these two configurations typically 
differs on the order of about 100 keV. Because of their small energy 
difference but often large difference in total angular momentum, the 
excited state in such a pair can become a long-lived spin 
isomer~\cite{Dracoulis16a}.

With very few exceptions, the relative energy order
of these states follows the empirical 
Gallagher-Moszkowski (GM) rule~\cite{Gallagher58a} that depends on the 
relative orientation of the \textit{spin} contributions to $\Omega_p$ and 
$\Omega_n$ in the 2qp configuration. Decomposing 
$\Omega_q = \Lambda_q \pm \Sigma_q$ into the projection of orbital angular 
momentum $\Lambda_q$ and spin $\Sigma_q$ on the symmetry axis, the 
configuration with larger binding energy has aligned spins, i.e.\ it is 
the $K_>$ configuration when
\begin{align}
\label{eq:GM:par}
\Omega_p = \Lambda_p \pm \Sigma_p \quad 
\text{and} \quad \Omega_n = \Lambda_n \pm \Sigma_n \quad 
\Rightarrow \quad K = \Omega_n + \Omega_p
\end{align}
and the $K_<$ configuration when 
\begin{align}
\label{eq:GM:apa}
\Omega_p = \Lambda_p \pm \Sigma_p \quad 
\text{and} \quad \Omega_n = \Lambda_n \mp \Sigma_n \quad 
\Rightarrow \quad K = |\Omega_n - \Omega_p| \, .
\end{align}
Coupling the proton and neutron angular momenta to the other possible 
$K$ then produces the excited configuration.

In the context of nuclear DFT, the GM splitting has to originate 
from the relative size and sign of time-odd terms, which is most 
straightforwardly analysed in terms of the local Skyrme EDF
(\ref{eq:Skyrme_to}). In a fully time-reversal-breaking 
mean-field calculation of such band head, $\Lambda_q$ and $\Sigma_q$ can 
be associated with integrals over the current and spin densities
\begin{align}
\Lambda_q
& = \int \! d^3r \; \bm{e}_z \cdot \big[ \bm{r} \times \bm{j}_q(\bm{r}) \big] ,
    \\
\Sigma_q
& = \tfrac{1}{2} \int \! d^3r \, s_{q,z}(\bm{r}),
\end{align}
Because of polarisation effects in self-consistent calculations, 
the actual values of these objects will slightly  differ from 
those obtained for the same configuration in schematic models. 
In axial symmetry or in a 3d calculation where the solution 
has good $z$ signature, the corresponding 
$x$ and $y$ components are zero, see Sec.~\ref{alispin}, although 
all components of $\bm{s}_q(\bm{r})$ and $\bm{j}_q(\bm{r})$ 
contribute to the EDF.
Because of the coupling rules (\ref{eq:GM:par}) and (\ref{eq:GM:apa}) 
that depend on the relative orientation of $\Sigma_p$ and $\Sigma_n$, 
but not on the relative orientation of the $\Lambda_q$, the 
GM splitting can clearly be attributed to the time-odd mean fields 
that mediate the effective proton-neutron spin-spin interaction. 
For the Skyrme EDF~(\ref{eq:Skyrme_to}) it is contained in the terms 
that are built out of the densities $\bm{s}_q(\bm{r})$, 
$\bm{T}_q(\bm{r})$, and $\bm{F}_q(\bm{r})$. The effective 
current-current interaction is also different for the two
configurations, but the GM rules indicate that its effect has 
to be smaller than that of the spin-spin terms.

As was pointed out by \citeasnoun{Robledo14a}, the D1S parameterisation 
of the Gogny force systematically produces an energy 
splitting of opposite sign, a failure that the authors attribute to the 
contribution of the density-dependent term to the time-odd mean fields
(assuming that time-odd fields are obtained from the same generator as the 
time-even ones). The same systematic deficiency is found with standard 
Skyrme EDFs \cite{Bonneau24a}, but there are recent indications that results 
might be qualitatively different for extended Skyrme EDFs with density-dependent 
NLO terms~\cite{Stryjczyk25a}. 

The failure of most EDF models to reproduce the sign of 
the GM splitting of course directly affects the capacity of 
these models to predict the correct ground-state 
configuration of odd-odd nuclei.

There is an additional energy shift for states in rotational $K=0$ bands of 
odd-odd nuclei, called the Newby shift~\cite{Newby62a}. Various phenomenological 
central+tensor spin-spin interactions have been proposed to describe 
GM splittings and the Newby shift in schematic models~\cite{BPO.76,NKRSN.94,Ward189a}.
A similar energy splitting, but of opposite sign, is also observed between 
the 2qp states in even-even nuclei~\cite{Gallagher62a}. Contrary to the GM 
splitting, its sign is usually correctly described by nuclear EDFs.
Altogether, these data provide quite stringent constraints on the size and 
sign of time-odd terms that have not been exploited yet in the context of
nuclear EDF models.

\subsection{Isospin symmetry and isobaric analogue states}

The isobaric analogue states (IAS)~\cite{(And61),(Jan65a)} are states or narrow resonances in an $(N,Z)$ nucleus that are strongly excited in neutron or proton knock-out reactions on neighbouring isobars $(N+1,Z-1)$ or $(N-1,Z+1)$, respectively. Their discovery provided the first compelling evidence that isospin symmetry is present in medium-heavy and heavy nuclei. Theoretically, they can be interpreted as states in which a neutron or proton in the parent nucleus is converted into a proton or neutron in the descendant nucleus, without altering any other properties or quantum numbers.

At the level of description that uses correlated states, the study of IAS requires determining the parent, $|N\pm1,Z\mp1\rangle$, and descendant, $|N,Z\rangle$, states, evaluating the corresponding reaction cross section, and comparing the results with the assumption that
\begin{equation}
|N,Z\rangle\simeq{\cal{N}}\hat{T}_\mp|N\pm1,Z\mp1\rangle,
\end{equation}
where $\hat{T}_\mp$ is the isosping raising ($+$) or lowering ($-$) operator and ${\cal{N}}$ is a normalisation constant.

At this point, we note that a rigorous discussion of the IAS relies on a precise definition of the operators $\hat{T}_\mp$, and that definitions in the literature vary because isospin symmetry is (weakly) broken. Therefore, the safest and most obvious definition should be universal and not depend on states $|N\pm1,Z\mp1\rangle$ or $|N,Z\rangle$. One such definition uses a fixed basis of single-particle states labelled by the space-spin quantum numbers $\alpha$ (e.g., of the harmonic oscillator), each equipped with an additional isospin quantum number $\tau$ equal to either neutron ($\tau=n$) or proton ($\tau=p$). In this case, we have,
\begin{equation}
\label{Tpm}
\hat{T}_+=\sum_\alpha a^+_{\alpha{}n} a_{\alpha{}p}, \quad
\hat{T}_-=\sum_\alpha a^+_{\alpha{}p} a_{\alpha{}n}, \quad
\hat{T}^\dagger_+=\hat{T}_-, \quad
[\hat{T}_+,\hat{T}_-]=2\hat{T}_z=\hat{N}-\hat{Z},
\end{equation}
that is, the isospin operators obey the exact SU(2) algebra.

At the mean-field or DFT level, of interest here, the IAS study can be further simplified. We focus on the case where both $N$ and $Z$ are odd; that is, $|N,Z\rangle$ is a state of an odd-odd nucleus, and $|N\pm1,Z\mp1\rangle$ are the states of its even-even isobaric neighbours.

First, let us consider the false-vacuum state $|\Phi_{N,Z}^{\text{vac}}\rangle$ of the odd-odd $(N,Z)$ nucleus, which in its canonical basis~\cite{(Rin80b),(Sch19b)} can be represented as follows,
\begin{equation}
\label{false-vacuum}
|\Phi_{N,Z}^{\text{vac}}\rangle=
\prod_{\alpha>0{}n}\left(u_{\alpha{}n}+v_{\alpha{}n}a^\dagger_{\alpha{}n}a^\dagger_{{\tilde\alpha}{}n}\right)
\prod_{\beta>0 {}p}\left(u_{\beta {}p}+v_{\beta {}p}a^\dagger_{\beta {}p}a^\dagger_{{\tilde\beta }{}p}\right)
|0\rangle,
\end{equation}
where $|\Phi_{N,Z}^{\text{vac}}\rangle$ is assumed to be a product of the neutron and proton states, and $\alpha>0$ and $\beta>0$  denote the products over one state of each neutron $(a^\dagger_{\alpha{}n},a^\dagger_{{\tilde\alpha}{}n})$ or proton $(a^\dagger_{\beta {}p},a^\dagger_{{\tilde\beta }{}p})$ canonical pair. Note that for states that break time reversal, the canonical pairs do not correspond to time-reversed states, but the single-(quasi)particle spectrum is nevertheless always composed of canonical pairs.

Already at this point, we observe that for the broken isospin symmetry, that is, for the Coulomb interaction included or for asymmetric systems $N\neq{}Z$, the neutron and proton canonical states are different, and thus the canonical bases are inappropriate for use in the definition of the isospin operators in equation~(\ref{Tpm}). In addition, the canonical bases for self-consistent doubly blocked states in odd-odd nuclei,
\begin{equation}
\label{doubly_blocked}\hspace*{-2mm}
|\Phi_{N,Z}\rangle=
a^\dagger_{\alpha'_0n}\!\!\prod_{\alpha'\neq\alpha'_0>0{}n}\!\!\left(u_{\alpha'{}n}+v_{\alpha'{}n}a^\dagger_{\alpha'{}n}a^\dagger_{{\tilde\alpha}'{}n}\right)
a^\dagger_{\beta' _0p}\!\!\prod_{\beta' \neq\beta' _0>0 {}p}\!\!\left(u_{\beta' {}p}+v_{\beta' {}p}a^\dagger_{\beta' {}p}a^\dagger_{{\tilde\beta' }{}p}\right),
\end{equation}
are different than those for the false-vacuum state~(\ref{false-vacuum}), and differ for every blocked state $\alpha'_0n$ and ${\beta'_0p}$. As a final point, the canonical bases for the even-even isobaric neighbours, $|\Phi_{N+1,Z-1}^{\text{vac}}\rangle$ and $|\Phi_{N-1,Z+1}^{\text{vac}}\rangle$, differ again.

We observe that, in general, analyses of isospin symmetry based on the canonical (or BCS) bases for the quasiparticle states are not justified.\footnote{An easy exercise of reducing this conclusion to single-particle bases of unpaired states is left to the reader.} On the one hand, correct implementations of such analyses performed directly in the external (harmonic oscillator) basis within equation~(\ref{Tpm}) already exist~\cite{(Sat12a),(Bac19)}. On the other hand, such analyses and conclusions based on proton-neutron mixed single-particle~\cite{(Sat13e),(She14b)} or quasiparticle~\cite{(Per04c),(Fue26)} states remain an interesting subject for future research. Note that the latter case corresponds to states that are not products of neutron and proton states. However, even for the isospin-symmetry-broken solutions, they still form proton-neutron mixed canonical pairs.

To proceed, one can nevertheless make a bold simplifying assumption. Analytic expression giving the $T=1$ IAS state in an odd-odd $N=Z$ nucleus $|\Phi_{N,Z}^{\text{IAS}}\rangle$ was derived by~\citeasnoun{(Bac25a)} under the  assumption that the $|\Phi_{N\pm1,Z\mp1}^{\text{vac}}\rangle=\prod_{\alpha>0n}(u_{\alpha{}n}^\pm + v_{\alpha{}n}^\pm a^\dagger_{\alpha{}n} a^\dagger_{{\tilde\alpha}n})\prod_{\beta>0p}(u_{\beta{}p}^\pm + v_{\beta{}p}^\pm a^\dagger_{\beta{}p} a^\dagger_{{\tilde\beta}p} )|0\rangle$  canonical bases are identical, that is,
 \begin{eqnarray}\label{PsiIAS}
\label{IAS}\hspace*{-25mm}
   |\Phi_{N,Z}^{\text{IAS}}\rangle\!\!&=&\sum_{\alpha_0>0} u_{\alpha_0p}^+v_{\alpha_0n}^+{\cal S}^{T=1}_{\alpha_0} 
     \prod_{\alpha>0\neq\alpha_0}(u_{\alpha{}n}^+ + v_{\alpha{}n}^+a^\dagger_{\alpha{}n} a^\dagger_{{\tilde\alpha}n})
     \prod_{\beta>0\neq\alpha_0}(u_{\beta{}p}^+ + v_{\beta{}p}^+a^\dagger_{\beta{}p} a^\dagger_{{\tilde\beta}p})
     |0\rangle \nonumber \\\hspace*{-25mm}
   \!\!&=&\sum_{\alpha_0>0} u_{\alpha_0n}^-v_{\alpha_0p}^-{\cal S}^{T=1}_{\alpha_0} 
     \prod_{\alpha>0\neq\alpha_0}(u_{\alpha{}n}^- + v_{\alpha{}n}^-a^\dagger_{\alpha{}n} a^\dagger_{{\tilde\alpha}n})
     \prod_{\beta>0\neq\alpha_0}(u_{\beta{}p}^- + v_{\beta{}p}^-a^\dagger_{\beta{}p} a^\dagger_{{\tilde\beta}p})
     |0\rangle \nonumber \\\hspace*{-25mm}
\end{eqnarray} 
where
\begin{equation}
   {\cal S}^{T=1}_{\alpha_0} =  \frac{1}{\sqrt{2}} 
   \left(a^\dagger_{\alpha_0p}a^\dagger_{{\tilde\alpha}_0n} + 
  a^\dagger_{\alpha_0n}a^\dagger_{{\tilde\alpha}_0p}
   \right)
\end{equation}
represents a $T=1$ proton-neutron pair.
We note that the equality of two forms of $|\Phi_{N,Z}^{\text{IAS}}\rangle$ given in equation~(\ref{PsiIAS}) is based on an assumed symmetry of the occupation factors,
$v_{\alpha p}^-=v_{\alpha n}^+$ and $v_{\alpha n}^-=v_{\alpha p}^+$, in the even-even $T_z=\pm 1$ mirror isobars $|\Phi_{N\pm1,Z\mp1}^{\text{vac}}\rangle$.

As concluded by~\citeasnoun{(Bac25a)}, the IAS state
is not a product state associated with a doubly
blocked even-even vacuum~(\ref{doubly_blocked}), but rather a coherent superposition of blocked $T = 1$ proton-neutron pairs atop the
even-even quasiparticle vacua. Therefore, determining which properties of the doubly blocked state or the IAS better align with experimental data remains an important challenge for future research.

\section{New theoretical developments\label{Theory}}

\subsection{Gaussian Overlap Approximation for odd nuclei}

Based on the alispin idea, see~\citeasnoun{(Sch10b)} and section~\ref{alispin}, \citeasnoun{(Roh15a),(Roh20)} derived basic equations of the generator coordinate method (GCM) in the Gaussian overlap approximation (GOA) for odd nuclei. The fundamental principle of that derivation was that for the collective motion of odd nuclei, the phase space of collective degrees of freedom must be supplemented by that of the angular-momentum orientation of the odd particle, which is described by a two-dimensional alispinor of two Kramers-degenerate states s.p.\ states. As in the well-known GOA for even-even nuclei~\cite{(Une76a)}, the Collective Bohr Hamiltonian is defined by the Hamiltonian and overlap kernels between the blocked states and takes the form of a 2$\times$2 differential operator acting in the alispinor space. 

So far, no practical implementations of these ideas exist, neither in simplified collective dynamics nor in the full 5D collective space of the Bohr Hamiltonian.

\subsection{Adiabatic Time-Dependent Hartree-Fock(-Bogoliubov) Approximation for odd nuclei\label{atdhfb}}

The standard adiabatic approximation for the collective motion of even-even nuclei depends on the assumed slowness of the time-dependent evolution described by the time-dependent Hartree-Fock(-Bogoliubov) [TDHF(B)] method~\cite{(Rin80b),(Sch19b)}, which is expressed through the smallness of the time-odd correction to the quasi-stationary, time-even density matrix~\cite{(Bar78b)}. For odd nuclei, the stationary solution of the HF(B) equations yields a density matrix with both time-even and time-odd components, and the fundamental principle of the standard adiabatic approximation does not apply.

A consistent formulation of the adiabatic approximation for odd nuclei remains an open area for future research.

\section{Conclusions}

Making predictions is difficult, especially about the future\footnote{The quote is most famously attributed to American baseball legend Yogi Berra; Niels Bohr, the Nobel Prize-winning Danish physicist; Samuel Goldwyn, the Hollywood film studio executive; or K. K. Steincke, a Danish politician who recorded it as a known political pun in the Danish parliament during the late 1930s (long before Berra's widespread usage).}. Nevertheless, this White Paper attempts to outline future prospects for nuclear DFT studies of odd and odd-odd nuclei, a subject that is gaining important momentum. We present numerous open problems and possible study directions, hoping that they can become a source of inspiration for nuclear theorists in their scientific careers and grant applications.

\section*{Acknowledgments}
The authors thank the \textit{Espace de Structure et de r{\'e}actions Nucl{\'e}aires Th{\'e}orique} (ESNT) Saclay, France, for its kind hospitality during the workshop on ``The microscopic modelling of odd-mass and odd-odd nuclei'' that stimulated the writing of this paper.

This work was partially supported by the STFC Grant Nos.~ST/P003885/1 and ~ST/V001035/1, and by a Leverhulme Trust Research Project Grant.
This work was supported in part by Fonds de la Recherche Scientifique (F.R.S.-FNRS) under the MIS Project nr. 40028446. W.R. is a Research Associate of the F.R.S.-FNRS. 
The work of K.N.\ was supported by JSPS KAKENHI Grant No. JP25K07293.
The work of A.A.\ was supported by the U.S.\ Department of Energy, Office of Science, Office of Nuclear Physics under Award No.\ DE-SC0013037.
The work of L.M.R.\ was supported by the Spanish Agencia Estatal de Investigacion (AEI) of the Ministry of Science and Innovation under Grant Nos.\ PID2021-127890NB-I00 and PID2024-159559NB-C21.
The work of B.B.\ has received funding from the European Research Council under the European Union’s Horizon Europe Research and Innovation Programme (Grant Agreement No.\ 101162059).
The work of R.R.-G.\ is funded by Nazarbayev University under the Faculty Development Competitive Research Grants Program (FDCRGP) for 2025-2027, Grant 040225FD4712.  The work of W.Y. was performed under the auspices of the U.S. Department of Energy by Lawrence Livermore National Laboratory under Contract No. DE-AC52-07NA27344.
The work of M.K. was supported by the Research Council of Finland (The Centre of Excellence in Neutron-Star Physics, Grant No. 374066).

We thank the CSC-IT Centre for Science Ltd., Finland, and the IFT Computer Centre at the University of Warsaw, Poland, for allocating computational resources. This project was partly undertaken on the Viking Cluster, which is a high-performance computing facility provided by the University of York. We are grateful for computational support from the University of York High-Performance Computing service, Viking, and the Research Computing team.     
Part of the computations were performed using high-performance computing resources from CC-IN2P3 of the CNRS.  
S.P.\ acknowledges the computational resources made available by CCRT (Topaze supercomputer).
Our research also benefited from computational resources on the Tier-1 supercomputers Zenobe and Lucia of the F{\'e}d{\'e}ration Wallonie-Bruxelles, infrastructure funded by the Walloon Region under grant agreement nr 1117545. Further computational resources have been provided by the clusters Consortium des {\'E}quipements de Calcul Intensif (C{\'E}CI), funded by F.R.S.-FNRS under Grant No. 2.5020.11 and by the Walloon Region.

We thank Grammarly\textsuperscript{\textregistered} for its support with English writing.

\bibliographystyle{jphysicsB-withTitles}
% Odd_White_Paper added by LR
%\begin{thebibliography}{10}
%\end{thebibliography}

\end{document}